\documentclass[twocolumnprb,aps,prb,amsmath,amssymb,showpacs,lengthcheck]{revtex4-2} 

\usepackage{color}
\usepackage{graphicx}
\newcommand{\bvec}{\mathbf}

\begin{document}

\title{Optical amplification in a CDW-phase of a quasi-two dimensional material}
\author{Stephan Michael}
\author{Hans Christian Schneider}
\affiliation{Department of Physics and Research Center OPTIMAS, TU Kaiserslautern, P.O. Box 3049, 67653 Kaiserslautern, Germany}

\begin{abstract}
We study theoretically the quenching of a charge-density-wave phase in a model system of a quasi-two dimensional material, which includes both electron-hole and electron-phonon interactions leading, respectively to excitonic and coherent-phonon contributions. We discuss how interaction processes affect anomalous expectation values and present a microscopic dynamical picture of the quenching of the phase. 
We use projection techniques to illustrate the time-dependent appearance of additional bands and anomalous expectation values. We propose an optical amplification effect with a high modulation frequency in the mid-infrared regime.
\end{abstract}

\maketitle

\section{Introduction}\label{Introduction}

Many aspects of the charge-density-wave (CDW) phase in two-dimensional materials like transition-metal dichalcogenides (TMDC) have been studied in some detail,\cite{gruner1988dynamics,kohn1967excitonic,aebi2001search,chan1973spin,hughes1977structural,rossnagel2011origin,chhowalla2013chemistry} but their origin is still under debate.~\cite{zhu2015classification,flicker2015charge,cho2016unconventional,leroux2018traces,nakata2018anisotropic,ueda2020correlation} TMDCs are layered materials with the metal atoms sandwiched between the chalcogen atoms and a van der Waals gap with reduced interlayer bonding between the layers. Therefore, these materials intrinsically have a two-dimensional character and the bulk materials are also called quasi two-dimensional.\cite{aebi2001search} For example, CDW transitions in single-layer $TiSe_{2}$\cite{chen2015charge} as well as three-dimensional structures with a hidden two-dimensional order\cite{chen2016hidden} were reported.

Materials like 1\textit{T}-TaS$_{2}$, 2\textit{H}-TaSe$_{2}$, and 1\textit{T}-TiSe$_{2}$ belong to the most studied CDW compounds.~\cite{rossnagel2011origin,porer2014non,zhang2016unveiling} In particular for 1T-TiSe$_{2}$ there are arguments that an electron-lattice deformation~\cite{rossnagel2002charge,karam2018strongly,hughes1977structural,hellgren2017critical} or an excitonic insulator mechanism~\cite{monney2009spontaneous,monney2010temperature,cazzaniga2012ab,monney2012electron,li2007semimetal,monney2011exciton,kogar2017signatures,chen2017reproduction,pasquier2018excitonic,golez2016photoinduced} is responsible, but the prevailing explanation is a combination of both.~\cite{kidd2002electron,van2010exciton,monney2012electron,phan2013exciton,watanabe2015charge,bianco2015electronic,monney2016revealing,kogar2017signatures,chen2017reproduction,pasquier2018excitonic,karam2018strongly,maschek2016superconductivity,singh2017stable,hellgren2017critical,lian2020ultrafast}  Besides excitonic and Peierls insulators,  other metal-insulator transitions such as Mott insulators have been extensively investigated, too.~\cite{cheiwchanchamnangij2012quasiparticle,ghatak2011nature,moon2018soft,chernikov2015population} 

Progress in time- and angle-resolved photoemission spectroscopy (trARPES) has made it possible to investigate the ultrafast non-equilibrium material response after optical excitation.~\cite{rohwer2011collapse,sobota2014ultrafast,eich2014time,smallwood2016ultrafast,lin2021coherent} Besides transition-metal dichalcogenides (TMDC),~\cite{chhowalla2013chemistry,voiry2015phase,steinhoff2016nonequilibrium} also graphene,~\cite{gierz2017probing,gierz2013snapshots,stroucken2013screening,mihnev2016microscopic,kadi2015impact} and other two-dimensional materials,~\cite{bhimanapati2015recent,xia2014two,chernikov2015population,chen2016ultrafast,steinhoff2017excitons} have been investigated.

Changing the symmetry of a material via optically induced phase transitions offers new ways to manipulate material properties on ultrafast timescales.~\cite{wall2012ultrafast,mizokawa2009local} Research about the ultrafast response of two dimensional materials like TMDC in connection with their rich electronic phase diagrams, e.g., superconductivity~\cite{sykora2009coexistence,kim2015mechanism} or CDW phases, may well be important for understanding the basic physics for the design of future ultrafast (optoelectronic or optospintronic) devices.~\cite{ritschel2015orbital,gu2016ultrafast} In this context a lot of current research around the optical excitation and melting of CDW phases was done.\cite{mathias2016self,yang2020bypassing,tanabe2018nonequilibrium,burian2020structurally,chavez2019charge,werdehausen2018photo,perfetto2020time,mathias2016self} Furthermore, spin-selective excitations,\cite{kohler2020formation} strain effects on CDWs\cite{cohen2019effect} and also $TiTe_{2}$/$TiSe_{2}$ Moire Bilayer \cite{lin2021coherent} were investigated and $TiSe_{2}$ was suggested as a saturable absorber.\cite{wei2019ultrafast} Also, research in initialization methods,\cite{karlsson2018generalized, tuovinen2019adiabatic} first attempts to describe pump/probe experiments\cite{freericks2017theoretical} or the collective excitations of an exciton insulator in a cavity\cite{lenk2020collective} was published. 

Recently, there has been a lot of interest in novel optoelectronic materials and photonic devices and also progresses on TMDC lasers are reported. The discussion focuses mainly on materials like $MoTe_{2}$, $MoS_{2}$, $MoSe_{2}$, $WS_{2}$, $WSe_{2}$, $WTe_{2}$ with band gaps in the eV regime,\cite{zhao2020strong,wang2020threshold,wutwo,perea2020microscopic,meckbach2020ultrafast,chen2015q,gies2021atomically,hahn2021influence,lohof2018prospects,li2019optical} because those materials are the most promising candidates for photonic devices. They possess large exciton binding energies and the maximum achievable gain of those materials exceeds ordinary semiconductor materials like GaAs.\cite{lohof2018prospects} However, the potential of CDW-phases in transition-metal dichalcogenides (TMDC) with band gaps in the infrared or mid-infrared regime has not yet been investigated to the best of our knowledge. These phases exist in a wide temperature range,~\cite{coleman1988scanning} but increased research in the material design (e.g. material composition, sourounding layers, strain) of two-dimensional materials might further improve the temperature stability of such phases (and potential devices) and even if it turns out that they do not achieve comparable performance to materials like Mott insulators, they might be superior to ordinary infrared or mid-infrared photonic devices like quantum-cascade lasers. Especially, the fast kinetic and spectral response of those materials and the associated fading-in and fading-out of bands during a CDW phase transition could be an opportunity to develop new ultra-fast devices (with e.g. high modulation frequencies).

In this paper, we study theoretically the quenching of a CDW phase in a quasi-two dimensional material due to optical excitation. Our model of the quasi-2D material is designed to display important properties of a phase-change material by including electron-hole and electron-phonon coupling and the accompanying excitonic and coherent-phonon effects. By using a tight-binding band structure, the model avoids the complexity of an ab-initio description of the coupled carrier-carrier and carrier-phonon interactions. The CDW transition is realized in the model by the appearance of anomalous expectation values, and we show how interaction processes affect these and other quantities during the quenching of the phase. We describe both normal and CDW phase consistently in one Brillouin zone. By using projection techniques we can visualize the time-dependent fading-in and fading-out of additional bands and anomalous expectation values during the phase change dynamics. Our theoretical approach suggests the possibility of an high-frequency optical amplifier in the mid-infrared regime based on suitable CDW materials. 

The paper is organized as follows. In Sect.~\ref{Model} we introduce a model composed of a tight-binding band structure where the quasiparticle band dynamics is described by an excitonic and a lattice contribution. Afterwards we set up the equations of motion for the non-equilibrium carrier dynamics including the optical excitation, the carrier-carrier as well as the carrier-phonon scattering contribution. Numerical results are presented in Sect.~\ref{Results}. We discuss first the self-consistent mean-field result for the equilibrium CDW phase, and then the non-equilibrium dynamics that occurs after an impulsive optical excitation and leads to a quenching of the phase. We further investigate the potential of the setup for optical amplification. We conclude the paper in Sect.~\ref{Conclusion}.

\section{Model}\label{Model}

\subsection{Tight-binding model}\label{TightBindingModel}

We investigate a quasi-two dimensional model system that has the potential to establish a CDW phase based on a normal-phase with an indirect band overlap on the Fermi surface, i.e., the normal phase displays a band nesting effect that might favor a charge-density wave instability. The model shares important properties with a transition metal dichalcogenide.

The band structure of transition-metal dichalcogenides can be calculated with good accuracy from optimized tight-binding models. For the prototypical material TiSe$_2$ a tight-binding hamiltonian has been used in Refs.~\onlinecite{van2010exciton,yoshida1980electron} to describe the band structure of the normal phase. Based on these results, we describe the band structure of such a transition metal dichalcogenide by taking into account the three $t_{2g}$ orbitals from the transition metal atom together with six p-orbitals from the two chalcogen atoms in the unit cell of the normal phase, and no spin-orbit coupling. One obtains from such a tight-binding Hamiltonian, the band structure in Figure~\ref{figure1}(a). The on-site energies and tight-binding coupling matrix elements are collected in Appendix~\ref{AppendixA}.

As our goal is the modeling of the \emph{field-induced dynamics} we introduce the following simplification to the single particle properties that allows us to retain  the important characteristics of the carrier and band-dynamics close to the relevant high-symmetry points.
Instead of the tight-binding approach that leads to the rather realistic band structure shown in Fig.~\ref{figure1}(a) we thus take only the bands close to the Fermi energy into account, which, in the normal phase, is the highest chalcogen-like band and the lowest transition-metal-like band around the Fermi surface, see Figure~\ref{figure1}(b). The equations we derive and collect below are not limited to such a two-band model, but one can use a tight-binding hamiltonian with an arbitrary number of orbitals as a starting point.

\subsection{Quasiparticle band dynamics}\label{QuasiparticleBandDynamics}

Our goal is to describe the electronic dynamics that initiate and accompany a change to a CDW phase that arises as a higher-order commensurable phase from the normal phase. A commensurate CDW phase is usually described using a BZ of reduced size, which is obtained by backfolding the normal-phase BZ via ordering wave vectors~$\bvec{Q}$. If we assume a hexagonal crystal structure and the CDW resulting from a $2\times2$ reconstruction the ordering wave vectors are nesting vectors $\bvec{Q}_{\varGamma M}$ that point from the $\Gamma$-point to the three inequivalent $M$-points.
The CDW phase BZ of reduced size can be embedded in the BZ of the normal phase, and so that a sum over all $k$ vectors in the  normal phase BZ can be described by using a set of momentum vectors $\bvec{Q}_{\varGamma M}$ as an extension of the sum over $\bvec{k}$ from the CDW BZ.
Alternatively, one can introduce a label to transform the sum over all nesting vectors $\bvec{Q}_{\varGamma M}$ into the index $\eta$ and form a ``superindex'' $\lambda$ by including this label with the band index. We will use the second possibility in the following, which is equivalent to introducing a generalized matrix propagator for CDW systems in Ref.~\onlinecite{mattuck1968quantum}. 
Further, to describe the relevant band dynamics around the high-symmetry $\Gamma$- or $M$-points of the normal-phase and obtain a numerically tractable model, we restrict the generalized matrix in the two-band model to a four-dimensional matrix where $\bvec{Q}_{\varGamma M}$ is that respective ordering wave vector that points from $\varGamma$ to the associated $M$-point in the normal-phase BZ, see Figure~\ref{figure1}(b). In the present model system this results in the index $\eta=0,1$, where the index $\eta$ distinguishes the vectors $\bvec{k}+\eta\bvec{Q}_{\varGamma M}$.

\subsubsection{Effective Two-Band Hamiltonian in Orbital and Band Basis}\label{InitialHamiltonian}

Based on the description of the single-particle properties of our model given in the previous subsections, we choose the two-band  Hamiltonian of the normal phase in the following form 
\begin{equation}
H_{0} =\sum_{\lambda}\sum_{\bvec{k}}\varepsilon_{\lambda}(\bvec{k})c_{\lambda_,\bvec{k}}^{\dagger}c^{\phantom{\dagger}}_{\lambda,\bvec{k}}
\label{EquH0}
\end{equation}
with the superindex $\lambda = (\ell, \eta)$ where $\eta=0,1$ as described in Section~\ref{QuasiparticleBandDynamics}.
We further assume that the band designated by $\ell = p$ derives from the chalcogen atoms and has p orbital character, while the second d-like orbital $\ell =d$ derives from the transition metal atom and has d orbital character.
For definiteness and to distinguish this basis from the tight-binding orbital basis in section \ref{TightBindingModel} we call this the \emph{effective orbital basis}, and the creation and annihilation operators in \eqref{EquH0} refer to it.

To make the general Hamiltonian in the present model system in the effective orbital basis explicit, we display the matrix structure:
\begin{equation}
H = \left(\begin{array}{cccc}
H_{(d,0)(d,0)} & H_{(d,0)(p,0)} & H_{(d,0)(d,1)} & H_{(d,0)(p,1)} \\
H_{(p,0)(d,0)} & H_{(p,0)(p,0)} & H_{(p,0)(d,1)} & H_{(p,0)(p,1)} \\
H_{(d,1)(d,0)} & H_{(d,1)(p,0)} & H_{(d,1)(d,1)} & H_{(d,1)(p,1)} \\
H_{(p,1)(d,0)} & H_{(p,1)(p,0)} & H_{(p,1)(d,1)} & H_{(p,1)(p,1)} \\
\end{array}\right)
\label{eq:H-eta}
\end{equation}
Anomalous contributions are contained in elements $H_{(d,0)(p,1)}$. This also applies to the representation of a coherence matrix $\rho$ introduced below.

In the next step to describe the CDW phase in our approach, we include the electron-phonon interaction, in particular the coupling to a coherent phonon. Then contributions appear which couple the $\eta=0$ with the $\eta =1$ components. Diagonalizing these contributions together with $H_0$ leads to what we refer to as the ``band basis'', which we denote by a tilde on operators $\tilde{H}$ and the coherence matrix $\tilde{\rho}$, respectively. In this case we still have a superindex $\lambda = (b,\eta)$, which is now formed by the band index $b=\text{c},\text{v}$, and the transformed index $\eta$. 

\subsubsection{Excitonic contribution}\label{ExcitonicContribution}

In order to include the excitonic contribution of a CDW in our model, we apply a screened Hartree-Fock approximation including anomalous expectation values. In the dynamical calculations below, the Hartree-Fock approximation becomes time-dependent. Our approach is similar in spirit to other HF/BCS-like~\cite{kohn1967excitonic,kohn1970two,van2010exciton} and GW~\cite{cazzaniga2012ab} based approaches for the time-independent situation.

We assume that the Hartree-term is already included in the tight-binding Hamiltonian and need not be considered in the self-energy, but this assumption is not essential and could be relaxed, in which case all contributions stemming from the electrons in the normal-phase band structure would not be assumed to be contained in the tight-binding hamiltonian (Equ.~\eqref{EquH0}). The exchange contribution is described by
\begin{equation}
(\tilde{H}_{\text{exc}})_{\lambda_{1},\lambda_{2}}(\bvec{k_{1}})=-\sum_{\lambda_{3},\lambda_{4},\bvec{k}_{2}}\tilde{\rho}_{\lambda_{3},\lambda_{4}}(\bvec{k}_{2}) \tilde{W}_{\lambda_{1}\bvec{k}_{1},\lambda_{4}\bvec{k}_{2}}^{\lambda_{3}\bvec{k}_{2},\lambda_{2}\bvec{k}_{1}}
\label{eq:H-exc}
\end{equation}
where $\tilde{W}$ are the Coulomb-Matrix elements including screening effects which will be approximated by the $\omega \to 0$ static limit of Lindhard dielectric function, see Section~\ref{CoulombDynamics}, and $\tilde{\rho}_{\lambda_{3},\lambda_{4}} (\bvec{k}_{2})$ are the coherences in the corresponding band basis. Equation~\eqref{eq:H-exc} can be derived as the static limit of a $GW$ self energy~\cite{Binder-Koch}, and screened potential can be viewed as having a parametric time-dependence on the carrier distributions. 

The excitonic contributions driving the CDW are the anomalous electron-hole contributions including nesting vectors $Q_{\varGamma M}$ to Eq.~\eqref{eq:H-exc}, which are not present in the band structure of the normal phase. 

\subsubsection{Lattice contribution}\label{LatticeContributionl}

The mechanism that changes the band structure in our model is the lattice distortion induced by the electron-phonon interaction, which is described by coherent phonons.\cite{PhysRevB.100.035431,rossi2002theory} The contributions of the electron-phonon coupling driving the formation of the CDW stem from ionic displacements between chalcogen and transition-metal atoms of adjacent unit cells. These ionic displacements transform the normal phase into the CDW phase with increased hybridization between electronic orbitals originating from different ordinary unit cells. We use the superindex $\lambda$ to specialize the coherent phonon contribution to the phonon mode characterizing the distortion consistent with the symmetry-broken CDW of the material, e.g., an A1g mode, in which the chalcogen and the transition-metal atoms of neighboring unit cells are displaced from each other. 

The electron-phonon matrix element describing the anomalous contribution driving the CDW transition is
\begin{align}
g_{q=0,\lambda_{1}=(l_{1},0),\lambda_{2}=(l_2,1)} (\bvec{k}) &= g_{0} ( \delta_{l_{1},p}\delta_{l_2,d} + \delta_{l_{1},d}\delta_{l_2,p)} ) \\
g_{q=0,\lambda_{1}=(l_{1},1),\lambda_{2}=(l_2,0)} (\bvec{k}) &= g_{0} ( \delta_{l_{1},p}\delta_{l_{2},d} + \delta_{l_{1},d}\delta_{l_{2},p} ) 
\end{align}
and the interaction of electrons with such a coherent phonon is described in the effective orbital basis by
\begin{eqnarray}
(H_{cp})_{\lambda_{1},\lambda_{2}}(\bvec{k}) & = & g_{q=0,\lambda_{1},\lambda_{2}}(\bvec{k})\big(B_{0}+B_{0}^{\dagger}\big)
%c_{\lambda_{1},\bvec{k}}^{\dagger}c_{\lambda_{2},\bvec{k}}
\end{eqnarray}
where $B_{0}=\left\langle b_{0}\right\rangle $ is the coherent phonon amplitude. For the equation of motion of the coherent phonon amplitude we obtain
\begin{equation}
\begin{split}
\frac{d}{dt}B_{0} & =  -\left(i\omega_{0}^{ph} +\gamma_{deph}^{P}\right)B_{0}  \\ 
 & \quad  +\frac{1}{i\hbar}\sum_{\lambda_{1},\lambda_{2},\bvec{k}}\left(\tilde{g}_{q=0,\lambda_{1},\lambda_{2}}(\bvec{k})\right)^{\dagger}\tilde{\rho}_{\lambda_{1},\lambda_{2}} (\bvec{k})    
\end{split}
\end{equation}
where the electron-phonon matrix element in the band basis is 
\begin{equation}
\tilde{g}_{q=0,\lambda_{1},\lambda_{2}}(\bvec{k})=U^{\dagger}g_{q=0,\lambda_{1},\lambda_{2}}(\bvec{k})U
\end{equation}
related to the matrix element in the effective orbital basis via a unitary transformation~$U$.

We want to emphasize that the coherent phonon amplitude is solely driven by anomalous contributions, which are not present in the band structure of the normal phase. This is in accordance with the concept that this phonon mode solely exists in the CDW phase and is related to the phonon softening during the phase transition.\cite{chan1973spin} However, the result of the initialization process is independent of any phonon softening and for the onset of a phase transition due to an optical excitation as calculated in the carrier dynamics, the softening of phonon modes is marginal, which is supported by the weak temperature or pressure dependence of CDW amplitude modes (e.g., A1g-CDW amplitude mode in Ref.~\onlinecite{PhysRevLett.91.136402}) in these regions. If the coherent phonon amplitude becomes finite, the permanent ionic displacement causes a change of the hybridization between electronic orbitals centered at different ions of adjacent unit cells, which enters the Hamiltonian $H_{0}$ as a correction of the tight-binding matrix-elements.

\subsubsection{Total band contribution}\label{TotalContribution}

For the total Hamiltonian including the excitonic and lattice contribution in the effective orbital basis (if necessary by unitary transformation) we obtain
\begin{equation}
\begin{split}
(H_{\text{tot}})_{\lambda_{1},\lambda_{2}}(\bvec{k})&= (H_{0})_{\lambda_{1},\lambda_{2}}(\bvec{k})+(H_{exc})_{\lambda_{1},\lambda_{2}}(\bvec{k}) \\ &\quad +(H_{cp})_{\lambda_{1},\lambda_{2}}(\bvec{k})
\label{Htot}
\end{split}
\end{equation}
In a more complete treatment, the undiagonalized Hamiltonian $H_{TB}$ could be used for the normal phase instead of $H_{0}$. In this case, the resulting hamiltonian and the dynamical equations derived below could still be expressed in the same formal structure using the superindex $\lambda$ and the blocks with different $\eta$ indices in hamiltonian~\eqref{Htot} would remain uncoupled.

A dynamical switching process is performed to establish a symmetry-broken initial state for the subsequent non-equilibrium dynamics. The coherent phonon mechanism for the phase transition has to be seeded, e.g., by an infinitesimal (orbital) charge transfer due to an excitonic contribution causing an infinitesimal spontaneous symmetry breaking. Also the anomalous contributions of the screened self-energy are seeded by an infinitesimal contribution of an ionic displacement causing an infinitesimal spontaneous symmetry breaking. This illustrates the mutual dependence of the excitonic and ionic contributions for the emergence of a CDW. Throughout the switching process we therefore apply the Hamiltonian
\begin{equation}
(H_{sw})_{\lambda_{1},\lambda_{2}}(\bvec{k})=(H_{tot})_{\lambda_{1},\lambda_{2}}(\bvec{k})+(H_{\epsilon})_{\lambda_{1},\lambda_{2}}
\end{equation}
including an infinitesimal off-diagonal symmetry-breaking contribution $H_{\epsilon}$ to probe the phase transition with matrix elements $\epsilon<10^{-7}$~meV which decay exponentially. As $H_{sw}$ and also $H_{tot}$ is time dependent its eigenvalues and eigenvectors are calculated for every step of the dynamical switching process or time-step of the subsequent dynamics, respectively. Thus, the electron-phonon and electron-electron matrix elements as well as the generalized coherences involve time-dependent basis states. In this time-dependent eigenbasis $\tilde{n}_{\lambda_{1}}(\bvec{k})=\tilde{\rho}_{\lambda_{1}\lambda_{1}}(\bvec{k})$ can be interpreted as the occupation of the state $\left|\lambda_{1}\bvec{k}\right\rangle $ at that time. For the subsequent non-equilibrium dynamics we assume that the polarization (coherences in the band basis) in the equations-of-motion die out faster than the dynamics of interest. As a response to the small stimulus in $H_{\epsilon}$ that mimics fluctuations and depending on the indirect band overlap on the Fermi surface (band nesting effect) of the band structure in the effective orbital basis (see e.g. Equ.~\ref{EquH0}) and the strength of the excitonic contribution (e.g. background dielectric constant) and lattice contribution (e.g. electron-phonon matrix elements) the system might establish the CDW phase or still favor the normal phase.

Folding and unfolding techniques are used to project band structures and wave vectors between super-cells and unit-cells.~\cite{PhysRevB.71.115215,boykin2007approximate,medeiros2015unfolding,medeiros2014effects,popescu2012extracting} We make use of this method to project the time-dependent results of our generalized Hamiltonian onto a unit cell and a super-cell representation, respectively. In particular, the appearance of ghost bands in the unit-cell visualizes the break-up of the commensurable phase subspace and illustrates the appearance of anomalous expectation values and interaction processes.

\subsection{Carrier dynamics via equation of motion technique}\label{CarrierDynamics}

We derive the equations of motion for the relevant correlation functions including microscopic carrier and band dynamics for a two-dimensional prototypical tight-binding model with nonisotropic band dispersion. For the following microscopic contributions to the carrier dynamics it also applies that anomalous dipole matrix elements, carrier-phonon matrix elements and Coulomb matrix elements vanish in the normal phase and only emerge due to the appearance of the CDW phase.

\subsubsection{Optical excitation}\label{OpticalExcitation}

We model the optical excitation in accordance to the optical part of the semiconductor Bloch equations between the (in the effective orbital basis chalcogen-like) valence and the (in the effective orbital basis transition-metal-like) conduction band in the time-dependent band basis for the polarizations by
\begin{equation}
\begin{split}
\frac{d}{dt}\tilde{p}_{\lambda_{1},\lambda_{2}} (\bvec{k}) & =  -\big(i\tilde{\omega}_{\lambda_{1}\bvec{k},\lambda_{2}\bvec{k}} + \gamma_{\text{deph}}^{p}\big)\tilde{p}_{\lambda_{1},\lambda_{2}}(\bvec{k}) \\
& \qquad -i\tilde{\varOmega}_{\lambda_{1}\bvec{k},\lambda_{2}\bvec{k}}\big(\tilde{n}_{\lambda_{1}}(\bvec{k})-\tilde{n}_{\lambda_{2}}(\bvec{k})\big)
\end{split}
\end{equation}
and for the occupations by 
\begin{equation}
\frac{d}{dt}\tilde{n}_{\lambda_{1}}(\bvec{k})=-\sum_{\lambda_{2}}\big(i\tilde{\varOmega}_{\lambda_{1}\bvec{k},\lambda_{2}\bvec{k}}\tilde{p}_{\lambda_{1},\lambda_{2}}(\bvec{k})+\text{h.c.}\big)
\end{equation}
where $\tilde{\varOmega}_{\lambda_{1}\bvec{k},\lambda_{2}\bvec{k}}=\hbar^{-1}\tilde{\mu}_{\lambda_{1}\bvec{k},\lambda_{2}\bvec{k}}E$ are Rabi energies, $E$ the time-dependent optical field and $\tilde{\mu}_{\lambda_{1}\bvec{k},\lambda_{2}\bvec{k}}$ the dipole matrix elements.

\subsubsection{Carrier-phonon scattering}\label{PhononDynamics}

The coherent phonon as discussed in Section~\ref{QuasiparticleBandDynamics} is the lowest order contribution using the language of dynamical correlation functions for the electron-phonon interaction, see Ref.~\onlinecite{rossi2002theory}. At the next order one finds scattering and dephasing terms. We focus on the carrier-phonon scattering terms and attach particular importance to the effect of cooling a quasi-equilibrium distribution after an optical excitation in the CDW phase around the Fermi surface. In doing so, we use a characteristic LO phonon mode which can be justified e.g. by arguments given in Ref.~\onlinecite{lucovsky1976reflectivity} to model inter- and intra-band scattering processes of such a cooling process. We obtain for the carrier-phonon scattering in Markov approximation
\begin{eqnarray}
\frac{d}{dt}\tilde{n}_{\lambda_{1}} (\bvec{k}_{1}) & = & \frac{2\pi}{\hbar}\sum_{\lambda_{2}\bvec{k}_{2}}\sum_{\pm}\Big(\sum_{\bvec{q}}\big| \tilde{M}_{\lambda_{1}\bvec{k}_{1},\lambda_{2}\bvec{k}_{2}}\left(\bvec{q}\right)\big|^{2}\Big) \times \notag \\ 
& & \big((\tilde{N}^{in}N_{+}^{ph}-\tilde{N}^{out}N_{-}^{ph}\big)\delta\left(\Delta\tilde{\varepsilon}\right)
\end{eqnarray}
with
\begin{eqnarray}
\tilde{N}^{in} & = & \big(1-\tilde{n}_{\lambda_{1}}(\bvec{k}_{1})\big)\tilde{n}_{\lambda_{2}}(\bvec{k}_{2}) \notag \\
\tilde{N}^{out} & = & \tilde{n}_{\lambda_{1}}(\bvec{k}_{1})\big(1-\tilde{n}_{\lambda_{2}}(\bvec{k}_{2})\big) \notag \\
N_{+}^{ph} & = & N_{LO}+\frac{1}{2}\pm\frac{1}{2} \notag \\
N_{-}^{ph} & = & N_{LO}+\frac{1}{2}\mp\frac{1}{2} \notag \\
\Delta\tilde{\varepsilon} & = & \tilde{\varepsilon}_{\lambda_{1}\bvec{k}_{1}}-\tilde{\varepsilon}_{\lambda_{2}\bvec{k}_{2}}\pm\hbar\omega_{0}^{ph} \notag
\end{eqnarray}
and 
\begin{equation}
\tilde{M}_{\lambda_{1}\bvec{k}_{1},\lambda_{2}\bvec{k}_{2}}\left(\bvec{q}\right)=M_{\bvec{q}}\left\langle \lambda_{1}\bvec{k}_{1}\left|e^{i\bvec{k}\bvec{r}}\right|\lambda_{2}\bvec{k}_{2}\right\rangle 
\label{eq:M-q}
\end{equation}
where $|\lambda_{1}\bvec{k}_{1}\rangle$ are the time-dependent eigenstates of the Hamiltonian in Eq.~\ref{Htot} and the effective Froehlich interaction $M_{\bvec{q}}$ includes the polar coupling constant $\alpha_{\text{pol}}$, the LO phonon energy $\hbar\omega_{LO}$ and an effective mass $m_{\text{eff}}$.~\cite{mahan1990many}

\subsubsection{Carrier-carrier scattering}\label{CoulombDynamics}

The carrier dynamics due to carrier-carrier scattering are part of the higher-order Coulomb contributions included in the equation of motion for the density matrix. The derivation of the Coulomb scattering is done starting from the Kadanoff-Baym equations by applying the second-order Born approximation for the self-energy. We focus on the carrier-carrier scattering terms and obtain the following equation of motion for the Coulomb scattering in Markov approximation 
\begin{equation}
\begin{split}
\frac{d}{dt}\tilde{n}_{\lambda_{1}}(\bvec{k}_{1})=\frac{2\pi}{\hbar}\sum_{\lambda_{2}\bvec{k}_{2},\lambda_{3}\bvec{k}_{3},\lambda_{4}\bvec{k}_{4}}\tilde{W}\big(\tilde{N}^{in}-\tilde{N}^{out}\big)\delta\left(\Delta\tilde{\varepsilon}\right)
\end{split}
\end{equation}
with 
\begin{eqnarray}
\tilde{N}^{in} & = & \left(1-\tilde{n}_{\lambda_{1}}(\bvec{k}_{1})\right)\tilde{n}_{\lambda_{2}}(\bvec{k}_{2})\left(1-\tilde{n}_{\lambda_{3}}(\bvec{k}_{3})\right)\tilde{n}_{\lambda_{4}}(\bvec{k}_{4})
\notag \\
\tilde{N}^{out} & = & \tilde{n}_{\lambda_{1}}(\bvec{k}_{1})\left(1-\tilde{n}_{\lambda_{2}}(\bvec{k}_{2})\right)\tilde{n}_{\lambda_{3}}(\bvec{k}_{3})\left(1-\tilde{n}_{\lambda_{4}}(\bvec{k}_{4})\right)
\notag \\
\Delta\tilde{\varepsilon} & = & \tilde{\varepsilon}_{\lambda_{1}\bvec{k}_{1}}-\tilde{\varepsilon}_{\lambda_{2}\bvec{k}_{2}}+\tilde{\varepsilon}_{\lambda_{3}\bvec{k}_{3}}-\tilde{\varepsilon}_{\lambda_{4}\bvec{k}_{4}} \notag
\end{eqnarray}
and the screened Coulomb-Matrix elements
\begin{equation}
\begin{split}
\tilde{W} =  \tilde{W}_{\lambda_{2}\bvec{k}_{2}\lambda_{3}\bvec{k}_{3}}^{\lambda_{1}\bvec{k}_{1}\lambda_{4}\bvec{k}_{4}}\left(\left(\tilde{W}_{\lambda_{2}\bvec{k}_{2}\lambda_{3}\bvec{k}_{3}}^{\lambda_{1}\bvec{k}_{1}\lambda_{4}\bvec{k}_{4}}\right)^{*}-\left(\tilde{W}_{\lambda_{2}\bvec{k}_{2}\lambda_{4}\bvec{k}_{4}}^{\lambda_{1}\bvec{k}_{1}\lambda_{3}\bvec{k}_{3}}\right)^{*}\right)
\end{split}
\end{equation}
with the direct and exchange contribution, respectively. The direct term is typically the dominant contribution and the exchange term is a minor correction to the direct term. The screened Coulomb-Matrix elements are
\begin{equation}
\tilde{W}_{\lambda_{2}\bvec{k}_{2}\lambda_{3}\bvec{k}_{3}}^{\lambda_{1}\bvec{k}_{1}\lambda_{4}\bvec{k}_{4}} = \sum_{q} w_{q} \tilde{I}_{\lambda_{1}\bvec{k}_{1},\lambda_{4}\bvec{k}_{4}} (\bvec{q})  \tilde{I}_{\lambda_{2}\bvec{k}_{2},\lambda_{3}\bvec{k}_{3}} \left( - \bvec{q} \right)
\label{eq:Coulomb-Matrix}
\end{equation}
with
\begin{equation}
\tilde{I}_{\lambda_{1}\bvec{k}_{1},\lambda_{4}\bvec{k}_{4}} \left( \bvec{q} \right) = \big\langle \lambda_{1}\bvec{k}_{1}\big|e^{i\bvec{q}\bvec{r}}\big|\lambda_{4}\bvec{k}_{4}\big\rangle
\label{eq:I-def}
\end{equation}
where $| \lambda_{1}\bvec{k}_{1}\rangle$ are the time-dependent eigenstates of the Hamiltonian in Equ.~\ref{Htot} and $w_{\bvec{q}}$ is the screened Coulomb potential
\begin{eqnarray}
w_{q}=\varepsilon^{-1}\left(\bvec{q}\right)v_{q}
\end{eqnarray}
where $v_{q}$ is the unscreened Coulomb potential including a background dielectric constant  $\epsilon_{b}$ and a normalization area $A$. The screening of the Coulomb interaction is time-dependent as the density of excited carriers changes and this effect is taken into account using the frequency independent limit of Lindhard dielectric function
\begin{eqnarray}
\varepsilon\left(\bvec{q}\right)=1-\frac{1}{A}\sum_{\lambda\bvec{k}}\tilde{V}_{\lambda(\bvec{k}-\bvec{q}),\lambda\bvec{k}}^{\lambda\bvec{k},\lambda(\bvec{k}-\bvec{q})}\frac{\tilde{n}_{\lambda}(\bvec{k}-\bvec{q})-\tilde{n}_{\lambda}(\bvec{k})}{\tilde{\varepsilon}_{\lambda\bvec{k}-\bvec{q}}-\tilde{\varepsilon}_{\lambda}(\bvec{k})}
\end{eqnarray}
where $\tilde{V}_{\lambda(\bvec{k}-\bvec{q}),\lambda\bvec{k}}^{\lambda\bvec{k},\lambda(\bvec{k}-\bvec{q})}$ are the Coulomb-matrix elements calculated from the unscreened Coulomb potential $v_{q}$.

\section{Results}\label{Results}

\subsection{CDW phase in equilibrium}\label{ResultsCDW}

\begin{figure}[tb]
\centering
\includegraphics[trim=0cm 0cm 0cm 0cm,clip,scale=0.45]{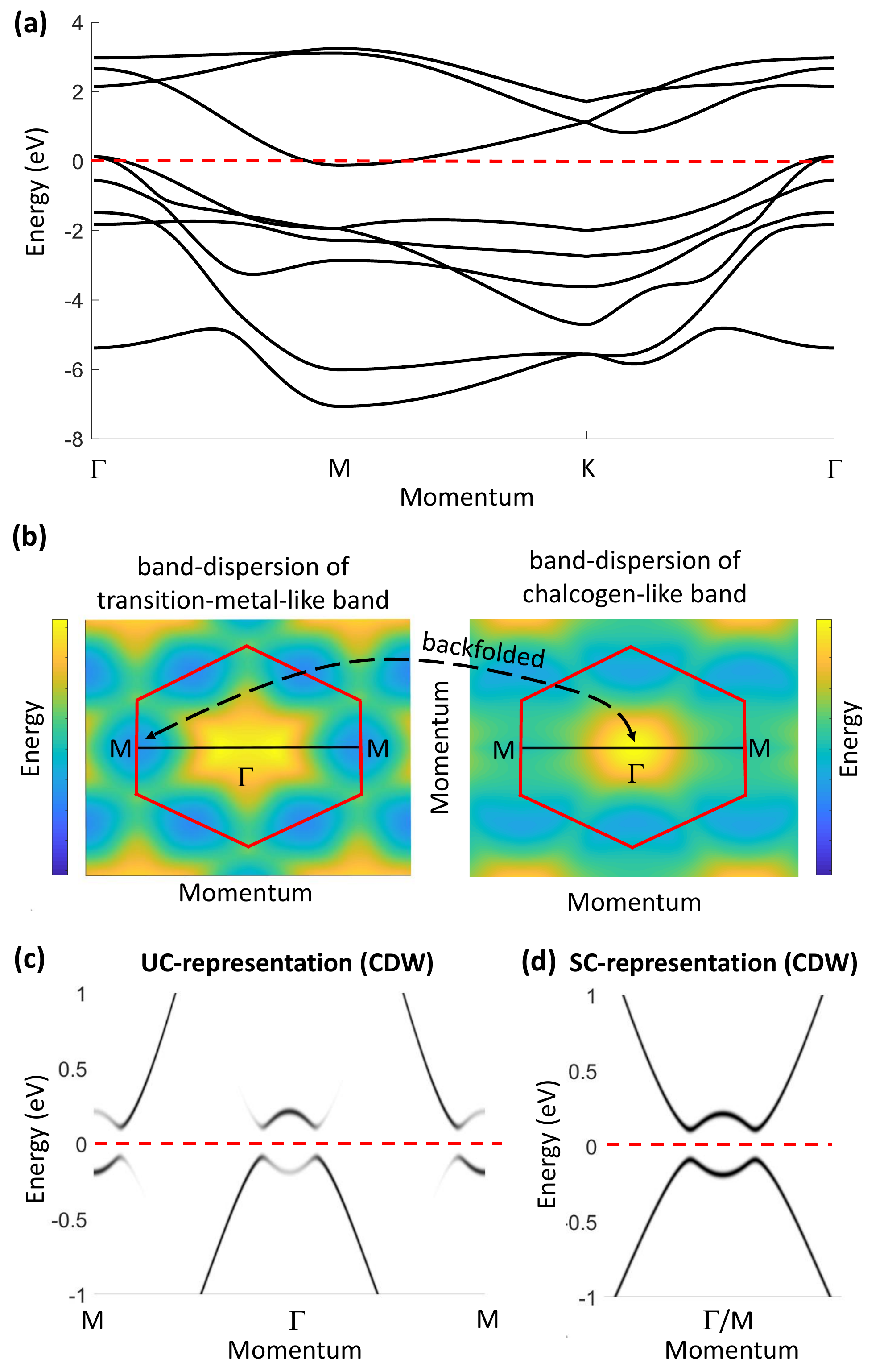}
\caption{(a) Model band structure of a prototypical transition-metal dichalcogenide around the Fermi surface (red dashed line) calculated with three t$_{\text{2g}}$ orbitals from the transition metal atom and six p-orbitals from the chalcogen atoms. An indirect band overlap between transition-metal bands on the M-point and dichalcogenide bands on the $\Gamma$-point is visible. (b) Band dispersion of the highest chalcogen-like band and the lowest transition-metal-like band. The red hexagon illustrates the unit cell of the normal phase. These bands are used to describe the characteristics of the carrier and band-dynamics close to the relevant high-symmetry points. The black dashed line illustrates the backfolding from the $\Gamma$ to one of the $M$ points of the normal phase BZ.
(c) Unit-cell representation of the bands in the CDW phase. (d) Super-cell representation of the bands in the CDW phase.}
\label{figure1}
\end{figure}

To study the formation of a CDW phase from a normal phase in a quasi-two dimensional material, in particular a transition metal dichalcogenide, the band structure of the normal phase is calculated via a tight-binding Hamiltonian. The details of the tight-binding Hamiltonian and parameters have been collected in Appendix~\ref{AppendixA}. After diagonalization of the tight-binding Hamiltonian, we obtain the band structure in Figure~\ref{figure1}.(a). 
The calculated band structure has an indirect band overlap on the Fermi surface (band nesting effect) and the band dispersion of the highest chalcogen-like band and the lowest transition-metal-like band are depicted in Figure~\ref{figure1}.(b). These bands are used to describe the characteristics of the carrier and band-dynamics close to the relevant high-symmetry points around the Fermi surface.

\begin{figure*}[tb]
\centering
\includegraphics[trim=0cm 0cm 0cm 0cm,clip,scale=0.5]{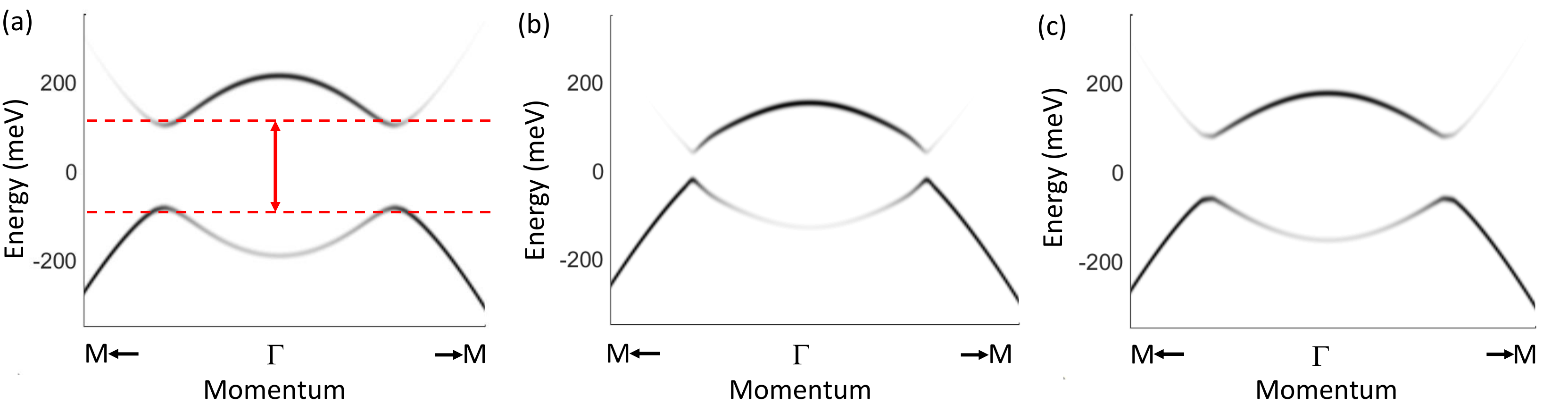}
\caption{(a) Band-structure in unit-cell representation of the CDW phase before the optical pulse. The transitions driven by the optical pulse is indicated by the red arrow. (b) Quenching of the insulator phase (CDW phase) during the optical excitation. (c) Band-structure of the quasi-equilibrium situation after optical excitation.}
\label{figure2}
\end{figure*}

\begin{figure}[tb]
\centering
\includegraphics[trim=0cm 0cm 0cm 0cm,clip,scale=0.5]{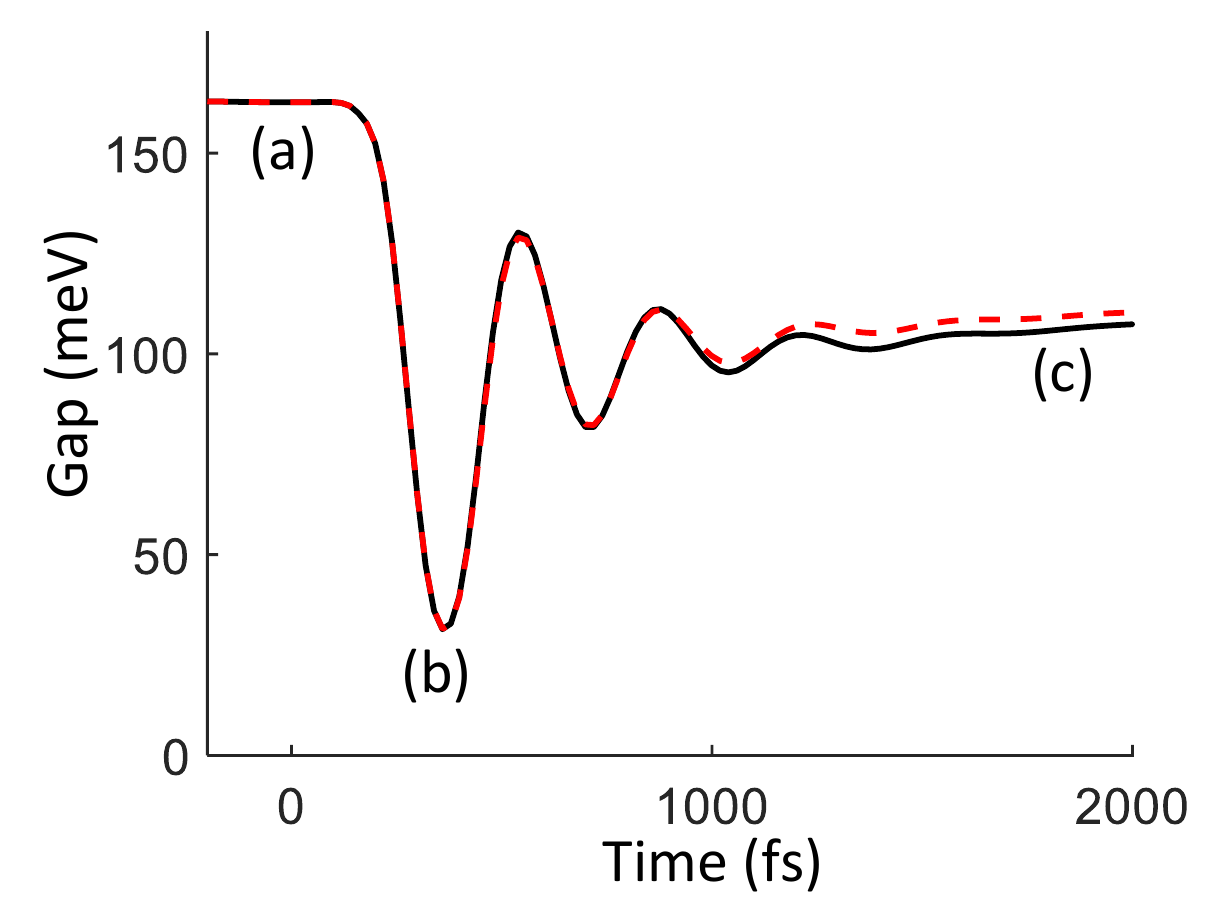}
\caption{Gap closing vs. time after optical excitation by a drive pulse (after 0 fs). The labels on the black line refer to the graphs in Fig.~\ref{figure2}. The effects of processes on longer timescales are visible by comparing the dashed red line with and the solid black line without those effects.}
\label{figure3}
\end{figure}

Starting from the normal phase, we obtain the symmetry-broken CDW-phase as described in Section~\ref{QuasiparticleBandDynamics} as the steady-state of an initilization process, during which the anomalous expectation values are allowed to develop. Depending on the parameters chosen, we may or may not obtain a symmetry broken phase from this initialization process. We choose here parameters for the lattice and excitonic contributions in our effective model so that a CDW phase with a Mexican-hat shaped band structure can develop. Our choice of paramaters does not deviate too far from actual material parameters, but we investigate parameter ranges given in brackets to study trends.  We choose a lattice temperature of $T_{L}=150$~$(150-300)$~K at which even materials like TiSe$_{2}$ are still in the CDW regime~\cite{coleman1988scanning} If the transition temperature of the prototypical CDW material is well above $300$~K, the qualitative results can also be transferred to room temperature conditions. In the quasiparticle band dynamics the lattice contribution is influenced by the electron-phonon matrix element for the anomalous contribution $g_{0}=1.6$~$(1.5-2.0)$~meV with $\hbar\omega_{0}^{ph}=12.4$~meV and $\gamma_{deph}^{P}=5.0$~ps$^{-1}$. An increase of $g_{0}$, for instance, enlarges the band gap in the CDW phase. The excitonic contribution is influenced by the background dielectric constant $\epsilon_{b}=8.0$~$(8.0-12.0)$. An increase of $\epsilon_{b}$ reduces e.g. the momentum dependent band gap and therefore the curvature of the Mexican hat band dispersion. Here, the excitonic and lattice contribution has a similar order of magnitude. 

The result of the equilibrium configuration obtained in the way just described is shown in unit-cell representation in Figure~\ref{figure1}(c) and in super-cell representation in Figure~\ref{figure1}(d). In the unit-cell representation a conduction band (band above the Fermi surface) at the $\Gamma$-point and a valence band (band below the Fermi surface) at the $M$-point appears. The appearance of such additional bands illustrates the formation of finite anomalous expectation values and interaction processes in the CDW-phase. We stress that we assume in general an anisotropic band structure, i.e., we perform the microscopic carrier and band dynamics in a two-dimensional k-space both for the band structure calculation and the dynamics described in the next subsection.

\subsection{Non-equilibrium gap dynamics}\label{ResultsDynamics}

We now turn to the excitation induced non-equilibrium dynamics. In the following, the band dynamics described above are still included and updated in time as described in Section~\ref{TotalContribution} so that, in particular, all calculated interaction matrix elements are time-dependent. We simulate the quenching of the phase by an optical drive pulse whose impact is illustrated in Fig.~\ref{figure2}. In Fig.~\ref{figure2}(a) we show a window of the band structure from Fig.~\ref{figure1}(c), i.e., the band-structure in unit-cell representation of the unexcited CDW phase together with the transition driven by the optical pulse depicted as a red arrow. The optical field couples bands below and above the Fermi energy via anomalous dipole matrix elements, which occur because in the CDW phase we have additional bands above the Fermi energy at the $\Gamma$ and below the Fermi energy around the $M$-point, respectively. 

During and after the pulse induced dynamics we also include carrier-carrier scattering as described in Sec.~\ref{CoulombDynamics} and carrier-phonon scattering as described in Section~\ref{PhononDynamics}. As stressed in connection with the choice of parameters for the equilibrium configuration, our goal is to study parametric dependencies, but to not deviate too far from actual material parameters. We choose the following numerical parameters and investigate the ranges given in brackets: The drive pulse(s) have a transition energy around the equilibrium band gap of the CDW phase (approximately $165$~meV), a $\sin^{2}$ envelope with FWHM of $200$~$(200-250)$~fs and induce a maximum Rabi energy of $\hbar\varOmega_{0}=8.0$~$(6-10)$~meV, where the Rabi energy $\varOmega_{\lambda_{1}\bvec{k},\lambda_{2}\bvec{k}}$ is band basis, momentum and time dependent, see Section~\ref{OpticalExcitation}. We assume a polarization dephasing of $\gamma_{\text{deph}}^{p}=5.0$~$(2.5-10.0)$~ps$^{-1}$ where we have checked that the influence of the different $\gamma_{\text{deph}}^{p}$  values has only a marginal influence on the carrier and band dynamics. For the cooling process due to the electron-phonon scattering, we use a characteristic LO phonon mode as described in Ref.~\onlinecite{lucovsky1976reflectivity} with the polar coupling constant $\alpha_{\text{pol}}=0.35$~$(0.2-0.6)$ and the LO phonon energy $\hbar\omega_{LO}=30.0$~$(10-40)$~meV. The effective mass $m_{\text{eff}}$ is determined by the averaged band dispersion of the effective mass of the Mexican-hat shaped bands in the CDW phase, cf.\ Eq.~\eqref{eq:M-q}.

In order to characterize the field-induced dynamics we first plot in Fig.~\ref{figure3} the time-dependent band gap around the $\Gamma$ point. The band structure corresponding to the unexcited system with a gap marked by (a) is shown in Fig.~\ref{figure2}(a). During the optical excitation electron-hole pairs are excited across the gap. This modifies the anomalous excitonic contributions and thus also the electron-lattice interaction, which affects the band dynamics and the band gap. Additionally, the excited electron-hole pairs induce carrier redistribution due to carrier-carrier and carrier-phonon scattering processes. The band and carrier dynamics influence each other and determine the material response due to the optical excitation. The field-induced dynamics show an oscillation of the gap up to about 1 ps and a subsequent slower relaxation. The band structure snapshots corresponding to the times marked ``(b)'' and ``(c)'' are shown in Fig.~\ref{figure2}(b) and (c), respectively. The quenching of the gap is clearly visible in Fig.~\ref{figure2}(b), but has already relaxed back to a quasi-equilibrium structure in Fig.~\ref{figure2}(c).

The features of the gap dynamics shown in Fig.~\ref{figure3} and the corresponding band structures in Fig.~\ref{figure2} reflect the different timescales of the spectral and kinetic response for carrier-carrier and carrier-phonon interaction. The carrier-carrier scattering processes as well as the buildup of screening and the excitonic contribution to the band dynamics are on an ultrafast timescale, while the lattice contribution to the band dynamics response and the carrier-phonon cooling process are slightly delayed. In Figure~\ref{figure2}(b) the bands are depicted for the case of maximum quenching of the insulator phase (CDW) during the optical excitation. Here, one has the highest probability of carrier cooling via inter-band carrier-phonon scattering. Figure~\ref{figure2}(c) illustrates the quasi-equilibrium situation in a partially quenched CDW phase that is reached after a few hundred femtoseconds up to a few picoseconds after the optical excitation. The transition into full thermal equilibrium is driven by interaction processes on longer timescales like non-radiative losses and spontaneous emission processes as described in Appendix~\ref{AppendixB}. These processes can be described by rates on the order of  $\gamma_{nr}=\gamma_{sp}=10^{-1}$~$(10^{-3}-10^{-1})$~ps$^{-1}$, so that their effect is marginal in the first picoseconds, as can be seen by comparing the dashed red line with and the solid black line without those effects in Figure~\ref{figure3}.

The different timescales of spectral and kinetic response for carrier-carrier and carrier-phonon dynamics, which determine the optical response discussed here, already indicate the possibility of a transient population inversion within the CDW-phase, which will be analyzed below.

\subsection{Scattering transitions with anomalous contributions}\label{ResultsDynamicsAnomalousContributions}

\begin{figure}[tb]
\centering
\includegraphics[trim=0cm 0cm 0cm 0cm,clip,scale=0.46]{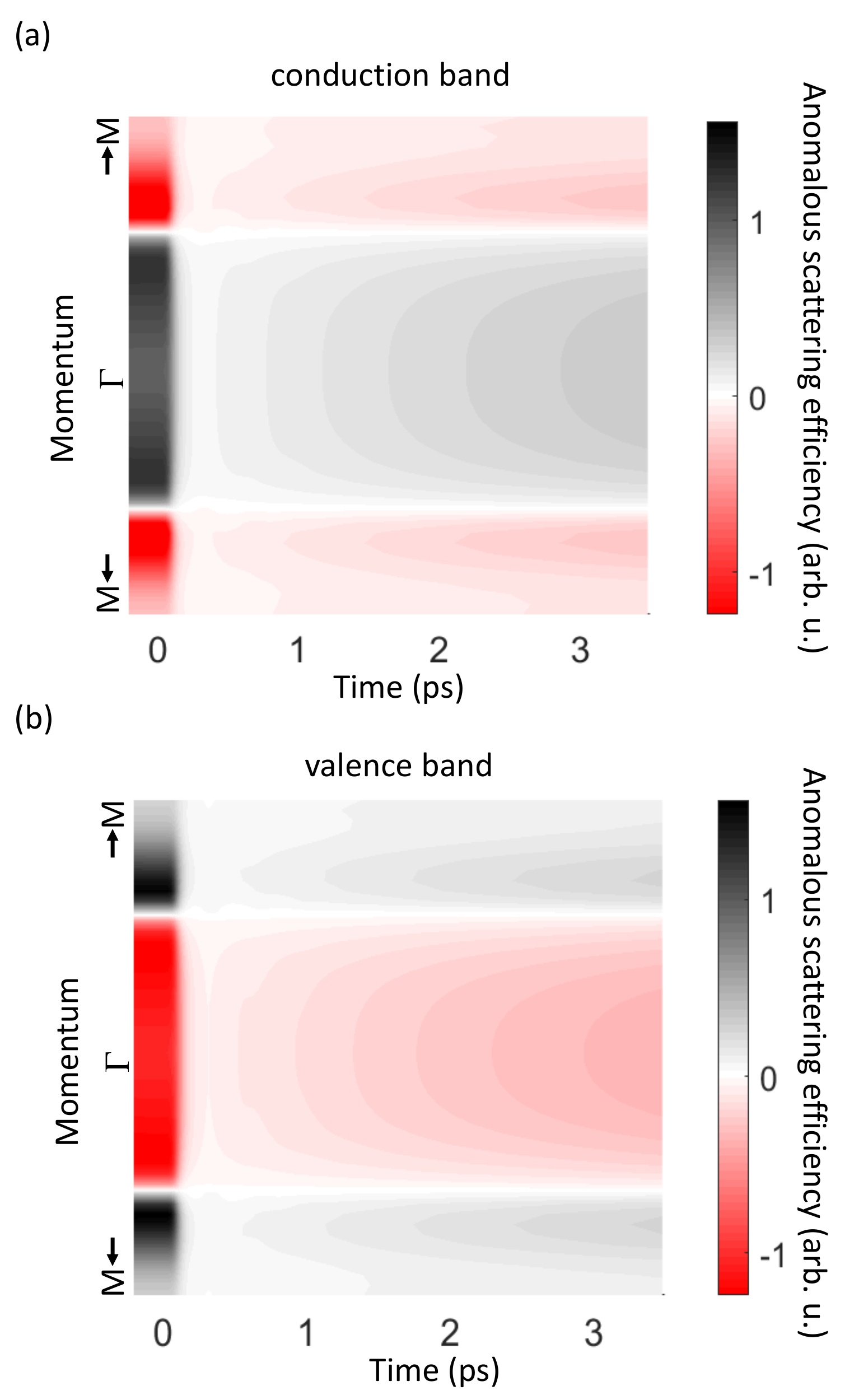}
\caption{Anomalous scattering (or charge-transfer) efficiency for the conduction band (a) and valence band (b) vs. time after quenching of the insulator phase (CDW phase) due to optical excitation (at 0 fs).}
\label{figure4}
\end{figure}

As the main focus of this paper is the description of scattering processes between states that are affected by anomalous expectation values, we illustrate some characteristics of these dynamics in here. While all scattering mechanisms are affected, we choose here the Coulomb scattering contribution. 

In the CDW representation, we separate out orbital indices corresponding to the embedded normal phase, i.e., $\eta =0$ in our notation, and call these $A$, whereas remaining orbital indices that occur due to the backfolding, i.e., $\eta =1$ in our present notation, are collected in set $B$. We now measure the efficiency of charge transfer between the subspaces $A$ and $B$ by the following expression
for Coulomb scattering affecting state $|\lambda\bvec{k}\rangle$
\begin{equation}
\begin{split}
\varGamma_{\lambda}^{A\rightarrow B} (\bvec{k}) = \frac{2\pi}{\hbar} \sum_{\lambda_{2}\bvec{k}_{2},\lambda_{3}\bvec{k}_{3},\lambda_{4}\bvec{k}_{4}} \Delta \tilde{p}_{\lambda\bvec{k},\lambda_{4}\bvec{k}_{4}}^{A} \times \\ \left| 
\tilde{W}_{\lambda_{2}\bvec{k}_{2},\lambda_{3}\bvec{k}_{3}}^{\lambda\bvec{k},\lambda_{4}\bvec{k}_{4}}
\right|^{2} \delta\left(\Delta\tilde{\varepsilon}\right)
\end{split}
\label{eq:gamma-ab}
\end{equation}
where
\begin{equation}
\Delta \tilde{p}_{\lambda\bvec{k},\lambda_{4}\bvec{k}_{4}}^{A}=
\sum_{\alpha\in A}
|\left\langle \alpha | \lambda \bvec{k} \right\rangle|^2
- |\left\langle \alpha| \lambda_{4}\bvec{k}_{4} \right\rangle|^2.
\label{eq:p-ab}
\end{equation}
Here, Eq.~\eqref{eq:p-ab} is the change of probability to find the charge located in orbitals belonging to the subspace $A$ before and after the scattering process from $\left|\lambda\bvec{k}\right\rangle$ to $\left|\lambda_{4}\bvec{k}_{4}\right\rangle$. We chose expression~\eqref{eq:p-ab} as a simple illustration of the impact of anomalous expectation values on scattering processes and therefore neglect the explicit influence of the time-dependent carrier-distribution, which would be superimposed with these effects. In particular, without anomalous expectation values, this expression vanishes, because we would have no coupling between the subspaces $A$ and $B$ by anomalous expectation values, i.e., the hamiltonian matrix~\eqref{eq:H-eta} would be block-diagonal. In this case, Eq.~\eqref{eq:I-def} and \eqref{eq:Coulomb-Matrix} show that $\Gamma^{A\to B}=0$. Conversely, the appearance of anomalous expectation values and anomalous interaction processes 
leads to an increased efficiency of charge transfer between subspace $A$ and subspace $B$.

Figure~\ref{figure4} depicts the charge-transfer efficiency $\varGamma_{\lambda}^{A\rightarrow B}$  between subspace $A$ and subspace $B$ in arbitrary units. The momentum space axis is the same as the one shown in Fig.~\ref{figure2}.
At times close to $t=0$, the system is in the CDW-phase. Here, the conduction band states $|\lambda \bvec{k} \rangle$ around the $\Gamma$-point have a pronounced orbital-character from orbitals which belong to subspace A. This causes the appearance of the conduction band at the $\Gamma$-point, see Figure~\ref{figure1}(c). Consequently, Fig.~\ref{figure4}(a) shows a positive transfer efficiency into subspace A, which can lead to a pronounced charge-transfer scattering as soon as scattering phase space becomes available. When one crosses over the Mexican hat minimum going from $\Gamma$ to $M$, a negative transfer efficiency due to the scattering processes with states around the $\Gamma$-point is visible. The opposite arguments apply to the valence bands: Fig.~\ref{figure4}(b) shows that a negative transfer efficiency into subspace A around the $\Gamma$-point and a positive transfer efficiency in a transition region occur.

For times beyond $0$~fs, i.e. after the optical excitation, Fig~\ref{figure4} shows a reduced anomalous scattering efficiency, which recovers on longer timescales. This is a consequence of the optically induced quenching of the CDW phase, which correlates with a quenching of anomalous interaction matrix elements. Generally speaking, anomalous interaction matrix elements like carrier-carrier and carrier-phonon anomalous matrix elements vanish in the band structure of the normal phase and only emerge due to the appearance of the CDW phase, so that the behavior discussed here for Coulomb scattering is typical of all scattering processes. These effects illustrated in Fig.~\ref{figure4} are very important for the carrier and band dynamics in a system with anomalous expectation values and are fully included in our approach.

\subsection{Optical Amplification}\label{ResultsOpticalAmplification}

\begin{figure}[tb]
\centering
\includegraphics[trim=0cm 0cm 0cm 0cm,clip,scale=0.38]{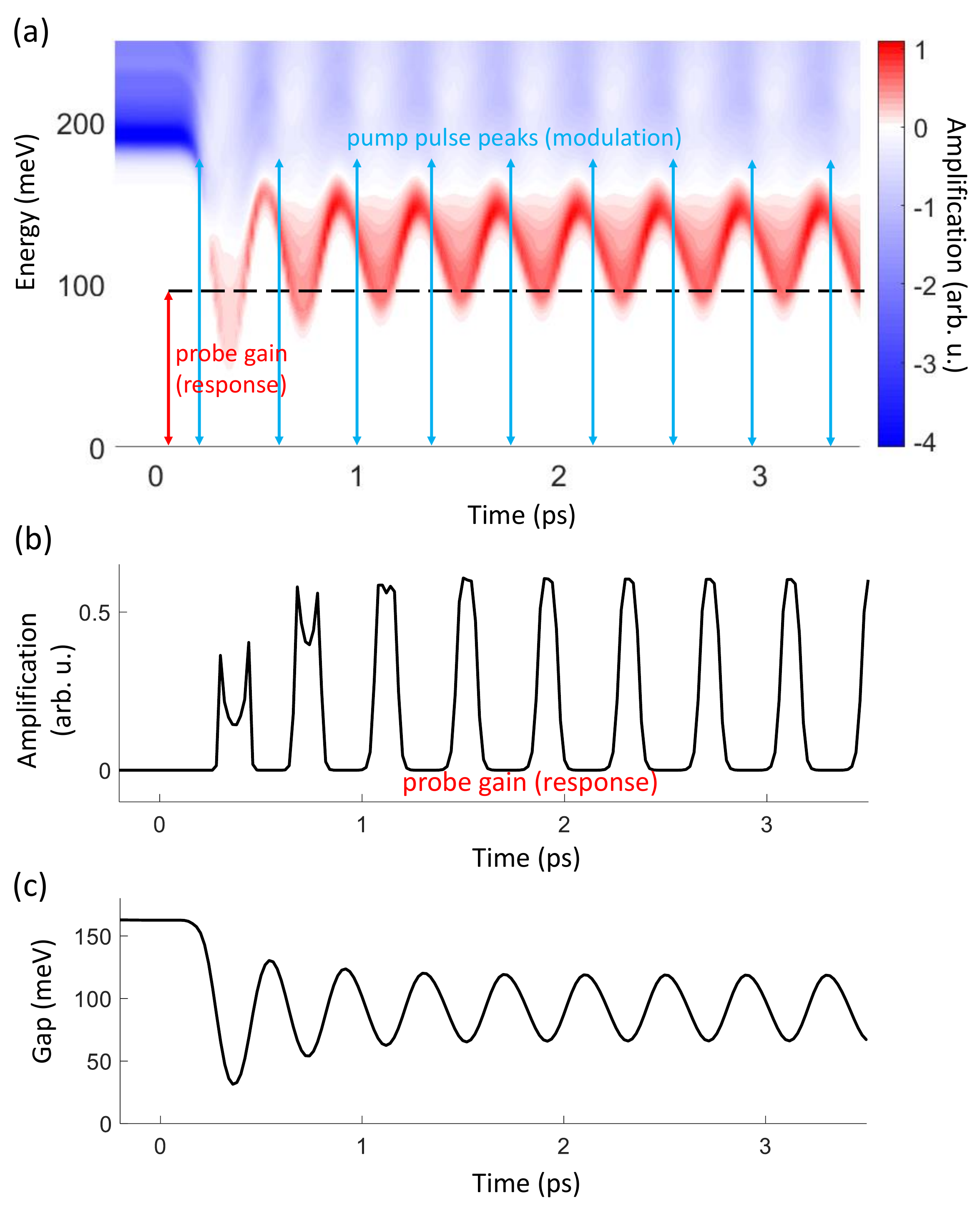}
\caption{(a) Small signal gain spectrum vs. time during periodical pump pulse peaks (modulation) in a quenched CDW-phase. (b) Small signal gain for a cw probe with a selected frequency (see black dashed line in (a)). (c) Gap closing vs. time during periodical pump pulse peaks (modulation).}
\label{figure5}
\end{figure}

So far we have applied our approach to the band dynamics and scattering transitions between time-dependent states for the case of impulsive excitation and subsequent relaxation. In this section we consider a modulation of the CDW material by a sequence of optical drive pulses. We wish to characterize the optical properties of the material under these driving conditions \emph{in the linear regime}. To do this in a meaningful way for a two-dimensional material we define a small-signal gain as is done for quantum wells as active material \cite{chow1999semiconductor} in, say, an optical amplifier: One imagines that the two-dimensional material with quantization area $A$ is stacked in $z$ direction with a layer density $N_L$ (suitable for two and quasi-two dimensional materials). The resulting structure is assumed to interact with a mode of the optical field, which leads to a confinement factor $\Gamma_c$. The small-signal gain is then defined 
\begin{equation}
\begin{split}
g(\omega) =  \alpha(\omega)\sum_{\lambda_{1}\neq\lambda_{2},\bvec{k}}&\left|\tilde{\mu}_{\lambda_{1}\bvec{k},\lambda_{2}\bvec{k}}\right|^{2} \left[\tilde{n}_{\lambda_{1}}(\bvec{k})-\tilde{n}_{\lambda_{2}}(\bvec{k})\right] \\
 & \times \mathcal{L}\Big(\frac{\tilde{\varepsilon}_{\lambda_{1}\bvec{k}}-\tilde{\varepsilon}_{\lambda_{2}\bvec{k}}}{\hbar}-\omega\Big) 
\end{split}
\label{eq:g-omega}
\end{equation}
where $\mathcal{L}$ denotes a Lorentzian lineshape function
and
\begin{equation}
\alpha(\omega) =\omega \Gamma_{c} N_{L} \big(c_{0}\varepsilon_{0}n_{b}\hbar\gamma_{\text{deph}}^{p}A\big)^{-1}
\end{equation}
In principle, the small-signal gain depends not only on concrete material specific properties, but also on the surrounding heterostructure. In the following, we avoid these complications necessary for the concrete design of a device such as an optical amplifier, and focus on the qualitative results for the optical amplification in a CDW-phase of a quasi-two dimensional material and provide only arbitrary units for the small-signal gain. 

Figure~\ref{figure5} shows the dynamics of the small-signal gain together with gap dynamics for a modulation due to a sequence of drive pulses. In Fig.~\ref{figure5}(a) the frequency and time-dependence of the small signal gain are shown as a color plot. The time dependence arises from drive pulses with the same characterisctics as before, which are now repeated as depicted by the blue arrows. The optical field creates carriers (electrons and holes) well away from the band gap, and trigger a carrier dynamics that --- with a time delay --- move the system out of the CDW phase so that optical gain becomes available at photon energies below the ``cold'' gap in the CDW phase. At photon energies well below 100 meV a band gap exists at all times and no amplification occurs at all, at energies ranging from 150 meV to almost 200 meV there is a pronounced amplification at almost all times. We therefore choose a fixed photon energy of about 100 meV for the probe gain because at this energy a pronounced amplification exists for short periods of time. This is illustrated in more detail in Fig.~\ref{figure5}(b) which shows the amplification at the chosen probe photon energy. A good contrast between maximum pulsed amplification and almost vanishing amplification between the pulses is achieved. Fig.~\ref{figure5}(c) shows the corresponding band gap dynamics due to the modulation by the drive pulses, which underlies the dynamics of the frequency-dependent gain in Fig.~\ref{figure5}(a). Note, however, that the absolute bottom of the band gap shown in Fig.~\ref{figure5}(c) is not optimal for the choice of probe photon energy because phase space and quenching of the anomalous matrix elements make the contributions from these k-space regions to Eq.~\ref{eq:g-omega} small. 

The theoretical study presented here suggests that this system could realize  a new kind of optical amplifier with the potential for high modulation frequencies in the mid-infrared regime. Our calculation show an intrinsic system response in the subpicosecond regime to a controlled optical excitation. We speculate that the inversion and modulation might also be induced by carrier injection processes and the probe gain might be utilized in a cavity to design an amplifier in the mid-infrared spectral region with high modulation frequencies.

However, there remain open questions that have to be investigated before such a device might be realized. Most importantly the material design of the active material that defines the strength of and the ratio between the excitonic and lattice contributions in the quasi-particle band-dynamics, i.e., the existence and the characteristics of such a CDW phase depends crucially on the proper material or alloy, and might be influenced by strain. Increased research in the material design might improve the performance of the prospective amplifier and also the already wide temperature range of such phases~\cite{coleman1988scanning}. 

\section{Conclusion}\label{Conclusion}

We investigated theoretically the formation of a CDW phase in a quasi-two dimensional material and the quenching of the phase due to optical excitation. We used a dynamical approach that includes excitonic and a lattice contributions in the quasi-particle band-dynamics and described the CDW phase embedded in the normal phase using generalized representation and projection techniques. The modeling of microscopic carrier dynamics included carrier-carrier scattering with a time-dependent screening and cooling due to carrier-phonon scattering. In the framework of our model, we were able to investigate the appearance of anomalous expectation values and interaction processes and the alteration of this properties induced by the quenching of the phase. We illustrated the time-dependent fading-in and fading-out of additional bands and anomalous expectation values as well as anomalous scattering contributions. Based on our theoretical approach, we propose a optical amplifier with the potential for high modulation frequencies in the mid-infrared regime.

%%%%%%%%%%%%%%%%%%%%%%%%%%%%%%%%%%%%%%%%%%%%%%%%%%%%%%%%%%%%%%%%%%%%

\begin{acknowledgments}
H.C.S acknowledges support from DFG through SFB/TRR 173 “Spin + X” %(project~B03)
\end{acknowledgments}

%%%%%%%%%%%%%%%%%%%%%%%%%%%%%%%%%%%%%%%%%%%%%%%%%%%%%%%%%%%%%%%%%%%%

\appendix 

\section{Tight-binding model}\label{AppendixA}

The prototypical band structure is calculated via diagonalization of a tight-binding Hamiltonian $H_{TB}$ similar to Ref.~\onlinecite{van2010exciton} to describe the band structure of the characteristic transition metal dichalcogenide. In the tight-binding Hamiltonian we consider the three t2g orbitals from the transition metal atom and six p-orbitals from the two chalcogen atoms in the unit cell of the normal-phase. We do not include a spin-orbit coupling in the calculation of the prototypical band structure. For example the lattice vectors $L_{1}=(3.54,0,0)$, $L_{2}=(-1.77,3.07,0)$ and $L_{3}=(0,0,6.01)$ in \AA{} would span a unit cell with the atom basis $B_{d}=(0,0,0)$ for the transition metal atom as well as $B_{p,1}=(0.33,-0.33,0.26)$ and $B_{p,2}=(-0.33, 0.33 -0.26)$ for the chalcogen atoms. Weak interactions between neighboring orbitals are usually neglected and the remaining interactions are expressed in terms of Slater-Koster integrals.\cite{slater1954simplified} The bond integrals between two orbitals are distinguished among $\sigma$, $\pi$ or eventually $\delta$ bondings. The resulting band structure around the high symmetry points of the small band-gap TMDCs is often highly unisotropic like in TiSe$_{2}$ as reported, e.g., in Ref.~\onlinecite{monney2010temperature}. We assume for the on-site energies (not normalized to the Fermi surface) $\epsilon^{t2g}_{0} =-11.0$ and $\epsilon^{p}_{0} =-14.8$ in eV and for the tight-binding coupling-elements dd$_{\sigma}=-0.7$, dd$_{\pi}=0.6$, dd$_{\delta}=-0.1$, pd$_{\sigma}=1.1$, pd$_{\pi}=0.9$, pp$_{\sigma}=1.0$ and pp$_{\pi}=-0.1$ in eV. In this prototypical tight-binding model, the nonisotropic band dispersion of the bands is nonparabolic and the Fermi surface cuts the indirect band-overlap between the chalcogen-like and transition-metal-like bands in the normal phase. The nesting vector (coincide with ordering wave vectors) is half the size of the normal-phase Brillouin zone (i.e. the ordering wave vectors $Q_{\varGamma M}$ pointing from the $\Gamma$-point to the $M$-points) and in the CDW phase we obtain a Mexican-hat shaped geometry with the Fermi surface between the small band-gap of the Mexican-hat shaped bands. 

\section{Interaction processes on longer timescales}\label{AppendixB}

After a quasi-equilibrium in the semiconductor CDW phase is reached, interaction processes on longer timescales are crucial for the further evolution of the dynamics. To take this into account, we include non-radiative losses and the effects of spontaneous emissions via
\begin{equation}
\begin{split}
\frac{d}{dt}\tilde{n}_{\lambda_{1}}(\bvec{k}) & = \gamma_{nr}\left(\tilde{f}_{\lambda_{1}}(\bvec{k})-\tilde{n}_{\lambda_{1}}(\bvec{k})\right) 
\\ & +\sum_{\lambda_{2}\neq\lambda_{1}}\frac{\left(\tilde{f}_{\lambda_{1}}(\bvec{k})-\tilde{n}_{\lambda_{1}}(\bvec{k})\right)\left(\tilde{f}_{\lambda_{2}}(\bvec{k})-\tilde{n}_{\lambda_{2}}(\bvec{k})\right)}{\tau_{sp}^{\lambda_{1}\bvec{k},\lambda_{2}\bvec{k}}} 
\end{split}
\end{equation}
where $\gamma_{nr}$ is the non-radiative loss rate, $f_{\lambda_{1}}(\bvec{k})$ is the equilibrium Fermi distribution for the current band structure and the spontaneous emission rate is approximated as 
\begin{eqnarray}
\tau_{sp}^{\lambda_{1}\bvec{k},\lambda_{2}\bvec{k}}=\left|\frac{\tilde{\mu}_{\lambda_{1}\bvec{k},\lambda_{2}\bvec{k}}}{\mu_{0}}\right|^{-2}\tau_{sp}^{\mu_{0}}
\end{eqnarray}
where $\gamma_{sp} = \left( \tau_{sp}^{\mu_{0}} \right) ^{-1}$ is the spontaneous emission rate regarding a normalized dipole matrix element $\mu_{0}$. %However, non-radiative losses might still be the major effect on longer timescales for a mid-infrared semiconductor band gap. 

%%%%%%%%%%%%%%%%%%%%%%%%%%%%%%%%%%%%%%%%%%%%%%%%%%%%%%%%%%%%%%%%%%%%

\bibliography{OptAmpCDW_v4}{}

\begin{thebibliography}{106}%
\makeatletter
\providecommand \@ifxundefined [1]{%
 \@ifx{#1\undefined}
}%
\providecommand \@ifnum [1]{%
 \ifnum #1\expandafter \@firstoftwo
 \else \expandafter \@secondoftwo
 \fi
}%
\providecommand \@ifx [1]{%
 \ifx #1\expandafter \@firstoftwo
 \else \expandafter \@secondoftwo
 \fi
}%
\providecommand \natexlab [1]{#1}%
\providecommand \enquote  [1]{``#1''}%
\providecommand \bibnamefont  [1]{#1}%
\providecommand \bibfnamefont [1]{#1}%
\providecommand \citenamefont [1]{#1}%
\providecommand \href@noop [0]{\@secondoftwo}%
\providecommand \href [0]{\begingroup \@sanitize@url \@href}%
\providecommand \@href[1]{\@@startlink{#1}\@@href}%
\providecommand \@@href[1]{\endgroup#1\@@endlink}%
\providecommand \@sanitize@url [0]{\catcode `\\12\catcode `\$12\catcode
  `\&12\catcode `\#12\catcode `\^12\catcode `\_12\catcode `\%12\relax}%
\providecommand \@@startlink[1]{}%
\providecommand \@@endlink[0]{}%
\providecommand \url  [0]{\begingroup\@sanitize@url \@url }%
\providecommand \@url [1]{\endgroup\@href {#1}{\urlprefix }}%
\providecommand \urlprefix  [0]{URL }%
\providecommand \Eprint [0]{\href }%
\providecommand \doibase [0]{https://doi.org/}%
\providecommand \selectlanguage [0]{\@gobble}%
\providecommand \bibinfo  [0]{\@secondoftwo}%
\providecommand \bibfield  [0]{\@secondoftwo}%
\providecommand \translation [1]{[#1]}%
\providecommand \BibitemOpen [0]{}%
\providecommand \bibitemStop [0]{}%
\providecommand \bibitemNoStop [0]{.\EOS\space}%
\providecommand \EOS [0]{\spacefactor3000\relax}%
\providecommand \BibitemShut  [1]{\csname bibitem#1\endcsname}%
\let\auto@bib@innerbib\@empty
%</preamble>
\bibitem [{\citenamefont {Gr\"uner}(1988)}]{gruner1988dynamics}%
  \BibitemOpen
  \bibfield  {author} {\bibinfo {author} {\bibfnamefont {G.}~\bibnamefont
  {Gr\"uner}},\ }\bibfield  {title} {\bibinfo {title} {The dynamics of
  charge-density waves},\ }\href {https://doi.org/10.1103/RevModPhys.60.1129}
  {\bibfield  {journal} {\bibinfo  {journal} {Rev. Mod. Phys.}\ }\textbf
  {\bibinfo {volume} {60}},\ \bibinfo {pages} {1129} (\bibinfo {year}
  {1988})}\BibitemShut {NoStop}%
\bibitem [{\citenamefont {Kohn}(1967)}]{kohn1967excitonic}%
  \BibitemOpen
  \bibfield  {author} {\bibinfo {author} {\bibfnamefont {W.}~\bibnamefont
  {Kohn}},\ }\bibfield  {title} {\bibinfo {title} {{Excitonic Phases}},\ }\href
  {https://doi.org/10.1103/PhysRevLett.19.439} {\bibfield  {journal} {\bibinfo
  {journal} {Phys. Rev. Lett.}\ }\textbf {\bibinfo {volume} {19}},\ \bibinfo
  {pages} {439} (\bibinfo {year} {1967})}\BibitemShut {NoStop}%
\bibitem [{\citenamefont {Aebi}\ \emph {et~al.}(2001)\citenamefont {Aebi},
  \citenamefont {Pillo}, \citenamefont {Berger},\ and\ \citenamefont
  {L{\'e}vy}}]{aebi2001search}%
  \BibitemOpen
  \bibfield  {author} {\bibinfo {author} {\bibfnamefont {P.}~\bibnamefont
  {Aebi}}, \bibinfo {author} {\bibfnamefont {T.}~\bibnamefont {Pillo}},
  \bibinfo {author} {\bibfnamefont {H.}~\bibnamefont {Berger}},\ and\ \bibinfo
  {author} {\bibfnamefont {F.}~\bibnamefont {L{\'e}vy}},\ }\bibfield  {title}
  {\bibinfo {title} {{On the search for Fermi surface nesting in quasi-2D
  materials}},\ }\href@noop {} {\bibfield  {journal} {\bibinfo  {journal}
  {Journal of electron spectroscopy and related phenomena}\ }\textbf {\bibinfo
  {volume} {117}},\ \bibinfo {pages} {433} (\bibinfo {year}
  {2001})}\BibitemShut {NoStop}%
\bibitem [{\citenamefont {Chan}\ and\ \citenamefont
  {Heine}(1973)}]{chan1973spin}%
  \BibitemOpen
  \bibfield  {author} {\bibinfo {author} {\bibfnamefont {S.-K.}\ \bibnamefont
  {Chan}}\ and\ \bibinfo {author} {\bibfnamefont {V.}~\bibnamefont {Heine}},\
  }\bibfield  {title} {\bibinfo {title} {{Spin density wave and soft phonon
  mode from nesting Fermi surfaces}},\ }\href@noop {} {\bibfield  {journal}
  {\bibinfo  {journal} {Journal of Physics F: Metal Physics}\ }\textbf
  {\bibinfo {volume} {3}},\ \bibinfo {pages} {795} (\bibinfo {year}
  {1973})}\BibitemShut {NoStop}%
\bibitem [{\citenamefont {Hughes}(1977)}]{hughes1977structural}%
  \BibitemOpen
  \bibfield  {author} {\bibinfo {author} {\bibfnamefont {H.~P.}\ \bibnamefont
  {Hughes}},\ }\bibfield  {title} {\bibinfo {title} {{Structural distortion in
  {TiSe$_{2}$} and related materials - a possible Jahn-Teller effect?}},\
  }\href@noop {} {\bibfield  {journal} {\bibinfo  {journal} {Journal of Physics
  C: Solid State Physics}\ }\textbf {\bibinfo {volume} {10}},\ \bibinfo {pages}
  {L319} (\bibinfo {year} {1977})}\BibitemShut {NoStop}%
\bibitem [{\citenamefont {Rossnagel}(2011)}]{rossnagel2011origin}%
  \BibitemOpen
  \bibfield  {author} {\bibinfo {author} {\bibfnamefont {K.}~\bibnamefont
  {Rossnagel}},\ }\bibfield  {title} {\bibinfo {title} {On the origin of
  charge-density waves in select layered transition-metal dichalcogenides},\
  }\href@noop {} {\bibfield  {journal} {\bibinfo  {journal} {Journal of
  Physics: Condensed Matter}\ }\textbf {\bibinfo {volume} {23}},\ \bibinfo
  {pages} {213001} (\bibinfo {year} {2011})}\BibitemShut {NoStop}%
\bibitem [{\citenamefont {Chhowalla}\ \emph {et~al.}(2013)\citenamefont
  {Chhowalla}, \citenamefont {Shin}, \citenamefont {Eda}, \citenamefont {Li},
  \citenamefont {Loh},\ and\ \citenamefont {Zhang}}]{chhowalla2013chemistry}%
  \BibitemOpen
  \bibfield  {author} {\bibinfo {author} {\bibfnamefont {M.}~\bibnamefont
  {Chhowalla}}, \bibinfo {author} {\bibfnamefont {H.~S.}\ \bibnamefont {Shin}},
  \bibinfo {author} {\bibfnamefont {G.}~\bibnamefont {Eda}}, \bibinfo {author}
  {\bibfnamefont {L.-J.}\ \bibnamefont {Li}}, \bibinfo {author} {\bibfnamefont
  {K.~P.}\ \bibnamefont {Loh}},\ and\ \bibinfo {author} {\bibfnamefont
  {H.}~\bibnamefont {Zhang}},\ }\bibfield  {title} {\bibinfo {title} {The
  chemistry of two-dimensional layered transition metal dichalcogenide
  nanosheets},\ }\href@noop {} {\bibfield  {journal} {\bibinfo  {journal}
  {Nature chemistry}\ }\textbf {\bibinfo {volume} {5}},\ \bibinfo {pages} {263}
  (\bibinfo {year} {2013})}\BibitemShut {NoStop}%
\bibitem [{\citenamefont {Zhu}\ \emph {et~al.}(2015)\citenamefont {Zhu},
  \citenamefont {Cao}, \citenamefont {Zhang}, \citenamefont {Plummer},\ and\
  \citenamefont {Guo}}]{zhu2015classification}%
  \BibitemOpen
  \bibfield  {author} {\bibinfo {author} {\bibfnamefont {X.}~\bibnamefont
  {Zhu}}, \bibinfo {author} {\bibfnamefont {Y.}~\bibnamefont {Cao}}, \bibinfo
  {author} {\bibfnamefont {J.}~\bibnamefont {Zhang}}, \bibinfo {author}
  {\bibfnamefont {E.}~\bibnamefont {Plummer}},\ and\ \bibinfo {author}
  {\bibfnamefont {J.}~\bibnamefont {Guo}},\ }\bibfield  {title} {\bibinfo
  {title} {Classification of charge density waves based on their nature},\
  }\href@noop {} {\bibfield  {journal} {\bibinfo  {journal} {Proceedings of the
  National Academy of Sciences}\ }\textbf {\bibinfo {volume} {112}},\ \bibinfo
  {pages} {2367} (\bibinfo {year} {2015})}\BibitemShut {NoStop}%
\bibitem [{\citenamefont {Flicker}\ and\ \citenamefont
  {Van~Wezel}(2015)}]{flicker2015charge}%
  \BibitemOpen
  \bibfield  {author} {\bibinfo {author} {\bibfnamefont {F.}~\bibnamefont
  {Flicker}}\ and\ \bibinfo {author} {\bibfnamefont {J.}~\bibnamefont
  {Van~Wezel}},\ }\bibfield  {title} {\bibinfo {title} {Charge order from
  orbital-dependent coupling evidenced by {NbSe$_{2}$}},\ }\href@noop {}
  {\bibfield  {journal} {\bibinfo  {journal} {Nature communications}\ }\textbf
  {\bibinfo {volume} {6}},\ \bibinfo {pages} {7034} (\bibinfo {year}
  {2015})}\BibitemShut {NoStop}%
\bibitem [{\citenamefont {Cho}\ \emph {et~al.}(2016)\citenamefont {Cho},
  \citenamefont {van~den Brink}, \citenamefont {Fehske}, \citenamefont
  {Becker},\ and\ \citenamefont {Sykora}}]{cho2016unconventional}%
  \BibitemOpen
  \bibfield  {author} {\bibinfo {author} {\bibfnamefont {D.-N.}\ \bibnamefont
  {Cho}}, \bibinfo {author} {\bibfnamefont {J.}~\bibnamefont {van~den Brink}},
  \bibinfo {author} {\bibfnamefont {H.}~\bibnamefont {Fehske}}, \bibinfo
  {author} {\bibfnamefont {K.~W.}\ \bibnamefont {Becker}},\ and\ \bibinfo
  {author} {\bibfnamefont {S.}~\bibnamefont {Sykora}},\ }\bibfield  {title}
  {\bibinfo {title} {Unconventional superconductivity and interaction induced
  {Fermi} surface reconstruction in the two-dimensional {Edwards} model},\
  }\href@noop {} {\bibfield  {journal} {\bibinfo  {journal} {Scientific
  reports}\ }\textbf {\bibinfo {volume} {6}},\ \bibinfo {pages} {22548}
  (\bibinfo {year} {2016})}\BibitemShut {NoStop}%
\bibitem [{\citenamefont {Leroux}\ \emph {et~al.}(2018)\citenamefont {Leroux},
  \citenamefont {Cario}, \citenamefont {Bosak},\ and\ \citenamefont
  {Rodi\`ere}}]{leroux2018traces}%
  \BibitemOpen
  \bibfield  {author} {\bibinfo {author} {\bibfnamefont {M.}~\bibnamefont
  {Leroux}}, \bibinfo {author} {\bibfnamefont {L.}~\bibnamefont {Cario}},
  \bibinfo {author} {\bibfnamefont {A.}~\bibnamefont {Bosak}},\ and\ \bibinfo
  {author} {\bibfnamefont {P.}~\bibnamefont {Rodi\`ere}},\ }\bibfield  {title}
  {\bibinfo {title} {Traces of charge density waves in {NbS}$_{2}$},\ }\href
  {https://doi.org/10.1103/PhysRevB.97.195140} {\bibfield  {journal} {\bibinfo
  {journal} {Phys. Rev. B}\ }\textbf {\bibinfo {volume} {97}},\ \bibinfo
  {pages} {195140} (\bibinfo {year} {2018})}\BibitemShut {NoStop}%
\bibitem [{\citenamefont {Nakata}\ \emph {et~al.}(2018)\citenamefont {Nakata},
  \citenamefont {Sugawara}, \citenamefont {Ichinokura}, \citenamefont {Okada},
  \citenamefont {Hitosugi}, \citenamefont {Koretsune}, \citenamefont {Ueno},
  \citenamefont {Hasegawa}, \citenamefont {Takahashi},\ and\ \citenamefont
  {Sato}}]{nakata2018anisotropic}%
  \BibitemOpen
  \bibfield  {author} {\bibinfo {author} {\bibfnamefont {Y.}~\bibnamefont
  {Nakata}}, \bibinfo {author} {\bibfnamefont {K.}~\bibnamefont {Sugawara}},
  \bibinfo {author} {\bibfnamefont {S.}~\bibnamefont {Ichinokura}}, \bibinfo
  {author} {\bibfnamefont {Y.}~\bibnamefont {Okada}}, \bibinfo {author}
  {\bibfnamefont {T.}~\bibnamefont {Hitosugi}}, \bibinfo {author}
  {\bibfnamefont {T.}~\bibnamefont {Koretsune}}, \bibinfo {author}
  {\bibfnamefont {K.}~\bibnamefont {Ueno}}, \bibinfo {author} {\bibfnamefont
  {S.}~\bibnamefont {Hasegawa}}, \bibinfo {author} {\bibfnamefont
  {T.}~\bibnamefont {Takahashi}},\ and\ \bibinfo {author} {\bibfnamefont
  {T.}~\bibnamefont {Sato}},\ }\bibfield  {title} {\bibinfo {title}
  {{Anisotropic band splitting in monolayer {NbSe$_{2}$}: Implications for
  superconductivity and charge density wave}},\ }\href@noop {} {\bibfield
  {journal} {\bibinfo  {journal} {npj 2D Materials and Applications}\ }\textbf
  {\bibinfo {volume} {2}},\ \bibinfo {pages} {12} (\bibinfo {year}
  {2018})}\BibitemShut {NoStop}%
\bibitem [{\citenamefont {Ueda}\ \emph {et~al.}(2021)\citenamefont {Ueda},
  \citenamefont {Porer}, \citenamefont {Mardegan}, \citenamefont {Parchenko},
  \citenamefont {Gurung}, \citenamefont {Fabrizi}, \citenamefont
  {Ramakrishnan}, \citenamefont {Boie}, \citenamefont {Neugebauer},
  \citenamefont {Burganov}, \citenamefont {Burian}, \citenamefont {Johnson},
  \citenamefont {Rossnagel},\ and\ \citenamefont
  {Staub}}]{ueda2020correlation}%
  \BibitemOpen
  \bibfield  {author} {\bibinfo {author} {\bibfnamefont {H.}~\bibnamefont
  {Ueda}}, \bibinfo {author} {\bibfnamefont {M.}~\bibnamefont {Porer}},
  \bibinfo {author} {\bibfnamefont {J.~R.~L.}\ \bibnamefont {Mardegan}},
  \bibinfo {author} {\bibfnamefont {S.}~\bibnamefont {Parchenko}}, \bibinfo
  {author} {\bibfnamefont {N.}~\bibnamefont {Gurung}}, \bibinfo {author}
  {\bibfnamefont {F.}~\bibnamefont {Fabrizi}}, \bibinfo {author} {\bibfnamefont
  {M.}~\bibnamefont {Ramakrishnan}}, \bibinfo {author} {\bibfnamefont
  {L.}~\bibnamefont {Boie}}, \bibinfo {author} {\bibfnamefont {M.~J.}\
  \bibnamefont {Neugebauer}}, \bibinfo {author} {\bibfnamefont
  {B.}~\bibnamefont {Burganov}}, \bibinfo {author} {\bibfnamefont
  {M.}~\bibnamefont {Burian}}, \bibinfo {author} {\bibfnamefont {S.~L.}\
  \bibnamefont {Johnson}}, \bibinfo {author} {\bibfnamefont {K.}~\bibnamefont
  {Rossnagel}},\ and\ \bibinfo {author} {\bibfnamefont {U.}~\bibnamefont
  {Staub}},\ }\bibfield  {title} {\bibinfo {title} {Correlation between
  electronic and structural orders in {1T}$-${TiSe$_{2}$}},\ }\href
  {https://doi.org/10.1103/PhysRevResearch.3.L022003} {\bibfield  {journal}
  {\bibinfo  {journal} {Phys. Rev. Research}\ }\textbf {\bibinfo {volume}
  {3}},\ \bibinfo {pages} {L022003} (\bibinfo {year} {2021})}\BibitemShut
  {NoStop}%
\bibitem [{\citenamefont {Chen}\ \emph
  {et~al.}(2015{\natexlab{a}})\citenamefont {Chen}, \citenamefont {Chan},
  \citenamefont {Fang}, \citenamefont {Zhang}, \citenamefont {Chou},
  \citenamefont {Mo}, \citenamefont {Hussain}, \citenamefont {Fedorov},\ and\
  \citenamefont {Chiang}}]{chen2015charge}%
  \BibitemOpen
  \bibfield  {author} {\bibinfo {author} {\bibfnamefont {P.}~\bibnamefont
  {Chen}}, \bibinfo {author} {\bibfnamefont {Y.-H.}\ \bibnamefont {Chan}},
  \bibinfo {author} {\bibfnamefont {X.-Y.}\ \bibnamefont {Fang}}, \bibinfo
  {author} {\bibfnamefont {Y.}~\bibnamefont {Zhang}}, \bibinfo {author}
  {\bibfnamefont {M.-Y.}\ \bibnamefont {Chou}}, \bibinfo {author}
  {\bibfnamefont {S.-K.}\ \bibnamefont {Mo}}, \bibinfo {author} {\bibfnamefont
  {Z.}~\bibnamefont {Hussain}}, \bibinfo {author} {\bibfnamefont {A.-V.}\
  \bibnamefont {Fedorov}},\ and\ \bibinfo {author} {\bibfnamefont {T.-C.}\
  \bibnamefont {Chiang}},\ }\bibfield  {title} {\bibinfo {title} {Charge
  density wave transition in single-layer titanium diselenide},\ }\href@noop {}
  {\bibfield  {journal} {\bibinfo  {journal} {Nature communications}\ }\textbf
  {\bibinfo {volume} {6}},\ \bibinfo {pages} {8943} (\bibinfo {year}
  {2015}{\natexlab{a}})}\BibitemShut {NoStop}%
\bibitem [{\citenamefont {Chen}\ \emph
  {et~al.}(2016{\natexlab{a}})\citenamefont {Chen}, \citenamefont {Chan},
  \citenamefont {Fang}, \citenamefont {Mo}, \citenamefont {Hussain},
  \citenamefont {Fedorov}, \citenamefont {Chou},\ and\ \citenamefont
  {Chiang}}]{chen2016hidden}%
  \BibitemOpen
  \bibfield  {author} {\bibinfo {author} {\bibfnamefont {P.}~\bibnamefont
  {Chen}}, \bibinfo {author} {\bibfnamefont {Y.-H.}\ \bibnamefont {Chan}},
  \bibinfo {author} {\bibfnamefont {X.-Y.}\ \bibnamefont {Fang}}, \bibinfo
  {author} {\bibfnamefont {S.-K.}\ \bibnamefont {Mo}}, \bibinfo {author}
  {\bibfnamefont {Z.}~\bibnamefont {Hussain}}, \bibinfo {author} {\bibfnamefont
  {A.-V.}\ \bibnamefont {Fedorov}}, \bibinfo {author} {\bibfnamefont
  {M.}~\bibnamefont {Chou}},\ and\ \bibinfo {author} {\bibfnamefont {T.-C.}\
  \bibnamefont {Chiang}},\ }\bibfield  {title} {\bibinfo {title} {{Hidden Order
  and Dimensional Crossover of the Charge Density Waves in {TiSe$_{2}$}}},\
  }\href@noop {} {\bibfield  {journal} {\bibinfo  {journal} {Scientific
  reports}\ }\textbf {\bibinfo {volume} {6}},\ \bibinfo {pages} {37910}
  (\bibinfo {year} {2016}{\natexlab{a}})}\BibitemShut {NoStop}%
\bibitem [{\citenamefont {Porer}\ \emph {et~al.}(2014)\citenamefont {Porer},
  \citenamefont {Leierseder}, \citenamefont {M{\'e}nard}, \citenamefont
  {Dachraoui}, \citenamefont {Mouchliadis}, \citenamefont {Perakis},
  \citenamefont {Heinzmann}, \citenamefont {Demsar}, \citenamefont
  {Rossnagel},\ and\ \citenamefont {Huber}}]{porer2014non}%
  \BibitemOpen
  \bibfield  {author} {\bibinfo {author} {\bibfnamefont {M.}~\bibnamefont
  {Porer}}, \bibinfo {author} {\bibfnamefont {U.}~\bibnamefont {Leierseder}},
  \bibinfo {author} {\bibfnamefont {J.-M.}\ \bibnamefont {M{\'e}nard}},
  \bibinfo {author} {\bibfnamefont {H.}~\bibnamefont {Dachraoui}}, \bibinfo
  {author} {\bibfnamefont {L.}~\bibnamefont {Mouchliadis}}, \bibinfo {author}
  {\bibfnamefont {I.}~\bibnamefont {Perakis}}, \bibinfo {author} {\bibfnamefont
  {U.}~\bibnamefont {Heinzmann}}, \bibinfo {author} {\bibfnamefont
  {J.}~\bibnamefont {Demsar}}, \bibinfo {author} {\bibfnamefont
  {K.}~\bibnamefont {Rossnagel}},\ and\ \bibinfo {author} {\bibfnamefont
  {R.}~\bibnamefont {Huber}},\ }\bibfield  {title} {\bibinfo {title}
  {Non-thermal separation of electronic and structural orders in a persisting
  charge density wave},\ }\href@noop {} {\bibfield  {journal} {\bibinfo
  {journal} {Nature materials}\ }\textbf {\bibinfo {volume} {13}},\ \bibinfo
  {pages} {857} (\bibinfo {year} {2014})}\BibitemShut {NoStop}%
\bibitem [{\citenamefont {Zhang}\ \emph {et~al.}(2018)\citenamefont {Zhang},
  \citenamefont {Yang}, \citenamefont {Lei}, \citenamefont {Lu}, \citenamefont
  {Li}, \citenamefont {Jia}, \citenamefont {Song}, \citenamefont {Sun},
  \citenamefont {Chen}, \citenamefont {Li},\ and\ \citenamefont
  {Li}}]{zhang2016unveiling}%
  \BibitemOpen
  \bibfield  {author} {\bibinfo {author} {\bibfnamefont {K.-W.}\ \bibnamefont
  {Zhang}}, \bibinfo {author} {\bibfnamefont {C.-L.}\ \bibnamefont {Yang}},
  \bibinfo {author} {\bibfnamefont {B.}~\bibnamefont {Lei}}, \bibinfo {author}
  {\bibfnamefont {P.}~\bibnamefont {Lu}}, \bibinfo {author} {\bibfnamefont
  {X.-B.}\ \bibnamefont {Li}}, \bibinfo {author} {\bibfnamefont {Z.-Y.}\
  \bibnamefont {Jia}}, \bibinfo {author} {\bibfnamefont {Y.-H.}\ \bibnamefont
  {Song}}, \bibinfo {author} {\bibfnamefont {J.}~\bibnamefont {Sun}}, \bibinfo
  {author} {\bibfnamefont {X.}~\bibnamefont {Chen}}, \bibinfo {author}
  {\bibfnamefont {J.-X.}\ \bibnamefont {Li}},\ and\ \bibinfo {author}
  {\bibfnamefont {S.-C.}\ \bibnamefont {Li}},\ }\bibfield  {title} {\bibinfo
  {title} {Unveiling the charge density wave inhomogeneity and pseudogap state
  in {1T}$-${TiSe$_{2}$}},\ }\href@noop {} {\bibfield  {journal} {\bibinfo
  {journal} {Science Bulletin}\ }\textbf {\bibinfo {volume} {63}},\ \bibinfo
  {pages} {426} (\bibinfo {year} {2018})}\BibitemShut {NoStop}%
\bibitem [{\citenamefont {Rossnagel}\ \emph {et~al.}(2002)\citenamefont
  {Rossnagel}, \citenamefont {Kipp},\ and\ \citenamefont
  {Skibowski}}]{rossnagel2002charge}%
  \BibitemOpen
  \bibfield  {author} {\bibinfo {author} {\bibfnamefont {K.}~\bibnamefont
  {Rossnagel}}, \bibinfo {author} {\bibfnamefont {L.}~\bibnamefont {Kipp}},\
  and\ \bibinfo {author} {\bibfnamefont {M.}~\bibnamefont {Skibowski}},\
  }\bibfield  {title} {\bibinfo {title} {Charge-density-wave phase transition
  in {1T}$-${TiSe$_{2}$}$:$ {Excitonic insulator versus band-type Jahn-Teller
  mechanism}},\ }\href {https://doi.org/10.1103/PhysRevB.65.235101} {\bibfield
  {journal} {\bibinfo  {journal} {Phys. Rev. B}\ }\textbf {\bibinfo {volume}
  {65}},\ \bibinfo {pages} {235101} (\bibinfo {year} {2002})}\BibitemShut
  {NoStop}%
\bibitem [{\citenamefont {Karam}\ \emph {et~al.}(2018)\citenamefont {Karam},
  \citenamefont {Hu},\ and\ \citenamefont {Blake}}]{karam2018strongly}%
  \BibitemOpen
  \bibfield  {author} {\bibinfo {author} {\bibfnamefont {T.~E.}\ \bibnamefont
  {Karam}}, \bibinfo {author} {\bibfnamefont {J.}~\bibnamefont {Hu}},\ and\
  \bibinfo {author} {\bibfnamefont {G.~A.}\ \bibnamefont {Blake}},\ }\bibfield
  {title} {\bibinfo {title} {{Strongly Coupled Electron--Phonon Dynamics in
  Few-Layer {TiSe$_{2}$} Exfoliates}},\ }\href@noop {} {\bibfield  {journal}
  {\bibinfo  {journal} {ACS Photonics}\ }\textbf {\bibinfo {volume} {5}},\
  \bibinfo {pages} {1228} (\bibinfo {year} {2018})}\BibitemShut {NoStop}%
\bibitem [{\citenamefont {Hellgren}\ \emph {et~al.}(2017)\citenamefont
  {Hellgren}, \citenamefont {Baima}, \citenamefont {Bianco}, \citenamefont
  {Calandra}, \citenamefont {Mauri},\ and\ \citenamefont
  {Wirtz}}]{hellgren2017critical}%
  \BibitemOpen
  \bibfield  {author} {\bibinfo {author} {\bibfnamefont {M.}~\bibnamefont
  {Hellgren}}, \bibinfo {author} {\bibfnamefont {J.}~\bibnamefont {Baima}},
  \bibinfo {author} {\bibfnamefont {R.}~\bibnamefont {Bianco}}, \bibinfo
  {author} {\bibfnamefont {M.}~\bibnamefont {Calandra}}, \bibinfo {author}
  {\bibfnamefont {F.}~\bibnamefont {Mauri}},\ and\ \bibinfo {author}
  {\bibfnamefont {L.}~\bibnamefont {Wirtz}},\ }\bibfield  {title} {\bibinfo
  {title} {{Critical Role of the Exchange Interaction for the Electronic
  Structure and Charge-Density-Wave Formation in {TiSe$_{2}$}}},\ }\href
  {https://doi.org/10.1103/PhysRevLett.119.176401} {\bibfield  {journal}
  {\bibinfo  {journal} {Phys. Rev. Lett.}\ }\textbf {\bibinfo {volume} {119}},\
  \bibinfo {pages} {176401} (\bibinfo {year} {2017})}\BibitemShut {NoStop}%
\bibitem [{\citenamefont {Monney}\ \emph {et~al.}(2009)\citenamefont {Monney},
  \citenamefont {Cercellier}, \citenamefont {Clerc}, \citenamefont {Battaglia},
  \citenamefont {Schwier}, \citenamefont {Didiot}, \citenamefont {Garnier},
  \citenamefont {Beck}, \citenamefont {Aebi}, \citenamefont {Berger},
  \citenamefont {Forr\'o},\ and\ \citenamefont
  {Patthey}}]{monney2009spontaneous}%
  \BibitemOpen
  \bibfield  {author} {\bibinfo {author} {\bibfnamefont {C.}~\bibnamefont
  {Monney}}, \bibinfo {author} {\bibfnamefont {H.}~\bibnamefont {Cercellier}},
  \bibinfo {author} {\bibfnamefont {F.}~\bibnamefont {Clerc}}, \bibinfo
  {author} {\bibfnamefont {C.}~\bibnamefont {Battaglia}}, \bibinfo {author}
  {\bibfnamefont {E.~F.}\ \bibnamefont {Schwier}}, \bibinfo {author}
  {\bibfnamefont {C.}~\bibnamefont {Didiot}}, \bibinfo {author} {\bibfnamefont
  {M.~G.}\ \bibnamefont {Garnier}}, \bibinfo {author} {\bibfnamefont
  {H.}~\bibnamefont {Beck}}, \bibinfo {author} {\bibfnamefont {P.}~\bibnamefont
  {Aebi}}, \bibinfo {author} {\bibfnamefont {H.}~\bibnamefont {Berger}},
  \bibinfo {author} {\bibfnamefont {L.}~\bibnamefont {Forr\'o}},\ and\ \bibinfo
  {author} {\bibfnamefont {L.}~\bibnamefont {Patthey}},\ }\bibfield  {title}
  {\bibinfo {title} {Spontaneous exciton condensation in {1T}$-${TiSe$_{2}$}$:$
  {BCS}--like approach},\ }\href {https://doi.org/10.1103/PhysRevB.79.045116}
  {\bibfield  {journal} {\bibinfo  {journal} {Phys. Rev. B}\ }\textbf {\bibinfo
  {volume} {79}},\ \bibinfo {pages} {045116} (\bibinfo {year}
  {2009})}\BibitemShut {NoStop}%
\bibitem [{\citenamefont {Monney}\ \emph {et~al.}(2010)\citenamefont {Monney},
  \citenamefont {Schwier}, \citenamefont {Garnier}, \citenamefont {Mariotti},
  \citenamefont {Didiot}, \citenamefont {Beck}, \citenamefont {Aebi},
  \citenamefont {Cercellier}, \citenamefont {Marcus}, \citenamefont
  {Battaglia}, \citenamefont {Berger},\ and\ \citenamefont
  {Titov}}]{monney2010temperature}%
  \BibitemOpen
  \bibfield  {author} {\bibinfo {author} {\bibfnamefont {C.}~\bibnamefont
  {Monney}}, \bibinfo {author} {\bibfnamefont {E.~F.}\ \bibnamefont {Schwier}},
  \bibinfo {author} {\bibfnamefont {M.~G.}\ \bibnamefont {Garnier}}, \bibinfo
  {author} {\bibfnamefont {N.}~\bibnamefont {Mariotti}}, \bibinfo {author}
  {\bibfnamefont {C.}~\bibnamefont {Didiot}}, \bibinfo {author} {\bibfnamefont
  {H.}~\bibnamefont {Beck}}, \bibinfo {author} {\bibfnamefont {P.}~\bibnamefont
  {Aebi}}, \bibinfo {author} {\bibfnamefont {H.}~\bibnamefont {Cercellier}},
  \bibinfo {author} {\bibfnamefont {J.}~\bibnamefont {Marcus}}, \bibinfo
  {author} {\bibfnamefont {C.}~\bibnamefont {Battaglia}}, \bibinfo {author}
  {\bibfnamefont {H.}~\bibnamefont {Berger}},\ and\ \bibinfo {author}
  {\bibfnamefont {A.~N.}\ \bibnamefont {Titov}},\ }\bibfield  {title} {\bibinfo
  {title} {Temperature-dependent photoemission on {1T}$-${TiSe$_{2}$}$:$
  {Interpretation} within the exciton condensate phase model},\ }\href
  {https://doi.org/10.1103/PhysRevB.81.155104} {\bibfield  {journal} {\bibinfo
  {journal} {Phys. Rev. B}\ }\textbf {\bibinfo {volume} {81}},\ \bibinfo
  {pages} {155104} (\bibinfo {year} {2010})}\BibitemShut {NoStop}%
\bibitem [{\citenamefont {Cazzaniga}\ \emph {et~al.}(2012)\citenamefont
  {Cazzaniga}, \citenamefont {Cercellier}, \citenamefont {Holzmann},
  \citenamefont {Monney}, \citenamefont {Aebi}, \citenamefont {Onida},\ and\
  \citenamefont {Olevano}}]{cazzaniga2012ab}%
  \BibitemOpen
  \bibfield  {author} {\bibinfo {author} {\bibfnamefont {M.}~\bibnamefont
  {Cazzaniga}}, \bibinfo {author} {\bibfnamefont {H.}~\bibnamefont
  {Cercellier}}, \bibinfo {author} {\bibfnamefont {M.}~\bibnamefont
  {Holzmann}}, \bibinfo {author} {\bibfnamefont {C.}~\bibnamefont {Monney}},
  \bibinfo {author} {\bibfnamefont {P.}~\bibnamefont {Aebi}}, \bibinfo {author}
  {\bibfnamefont {G.}~\bibnamefont {Onida}},\ and\ \bibinfo {author}
  {\bibfnamefont {V.}~\bibnamefont {Olevano}},\ }\bibfield  {title} {\bibinfo
  {title} {Ab initio many-body effects in {TiSe$_{2}$}: A possible excitonic
  insulator scenario from {GW} band-shape renormalization},\ }\href
  {https://doi.org/10.1103/PhysRevB.85.195111} {\bibfield  {journal} {\bibinfo
  {journal} {Phys. Rev. B}\ }\textbf {\bibinfo {volume} {85}},\ \bibinfo
  {pages} {195111} (\bibinfo {year} {2012})}\BibitemShut {NoStop}%
\bibitem [{\citenamefont {Monney}\ \emph {et~al.}(2012)\citenamefont {Monney},
  \citenamefont {Monney}, \citenamefont {Aebi},\ and\ \citenamefont
  {Beck}}]{monney2012electron}%
  \BibitemOpen
  \bibfield  {author} {\bibinfo {author} {\bibfnamefont {C.}~\bibnamefont
  {Monney}}, \bibinfo {author} {\bibfnamefont {G.}~\bibnamefont {Monney}},
  \bibinfo {author} {\bibfnamefont {P.}~\bibnamefont {Aebi}},\ and\ \bibinfo
  {author} {\bibfnamefont {H.}~\bibnamefont {Beck}},\ }\bibfield  {title}
  {\bibinfo {title} {Electron-hole fluctuation phase in {$1T$-TiSe${}_{2}$}},\
  }\href {https://doi.org/10.1103/PhysRevB.85.235150} {\bibfield  {journal}
  {\bibinfo  {journal} {Phys. Rev. B}\ }\textbf {\bibinfo {volume} {85}},\
  \bibinfo {pages} {235150} (\bibinfo {year} {2012})}\BibitemShut {NoStop}%
\bibitem [{\citenamefont {Li}\ \emph {et~al.}(2007)\citenamefont {Li},
  \citenamefont {Hu}, \citenamefont {Qian}, \citenamefont {Hsieh},
  \citenamefont {Hasan}, \citenamefont {Morosan}, \citenamefont {Cava},\ and\
  \citenamefont {Wang}}]{li2007semimetal}%
  \BibitemOpen
  \bibfield  {author} {\bibinfo {author} {\bibfnamefont {G.}~\bibnamefont
  {Li}}, \bibinfo {author} {\bibfnamefont {W.~Z.}\ \bibnamefont {Hu}}, \bibinfo
  {author} {\bibfnamefont {D.}~\bibnamefont {Qian}}, \bibinfo {author}
  {\bibfnamefont {D.}~\bibnamefont {Hsieh}}, \bibinfo {author} {\bibfnamefont
  {M.~Z.}\ \bibnamefont {Hasan}}, \bibinfo {author} {\bibfnamefont
  {E.}~\bibnamefont {Morosan}}, \bibinfo {author} {\bibfnamefont {R.~J.}\
  \bibnamefont {Cava}},\ and\ \bibinfo {author} {\bibfnamefont {N.~L.}\
  \bibnamefont {Wang}},\ }\bibfield  {title} {\bibinfo {title}
  {{Semimetal-to-Semimetal Charge Density Wave Transition in
  {1T-TiSe$_{2}$}}},\ }\href {https://doi.org/10.1103/PhysRevLett.99.027404}
  {\bibfield  {journal} {\bibinfo  {journal} {Phys. Rev. Lett.}\ }\textbf
  {\bibinfo {volume} {99}},\ \bibinfo {pages} {027404} (\bibinfo {year}
  {2007})}\BibitemShut {NoStop}%
\bibitem [{\citenamefont {Monney}\ \emph {et~al.}(2011)\citenamefont {Monney},
  \citenamefont {Battaglia}, \citenamefont {Cercellier}, \citenamefont {Aebi},\
  and\ \citenamefont {Beck}}]{monney2011exciton}%
  \BibitemOpen
  \bibfield  {author} {\bibinfo {author} {\bibfnamefont {C.}~\bibnamefont
  {Monney}}, \bibinfo {author} {\bibfnamefont {C.}~\bibnamefont {Battaglia}},
  \bibinfo {author} {\bibfnamefont {H.}~\bibnamefont {Cercellier}}, \bibinfo
  {author} {\bibfnamefont {P.}~\bibnamefont {Aebi}},\ and\ \bibinfo {author}
  {\bibfnamefont {H.}~\bibnamefont {Beck}},\ }\bibfield  {title} {\bibinfo
  {title} {{Exciton Condensation Driving the Periodic Lattice Distortion of
  {1T}$-${TiSe$_{2}$}}},\ }\href
  {https://doi.org/10.1103/PhysRevLett.106.106404} {\bibfield  {journal}
  {\bibinfo  {journal} {Phys. Rev. Lett.}\ }\textbf {\bibinfo {volume} {106}},\
  \bibinfo {pages} {106404} (\bibinfo {year} {2011})}\BibitemShut {NoStop}%
\bibitem [{\citenamefont {Kogar}\ \emph {et~al.}(2017)\citenamefont {Kogar},
  \citenamefont {Rak}, \citenamefont {Vig}, \citenamefont {Husain},
  \citenamefont {Flicker}, \citenamefont {Joe}, \citenamefont {Venema},
  \citenamefont {MacDougall}, \citenamefont {Chiang}, \citenamefont {Fradkin}
  \emph {et~al.}}]{kogar2017signatures}%
  \BibitemOpen
  \bibfield  {author} {\bibinfo {author} {\bibfnamefont {A.}~\bibnamefont
  {Kogar}}, \bibinfo {author} {\bibfnamefont {M.~S.}\ \bibnamefont {Rak}},
  \bibinfo {author} {\bibfnamefont {S.}~\bibnamefont {Vig}}, \bibinfo {author}
  {\bibfnamefont {A.~A.}\ \bibnamefont {Husain}}, \bibinfo {author}
  {\bibfnamefont {F.}~\bibnamefont {Flicker}}, \bibinfo {author} {\bibfnamefont
  {Y.~I.}\ \bibnamefont {Joe}}, \bibinfo {author} {\bibfnamefont
  {L.}~\bibnamefont {Venema}}, \bibinfo {author} {\bibfnamefont {G.~J.}\
  \bibnamefont {MacDougall}}, \bibinfo {author} {\bibfnamefont {T.~C.}\
  \bibnamefont {Chiang}}, \bibinfo {author} {\bibfnamefont {E.}~\bibnamefont
  {Fradkin}}, \emph {et~al.},\ }\bibfield  {title} {\bibinfo {title}
  {Signatures of exciton condensation in a transition metal dichalcogenide},\
  }\href@noop {} {\bibfield  {journal} {\bibinfo  {journal} {Science}\ }\textbf
  {\bibinfo {volume} {358}},\ \bibinfo {pages} {1314} (\bibinfo {year}
  {2017})}\BibitemShut {NoStop}%
\bibitem [{\citenamefont {Chen}\ \emph {et~al.}(2018)\citenamefont {Chen},
  \citenamefont {Singh}, \citenamefont {Lin},\ and\ \citenamefont
  {Pereira}}]{chen2017reproduction}%
  \BibitemOpen
  \bibfield  {author} {\bibinfo {author} {\bibfnamefont {C.}~\bibnamefont
  {Chen}}, \bibinfo {author} {\bibfnamefont {B.}~\bibnamefont {Singh}},
  \bibinfo {author} {\bibfnamefont {H.}~\bibnamefont {Lin}},\ and\ \bibinfo
  {author} {\bibfnamefont {V.~M.}\ \bibnamefont {Pereira}},\ }\bibfield
  {title} {\bibinfo {title} {{Reproduction of the Charge Density Wave Phase
  Diagram in {1T}$-${TiSe$_{2}$} Exposes its Excitonic Character}},\ }\href
  {https://doi.org/10.1103/PhysRevLett.121.226602} {\bibfield  {journal}
  {\bibinfo  {journal} {Phys. Rev. Lett.}\ }\textbf {\bibinfo {volume} {121}},\
  \bibinfo {pages} {226602} (\bibinfo {year} {2018})}\BibitemShut {NoStop}%
\bibitem [{\citenamefont {Pasquier}\ and\ \citenamefont
  {Yazyev}(2018)}]{pasquier2018excitonic}%
  \BibitemOpen
  \bibfield  {author} {\bibinfo {author} {\bibfnamefont {D.}~\bibnamefont
  {Pasquier}}\ and\ \bibinfo {author} {\bibfnamefont {O.~V.}\ \bibnamefont
  {Yazyev}},\ }\bibfield  {title} {\bibinfo {title} {Excitonic effects in
  two-dimensional {TiSe$_{2}$} from hybrid density functional theory},\ }\href
  {https://doi.org/10.1103/PhysRevB.98.235106} {\bibfield  {journal} {\bibinfo
  {journal} {Phys. Rev. B}\ }\textbf {\bibinfo {volume} {98}},\ \bibinfo
  {pages} {235106} (\bibinfo {year} {2018})}\BibitemShut {NoStop}%
\bibitem [{\citenamefont {Gole\ifmmode~\check{z}\else \v{z}\fi{}}\ \emph
  {et~al.}(2016)\citenamefont {Gole\ifmmode~\check{z}\else \v{z}\fi{}},
  \citenamefont {Werner},\ and\ \citenamefont
  {Eckstein}}]{golez2016photoinduced}%
  \BibitemOpen
  \bibfield  {author} {\bibinfo {author} {\bibfnamefont {D.}~\bibnamefont
  {Gole\ifmmode~\check{z}\else \v{z}\fi{}}}, \bibinfo {author} {\bibfnamefont
  {P.}~\bibnamefont {Werner}},\ and\ \bibinfo {author} {\bibfnamefont
  {M.}~\bibnamefont {Eckstein}},\ }\bibfield  {title} {\bibinfo {title}
  {Photoinduced gap closure in an excitonic insulator},\ }\href
  {https://doi.org/10.1103/PhysRevB.94.035121} {\bibfield  {journal} {\bibinfo
  {journal} {Phys. Rev. B}\ }\textbf {\bibinfo {volume} {94}},\ \bibinfo
  {pages} {035121} (\bibinfo {year} {2016})}\BibitemShut {NoStop}%
\bibitem [{\citenamefont {Kidd}\ \emph {et~al.}(2002)\citenamefont {Kidd},
  \citenamefont {Miller}, \citenamefont {Chou},\ and\ \citenamefont
  {Chiang}}]{kidd2002electron}%
  \BibitemOpen
  \bibfield  {author} {\bibinfo {author} {\bibfnamefont {T.~E.}\ \bibnamefont
  {Kidd}}, \bibinfo {author} {\bibfnamefont {T.}~\bibnamefont {Miller}},
  \bibinfo {author} {\bibfnamefont {M.~Y.}\ \bibnamefont {Chou}},\ and\
  \bibinfo {author} {\bibfnamefont {T.-C.}\ \bibnamefont {Chiang}},\ }\bibfield
   {title} {\bibinfo {title} {{Electron-Hole Coupling and the Charge Density
  Wave Transition in {TiSe$_{2}$}}},\ }\href
  {https://doi.org/10.1103/PhysRevLett.88.226402} {\bibfield  {journal}
  {\bibinfo  {journal} {Phys. Rev. Lett.}\ }\textbf {\bibinfo {volume} {88}},\
  \bibinfo {pages} {226402} (\bibinfo {year} {2002})}\BibitemShut {NoStop}%
\bibitem [{\citenamefont {van Wezel}\ \emph {et~al.}(2010)\citenamefont {van
  Wezel}, \citenamefont {Nahai-Williamson},\ and\ \citenamefont
  {Saxena}}]{van2010exciton}%
  \BibitemOpen
  \bibfield  {author} {\bibinfo {author} {\bibfnamefont {J.}~\bibnamefont {van
  Wezel}}, \bibinfo {author} {\bibfnamefont {P.}~\bibnamefont
  {Nahai-Williamson}},\ and\ \bibinfo {author} {\bibfnamefont {S.~S.}\
  \bibnamefont {Saxena}},\ }\bibfield  {title} {\bibinfo {title}
  {Exciton-phonon-driven charge density wave in {TiSe$_{2}$}},\ }\href
  {https://doi.org/10.1103/PhysRevB.81.165109} {\bibfield  {journal} {\bibinfo
  {journal} {Phys. Rev. B}\ }\textbf {\bibinfo {volume} {81}},\ \bibinfo
  {pages} {165109} (\bibinfo {year} {2010})}\BibitemShut {NoStop}%
\bibitem [{\citenamefont {Phan}\ \emph {et~al.}(2013)\citenamefont {Phan},
  \citenamefont {Becker},\ and\ \citenamefont {Fehske}}]{phan2013exciton}%
  \BibitemOpen
  \bibfield  {author} {\bibinfo {author} {\bibfnamefont {V.-N.}\ \bibnamefont
  {Phan}}, \bibinfo {author} {\bibfnamefont {K.~W.}\ \bibnamefont {Becker}},\
  and\ \bibinfo {author} {\bibfnamefont {H.}~\bibnamefont {Fehske}},\
  }\bibfield  {title} {\bibinfo {title} {Exciton condensation due to
  electron-phonon interaction},\ }\href
  {https://doi.org/10.1103/PhysRevB.88.205123} {\bibfield  {journal} {\bibinfo
  {journal} {Phys. Rev. B}\ }\textbf {\bibinfo {volume} {88}},\ \bibinfo
  {pages} {205123} (\bibinfo {year} {2013})}\BibitemShut {NoStop}%
\bibitem [{\citenamefont {Watanabe}\ \emph {et~al.}(2015)\citenamefont
  {Watanabe}, \citenamefont {Seki},\ and\ \citenamefont
  {Yunoki}}]{watanabe2015charge}%
  \BibitemOpen
  \bibfield  {author} {\bibinfo {author} {\bibfnamefont {H.}~\bibnamefont
  {Watanabe}}, \bibinfo {author} {\bibfnamefont {K.}~\bibnamefont {Seki}},\
  and\ \bibinfo {author} {\bibfnamefont {S.}~\bibnamefont {Yunoki}},\
  }\bibfield  {title} {\bibinfo {title} {Charge-density wave induced by
  combined electron-electron and electron-phonon interactions in
  {1T}$-${TiSe$_{2}$}$:$ {A} variational {Monte Carlo} study},\ }\href
  {https://doi.org/10.1103/PhysRevB.91.205135} {\bibfield  {journal} {\bibinfo
  {journal} {Phys. Rev. B}\ }\textbf {\bibinfo {volume} {91}},\ \bibinfo
  {pages} {205135} (\bibinfo {year} {2015})}\BibitemShut {NoStop}%
\bibitem [{\citenamefont {Bianco}\ \emph {et~al.}(2015)\citenamefont {Bianco},
  \citenamefont {Calandra},\ and\ \citenamefont
  {Mauri}}]{bianco2015electronic}%
  \BibitemOpen
  \bibfield  {author} {\bibinfo {author} {\bibfnamefont {R.}~\bibnamefont
  {Bianco}}, \bibinfo {author} {\bibfnamefont {M.}~\bibnamefont {Calandra}},\
  and\ \bibinfo {author} {\bibfnamefont {F.}~\bibnamefont {Mauri}},\ }\bibfield
   {title} {\bibinfo {title} {Electronic and vibrational properties of
  {TiSe$_{2}$} in the charge-density-wave phase from first principles},\ }\href
  {https://doi.org/10.1103/PhysRevB.92.094107} {\bibfield  {journal} {\bibinfo
  {journal} {Phys. Rev. B}\ }\textbf {\bibinfo {volume} {92}},\ \bibinfo
  {pages} {094107} (\bibinfo {year} {2015})}\BibitemShut {NoStop}%
\bibitem [{\citenamefont {Monney}\ \emph {et~al.}(2016)\citenamefont {Monney},
  \citenamefont {Puppin}, \citenamefont {Nicholson}, \citenamefont {Hoesch},
  \citenamefont {Chapman}, \citenamefont {Springate}, \citenamefont {Berger},
  \citenamefont {Magrez}, \citenamefont {Cacho}, \citenamefont {Ernstorfer},\
  and\ \citenamefont {Wolf}}]{monney2016revealing}%
  \BibitemOpen
  \bibfield  {author} {\bibinfo {author} {\bibfnamefont {C.}~\bibnamefont
  {Monney}}, \bibinfo {author} {\bibfnamefont {M.}~\bibnamefont {Puppin}},
  \bibinfo {author} {\bibfnamefont {C.~W.}\ \bibnamefont {Nicholson}}, \bibinfo
  {author} {\bibfnamefont {M.}~\bibnamefont {Hoesch}}, \bibinfo {author}
  {\bibfnamefont {R.~T.}\ \bibnamefont {Chapman}}, \bibinfo {author}
  {\bibfnamefont {E.}~\bibnamefont {Springate}}, \bibinfo {author}
  {\bibfnamefont {H.}~\bibnamefont {Berger}}, \bibinfo {author} {\bibfnamefont
  {A.}~\bibnamefont {Magrez}}, \bibinfo {author} {\bibfnamefont
  {C.}~\bibnamefont {Cacho}}, \bibinfo {author} {\bibfnamefont
  {R.}~\bibnamefont {Ernstorfer}},\ and\ \bibinfo {author} {\bibfnamefont
  {M.}~\bibnamefont {Wolf}},\ }\bibfield  {title} {\bibinfo {title} {Revealing
  the role of electrons and phonons in the ultrafast recovery of charge density
  wave correlations in {1T}$-${TiSe$_{2}$}},\ }\href
  {https://doi.org/10.1103/PhysRevB.94.165165} {\bibfield  {journal} {\bibinfo
  {journal} {Phys. Rev. B}\ }\textbf {\bibinfo {volume} {94}},\ \bibinfo
  {pages} {165165} (\bibinfo {year} {2016})}\BibitemShut {NoStop}%
\bibitem [{\citenamefont {Maschek}\ \emph {et~al.}(2016)\citenamefont
  {Maschek}, \citenamefont {Rosenkranz}, \citenamefont {Hott}, \citenamefont
  {Heid}, \citenamefont {Merz}, \citenamefont {Zocco}, \citenamefont {Said},
  \citenamefont {Alatas}, \citenamefont {Karapetrov}, \citenamefont {Zhu},
  \citenamefont {van Wezel},\ and\ \citenamefont
  {Weber}}]{maschek2016superconductivity}%
  \BibitemOpen
  \bibfield  {author} {\bibinfo {author} {\bibfnamefont {M.}~\bibnamefont
  {Maschek}}, \bibinfo {author} {\bibfnamefont {S.}~\bibnamefont {Rosenkranz}},
  \bibinfo {author} {\bibfnamefont {R.}~\bibnamefont {Hott}}, \bibinfo {author}
  {\bibfnamefont {R.}~\bibnamefont {Heid}}, \bibinfo {author} {\bibfnamefont
  {M.}~\bibnamefont {Merz}}, \bibinfo {author} {\bibfnamefont {D.~A.}\
  \bibnamefont {Zocco}}, \bibinfo {author} {\bibfnamefont {A.~H.}\ \bibnamefont
  {Said}}, \bibinfo {author} {\bibfnamefont {A.}~\bibnamefont {Alatas}},
  \bibinfo {author} {\bibfnamefont {G.}~\bibnamefont {Karapetrov}}, \bibinfo
  {author} {\bibfnamefont {S.}~\bibnamefont {Zhu}}, \bibinfo {author}
  {\bibfnamefont {J.}~\bibnamefont {van Wezel}},\ and\ \bibinfo {author}
  {\bibfnamefont {F.}~\bibnamefont {Weber}},\ }\bibfield  {title} {\bibinfo
  {title} {Superconductivity and hybrid soft modes in {TiSe$_{2}$}},\ }\href
  {https://doi.org/10.1103/PhysRevB.94.214507} {\bibfield  {journal} {\bibinfo
  {journal} {Phys. Rev. B}\ }\textbf {\bibinfo {volume} {94}},\ \bibinfo
  {pages} {214507} (\bibinfo {year} {2016})}\BibitemShut {NoStop}%
\bibitem [{\citenamefont {Singh}\ \emph {et~al.}(2017)\citenamefont {Singh},
  \citenamefont {Hsu}, \citenamefont {Tsai}, \citenamefont {Pereira},\ and\
  \citenamefont {Lin}}]{singh2017stable}%
  \BibitemOpen
  \bibfield  {author} {\bibinfo {author} {\bibfnamefont {B.}~\bibnamefont
  {Singh}}, \bibinfo {author} {\bibfnamefont {C.-H.}\ \bibnamefont {Hsu}},
  \bibinfo {author} {\bibfnamefont {W.-F.}\ \bibnamefont {Tsai}}, \bibinfo
  {author} {\bibfnamefont {V.~M.}\ \bibnamefont {Pereira}},\ and\ \bibinfo
  {author} {\bibfnamefont {H.}~\bibnamefont {Lin}},\ }\bibfield  {title}
  {\bibinfo {title} {Stable charge density wave phase in a {1T}$-${TiSe$_{2}$}
  monolayer},\ }\href {https://doi.org/10.1103/PhysRevB.95.245136} {\bibfield
  {journal} {\bibinfo  {journal} {Phys. Rev. B}\ }\textbf {\bibinfo {volume}
  {95}},\ \bibinfo {pages} {245136} (\bibinfo {year} {2017})}\BibitemShut
  {NoStop}%
\bibitem [{\citenamefont {Lian}\ \emph {et~al.}(2020)\citenamefont {Lian},
  \citenamefont {Zhang}, \citenamefont {Hu}, \citenamefont {Guan},\ and\
  \citenamefont {Meng}}]{lian2020ultrafast}%
  \BibitemOpen
  \bibfield  {author} {\bibinfo {author} {\bibfnamefont {C.}~\bibnamefont
  {Lian}}, \bibinfo {author} {\bibfnamefont {S.-J.}\ \bibnamefont {Zhang}},
  \bibinfo {author} {\bibfnamefont {S.-Q.}\ \bibnamefont {Hu}}, \bibinfo
  {author} {\bibfnamefont {M.-X.}\ \bibnamefont {Guan}},\ and\ \bibinfo
  {author} {\bibfnamefont {S.}~\bibnamefont {Meng}},\ }\bibfield  {title}
  {\bibinfo {title} {Ultrafast charge ordering by self-amplified
  exciton--phonon dynamics in {TiSe$_{2}$}},\ }\href@noop {} {\bibfield
  {journal} {\bibinfo  {journal} {Nature communications}\ }\textbf {\bibinfo
  {volume} {11}},\ \bibinfo {pages} {43} (\bibinfo {year} {2020})}\BibitemShut
  {NoStop}%
\bibitem [{\citenamefont {Cheiwchanchamnangij}\ and\ \citenamefont
  {Lambrecht}(2012)}]{cheiwchanchamnangij2012quasiparticle}%
  \BibitemOpen
  \bibfield  {author} {\bibinfo {author} {\bibfnamefont {T.}~\bibnamefont
  {Cheiwchanchamnangij}}\ and\ \bibinfo {author} {\bibfnamefont {W.~R.}\
  \bibnamefont {Lambrecht}},\ }\bibfield  {title} {\bibinfo {title}
  {Quasiparticle band structure calculation of monolayer, bilayer, and bulk
  {MoS$_{2}$}},\ }\href@noop {} {\bibfield  {journal} {\bibinfo  {journal}
  {Phys. Rev. B}\ }\textbf {\bibinfo {volume} {85}},\ \bibinfo {pages} {205302}
  (\bibinfo {year} {2012})}\BibitemShut {NoStop}%
\bibitem [{\citenamefont {Ghatak}\ \emph {et~al.}(2011)\citenamefont {Ghatak},
  \citenamefont {Pal},\ and\ \citenamefont {Ghosh}}]{ghatak2011nature}%
  \BibitemOpen
  \bibfield  {author} {\bibinfo {author} {\bibfnamefont {S.}~\bibnamefont
  {Ghatak}}, \bibinfo {author} {\bibfnamefont {A.~N.}\ \bibnamefont {Pal}},\
  and\ \bibinfo {author} {\bibfnamefont {A.}~\bibnamefont {Ghosh}},\ }\bibfield
   {title} {\bibinfo {title} {Nature of electronic states in atomically thin
  {MoS$_{2}$} field-effect transistors},\ }\href@noop {} {\bibfield  {journal}
  {\bibinfo  {journal} {Acs Nano}\ }\textbf {\bibinfo {volume} {5}},\ \bibinfo
  {pages} {7707} (\bibinfo {year} {2011})}\BibitemShut {NoStop}%
\bibitem [{\citenamefont {Moon}\ \emph {et~al.}(2018)\citenamefont {Moon},
  \citenamefont {Bae}, \citenamefont {Joo}, \citenamefont {Choi}, \citenamefont
  {Han}, \citenamefont {Lim},\ and\ \citenamefont {Lee}}]{moon2018soft}%
  \BibitemOpen
  \bibfield  {author} {\bibinfo {author} {\bibfnamefont {B.~H.}\ \bibnamefont
  {Moon}}, \bibinfo {author} {\bibfnamefont {J.~J.}\ \bibnamefont {Bae}},
  \bibinfo {author} {\bibfnamefont {M.-K.}\ \bibnamefont {Joo}}, \bibinfo
  {author} {\bibfnamefont {H.}~\bibnamefont {Choi}}, \bibinfo {author}
  {\bibfnamefont {G.~H.}\ \bibnamefont {Han}}, \bibinfo {author} {\bibfnamefont
  {H.}~\bibnamefont {Lim}},\ and\ \bibinfo {author} {\bibfnamefont {Y.~H.}\
  \bibnamefont {Lee}},\ }\bibfield  {title} {\bibinfo {title} {{Soft Coulomb
  gap and asymmetric scaling towards metal-insulator quantum criticality in
  multilayer {MoS$_{2}$}}},\ }\href@noop {} {\bibfield  {journal} {\bibinfo
  {journal} {Nat. Commun.}\ }\textbf {\bibinfo {volume} {9}},\ \bibinfo {pages}
  {2052} (\bibinfo {year} {2018})}\BibitemShut {NoStop}%
\bibitem [{\citenamefont {Chernikov}\ \emph {et~al.}(2015)\citenamefont
  {Chernikov}, \citenamefont {Ruppert}, \citenamefont {Hill}, \citenamefont
  {Rigosi},\ and\ \citenamefont {Heinz}}]{chernikov2015population}%
  \BibitemOpen
  \bibfield  {author} {\bibinfo {author} {\bibfnamefont {A.}~\bibnamefont
  {Chernikov}}, \bibinfo {author} {\bibfnamefont {C.}~\bibnamefont {Ruppert}},
  \bibinfo {author} {\bibfnamefont {H.~M.}\ \bibnamefont {Hill}}, \bibinfo
  {author} {\bibfnamefont {A.~F.}\ \bibnamefont {Rigosi}},\ and\ \bibinfo
  {author} {\bibfnamefont {T.~F.}\ \bibnamefont {Heinz}},\ }\bibfield  {title}
  {\bibinfo {title} {Population inversion and giant bandgap renormalization in
  atomically thin {WS$_{2}$} layers},\ }\href@noop {} {\bibfield  {journal}
  {\bibinfo  {journal} {Nature Photonics}\ }\textbf {\bibinfo {volume} {9}},\
  \bibinfo {pages} {466} (\bibinfo {year} {2015})}\BibitemShut {NoStop}%
\bibitem [{\citenamefont {Rohwer}\ \emph {et~al.}(2011)\citenamefont {Rohwer},
  \citenamefont {Hellmann}, \citenamefont {Wiesenmayer}, \citenamefont {Sohrt},
  \citenamefont {Stange}, \citenamefont {Slomski}, \citenamefont {Carr},
  \citenamefont {Liu}, \citenamefont {Avila}, \citenamefont {Kall{\"a}ne} \emph
  {et~al.}}]{rohwer2011collapse}%
  \BibitemOpen
  \bibfield  {author} {\bibinfo {author} {\bibfnamefont {T.}~\bibnamefont
  {Rohwer}}, \bibinfo {author} {\bibfnamefont {S.}~\bibnamefont {Hellmann}},
  \bibinfo {author} {\bibfnamefont {M.}~\bibnamefont {Wiesenmayer}}, \bibinfo
  {author} {\bibfnamefont {C.}~\bibnamefont {Sohrt}}, \bibinfo {author}
  {\bibfnamefont {A.}~\bibnamefont {Stange}}, \bibinfo {author} {\bibfnamefont
  {B.}~\bibnamefont {Slomski}}, \bibinfo {author} {\bibfnamefont
  {A.}~\bibnamefont {Carr}}, \bibinfo {author} {\bibfnamefont {Y.}~\bibnamefont
  {Liu}}, \bibinfo {author} {\bibfnamefont {L.~M.}\ \bibnamefont {Avila}},
  \bibinfo {author} {\bibfnamefont {M.}~\bibnamefont {Kall{\"a}ne}}, \emph
  {et~al.},\ }\bibfield  {title} {\bibinfo {title} {Collapse of long-range
  charge order tracked by time-resolved photoemission at high momenta},\
  }\href@noop {} {\bibfield  {journal} {\bibinfo  {journal} {Nature}\ }\textbf
  {\bibinfo {volume} {471}},\ \bibinfo {pages} {490} (\bibinfo {year}
  {2011})}\BibitemShut {NoStop}%
\bibitem [{\citenamefont {Sobota}\ \emph {et~al.}(2014)\citenamefont {Sobota},
  \citenamefont {Yang}, \citenamefont {Leuenberger}, \citenamefont {Kemper},
  \citenamefont {Analytis}, \citenamefont {Fisher}, \citenamefont {Kirchmann},
  \citenamefont {Devereaux},\ and\ \citenamefont {Shen}}]{sobota2014ultrafast}%
  \BibitemOpen
  \bibfield  {author} {\bibinfo {author} {\bibfnamefont {J.~A.}\ \bibnamefont
  {Sobota}}, \bibinfo {author} {\bibfnamefont {S.-L.}\ \bibnamefont {Yang}},
  \bibinfo {author} {\bibfnamefont {D.}~\bibnamefont {Leuenberger}}, \bibinfo
  {author} {\bibfnamefont {A.~F.}\ \bibnamefont {Kemper}}, \bibinfo {author}
  {\bibfnamefont {J.~G.}\ \bibnamefont {Analytis}}, \bibinfo {author}
  {\bibfnamefont {I.~R.}\ \bibnamefont {Fisher}}, \bibinfo {author}
  {\bibfnamefont {P.~S.}\ \bibnamefont {Kirchmann}}, \bibinfo {author}
  {\bibfnamefont {T.~P.}\ \bibnamefont {Devereaux}},\ and\ \bibinfo {author}
  {\bibfnamefont {Z.-X.}\ \bibnamefont {Shen}},\ }\bibfield  {title} {\bibinfo
  {title} {Ultrafast electron dynamics in the topological insulator
  {Bi$_{2}$Se$_{3}$} studied by time-resolved photoemission spectroscopy},\
  }\href@noop {} {\bibfield  {journal} {\bibinfo  {journal} {Journal of
  Electron Spectroscopy and Related Phenomena}\ }\textbf {\bibinfo {volume}
  {195}},\ \bibinfo {pages} {249} (\bibinfo {year} {2014})}\BibitemShut
  {NoStop}%
\bibitem [{\citenamefont {Eich}\ \emph {et~al.}(2014)\citenamefont {Eich},
  \citenamefont {Stange}, \citenamefont {Carr}, \citenamefont {Urbancic},
  \citenamefont {Popmintchev}, \citenamefont {Wiesenmayer}, \citenamefont
  {Jansen}, \citenamefont {Ruffing}, \citenamefont {Jakobs}, \citenamefont
  {Rohwer} \emph {et~al.}}]{eich2014time}%
  \BibitemOpen
  \bibfield  {author} {\bibinfo {author} {\bibfnamefont {S.}~\bibnamefont
  {Eich}}, \bibinfo {author} {\bibfnamefont {A.}~\bibnamefont {Stange}},
  \bibinfo {author} {\bibfnamefont {A.}~\bibnamefont {Carr}}, \bibinfo {author}
  {\bibfnamefont {J.}~\bibnamefont {Urbancic}}, \bibinfo {author}
  {\bibfnamefont {T.}~\bibnamefont {Popmintchev}}, \bibinfo {author}
  {\bibfnamefont {M.}~\bibnamefont {Wiesenmayer}}, \bibinfo {author}
  {\bibfnamefont {K.}~\bibnamefont {Jansen}}, \bibinfo {author} {\bibfnamefont
  {A.}~\bibnamefont {Ruffing}}, \bibinfo {author} {\bibfnamefont
  {S.}~\bibnamefont {Jakobs}}, \bibinfo {author} {\bibfnamefont
  {T.}~\bibnamefont {Rohwer}}, \emph {et~al.},\ }\bibfield  {title} {\bibinfo
  {title} {Time- and angle-resolved photoemission spectroscopy with optimized
  high-harmonic pulses using frequency-doubled {Ti:Sapphire lasers}},\
  }\href@noop {} {\bibfield  {journal} {\bibinfo  {journal} {Journal of
  Electron Spectroscopy and Related Phenomena}\ }\textbf {\bibinfo {volume}
  {195}},\ \bibinfo {pages} {231} (\bibinfo {year} {2014})}\BibitemShut
  {NoStop}%
\bibitem [{\citenamefont {Smallwood}\ \emph {et~al.}(2016)\citenamefont
  {Smallwood}, \citenamefont {Kaindl},\ and\ \citenamefont
  {Lanzara}}]{smallwood2016ultrafast}%
  \BibitemOpen
  \bibfield  {author} {\bibinfo {author} {\bibfnamefont {C.~L.}\ \bibnamefont
  {Smallwood}}, \bibinfo {author} {\bibfnamefont {R.~A.}\ \bibnamefont
  {Kaindl}},\ and\ \bibinfo {author} {\bibfnamefont {A.}~\bibnamefont
  {Lanzara}},\ }\bibfield  {title} {\bibinfo {title} {Ultrafast angle-resolved
  photoemission spectroscopy of quantum materials},\ }\href@noop {} {\bibfield
  {journal} {\bibinfo  {journal} {EPL (Europhysics Letters)}\ }\textbf
  {\bibinfo {volume} {115}},\ \bibinfo {pages} {27001} (\bibinfo {year}
  {2016})}\BibitemShut {NoStop}%
\bibitem [{\citenamefont {Lin}\ \emph {et~al.}(2021)\citenamefont {Lin},
  \citenamefont {He}, \citenamefont {Hlevyack}, \citenamefont {Chen},
  \citenamefont {Mo}, \citenamefont {Chou},\ and\ \citenamefont
  {Chiang}}]{lin2021coherent}%
  \BibitemOpen
  \bibfield  {author} {\bibinfo {author} {\bibfnamefont {M.-K.}\ \bibnamefont
  {Lin}}, \bibinfo {author} {\bibfnamefont {T.}~\bibnamefont {He}}, \bibinfo
  {author} {\bibfnamefont {J.~A.}\ \bibnamefont {Hlevyack}}, \bibinfo {author}
  {\bibfnamefont {P.}~\bibnamefont {Chen}}, \bibinfo {author} {\bibfnamefont
  {S.-K.}\ \bibnamefont {Mo}}, \bibinfo {author} {\bibfnamefont {M.-Y.}\
  \bibnamefont {Chou}},\ and\ \bibinfo {author} {\bibfnamefont {T.-C.}\
  \bibnamefont {Chiang}},\ }\bibfield  {title} {\bibinfo {title} {Coherent
  electronic band structure of {TiTe$_{2}$/TiSe$_{2}$} {Moir\'{e} Bilayer}},\
  }\href@noop {} {\bibfield  {journal} {\bibinfo  {journal} {ACS nano}\
  }\textbf {\bibinfo {volume} {15}},\ \bibinfo {pages} {3359} (\bibinfo {year}
  {2021})}\BibitemShut {NoStop}%
\bibitem [{\citenamefont {Voiry}\ \emph {et~al.}(2015)\citenamefont {Voiry},
  \citenamefont {Mohite},\ and\ \citenamefont {Chhowalla}}]{voiry2015phase}%
  \BibitemOpen
  \bibfield  {author} {\bibinfo {author} {\bibfnamefont {D.}~\bibnamefont
  {Voiry}}, \bibinfo {author} {\bibfnamefont {A.}~\bibnamefont {Mohite}},\ and\
  \bibinfo {author} {\bibfnamefont {M.}~\bibnamefont {Chhowalla}},\ }\bibfield
  {title} {\bibinfo {title} {Phase engineering of transition metal
  dichalcogenides},\ }\href@noop {} {\bibfield  {journal} {\bibinfo  {journal}
  {Chemical Society Reviews}\ }\textbf {\bibinfo {volume} {44}},\ \bibinfo
  {pages} {2702} (\bibinfo {year} {2015})}\BibitemShut {NoStop}%
\bibitem [{\citenamefont {Steinhoff}\ \emph {et~al.}(2016)\citenamefont
  {Steinhoff}, \citenamefont {Florian}, \citenamefont {R{\"o}sner},
  \citenamefont {Lorke}, \citenamefont {Wehling}, \citenamefont {Gies},\ and\
  \citenamefont {Jahnke}}]{steinhoff2016nonequilibrium}%
  \BibitemOpen
  \bibfield  {author} {\bibinfo {author} {\bibfnamefont {A.}~\bibnamefont
  {Steinhoff}}, \bibinfo {author} {\bibfnamefont {M.}~\bibnamefont {Florian}},
  \bibinfo {author} {\bibfnamefont {M.}~\bibnamefont {R{\"o}sner}}, \bibinfo
  {author} {\bibfnamefont {M.}~\bibnamefont {Lorke}}, \bibinfo {author}
  {\bibfnamefont {T.~O.}\ \bibnamefont {Wehling}}, \bibinfo {author}
  {\bibfnamefont {C.}~\bibnamefont {Gies}},\ and\ \bibinfo {author}
  {\bibfnamefont {F.}~\bibnamefont {Jahnke}},\ }\bibfield  {title} {\bibinfo
  {title} {Nonequilibrium carrier dynamics in transition metal dichalcogenide
  semiconductors},\ }\href@noop {} {\bibfield  {journal} {\bibinfo  {journal}
  {2D Materials}\ }\textbf {\bibinfo {volume} {3}},\ \bibinfo {pages} {031006}
  (\bibinfo {year} {2016})}\BibitemShut {NoStop}%
\bibitem [{\citenamefont {Gierz}(2017)}]{gierz2017probing}%
  \BibitemOpen
  \bibfield  {author} {\bibinfo {author} {\bibfnamefont {I.}~\bibnamefont
  {Gierz}},\ }\bibfield  {title} {\bibinfo {title} {Probing carrier dynamics in
  photo-excited graphene with time-resolved {ARPES}},\ }\href@noop {}
  {\bibfield  {journal} {\bibinfo  {journal} {Journal of Electron Spectroscopy
  and Related Phenomena}\ }\textbf {\bibinfo {volume} {219}},\ \bibinfo {pages}
  {53} (\bibinfo {year} {2017})}\BibitemShut {NoStop}%
\bibitem [{\citenamefont {Gierz}\ \emph {et~al.}(2013)\citenamefont {Gierz},
  \citenamefont {Petersen}, \citenamefont {Mitrano}, \citenamefont {Cacho},
  \citenamefont {Turcu}, \citenamefont {Springate}, \citenamefont {St{\"o}hr},
  \citenamefont {K{\"o}hler}, \citenamefont {Starke},\ and\ \citenamefont
  {Cavalleri}}]{gierz2013snapshots}%
  \BibitemOpen
  \bibfield  {author} {\bibinfo {author} {\bibfnamefont {I.}~\bibnamefont
  {Gierz}}, \bibinfo {author} {\bibfnamefont {J.~C.}\ \bibnamefont {Petersen}},
  \bibinfo {author} {\bibfnamefont {M.}~\bibnamefont {Mitrano}}, \bibinfo
  {author} {\bibfnamefont {C.}~\bibnamefont {Cacho}}, \bibinfo {author}
  {\bibfnamefont {I.~E.}\ \bibnamefont {Turcu}}, \bibinfo {author}
  {\bibfnamefont {E.}~\bibnamefont {Springate}}, \bibinfo {author}
  {\bibfnamefont {A.}~\bibnamefont {St{\"o}hr}}, \bibinfo {author}
  {\bibfnamefont {A.}~\bibnamefont {K{\"o}hler}}, \bibinfo {author}
  {\bibfnamefont {U.}~\bibnamefont {Starke}},\ and\ \bibinfo {author}
  {\bibfnamefont {A.}~\bibnamefont {Cavalleri}},\ }\bibfield  {title} {\bibinfo
  {title} {Snapshots of non-equilibrium {Dirac} carrier distributions in
  graphene},\ }\href@noop {} {\bibfield  {journal} {\bibinfo  {journal} {Nature
  materials}\ }\textbf {\bibinfo {volume} {12}},\ \bibinfo {pages} {1119}
  (\bibinfo {year} {2013})}\BibitemShut {NoStop}%
\bibitem [{\citenamefont {Stroucken}\ \emph {et~al.}(2013)\citenamefont
  {Stroucken}, \citenamefont {Gr\"onqvist},\ and\ \citenamefont
  {Koch}}]{stroucken2013screening}%
  \BibitemOpen
  \bibfield  {author} {\bibinfo {author} {\bibfnamefont {T.}~\bibnamefont
  {Stroucken}}, \bibinfo {author} {\bibfnamefont {J.~H.}\ \bibnamefont
  {Gr\"onqvist}},\ and\ \bibinfo {author} {\bibfnamefont {S.~W.}\ \bibnamefont
  {Koch}},\ }\bibfield  {title} {\bibinfo {title} {Screening and gap generation
  in bilayer graphene},\ }\href {https://doi.org/10.1103/PhysRevB.87.245428}
  {\bibfield  {journal} {\bibinfo  {journal} {Phys. Rev. B}\ }\textbf {\bibinfo
  {volume} {87}},\ \bibinfo {pages} {245428} (\bibinfo {year}
  {2013})}\BibitemShut {NoStop}%
\bibitem [{\citenamefont {Mihnev}\ \emph {et~al.}(2016)\citenamefont {Mihnev},
  \citenamefont {Kadi}, \citenamefont {Divin}, \citenamefont {Winzer},
  \citenamefont {Lee}, \citenamefont {Liu}, \citenamefont {Zhong},
  \citenamefont {Berger}, \citenamefont {De~Heer}, \citenamefont {Malic} \emph
  {et~al.}}]{mihnev2016microscopic}%
  \BibitemOpen
  \bibfield  {author} {\bibinfo {author} {\bibfnamefont {M.~T.}\ \bibnamefont
  {Mihnev}}, \bibinfo {author} {\bibfnamefont {F.}~\bibnamefont {Kadi}},
  \bibinfo {author} {\bibfnamefont {C.~J.}\ \bibnamefont {Divin}}, \bibinfo
  {author} {\bibfnamefont {T.}~\bibnamefont {Winzer}}, \bibinfo {author}
  {\bibfnamefont {S.}~\bibnamefont {Lee}}, \bibinfo {author} {\bibfnamefont
  {C.-H.}\ \bibnamefont {Liu}}, \bibinfo {author} {\bibfnamefont
  {Z.}~\bibnamefont {Zhong}}, \bibinfo {author} {\bibfnamefont
  {C.}~\bibnamefont {Berger}}, \bibinfo {author} {\bibfnamefont {W.~A.}\
  \bibnamefont {De~Heer}}, \bibinfo {author} {\bibfnamefont {E.}~\bibnamefont
  {Malic}}, \emph {et~al.},\ }\bibfield  {title} {\bibinfo {title} {Microscopic
  origins of the terahertz carrier relaxation and cooling dynamics in
  graphene},\ }\href@noop {} {\bibfield  {journal} {\bibinfo  {journal} {Nature
  communications}\ }\textbf {\bibinfo {volume} {7}},\ \bibinfo {pages} {11617}
  (\bibinfo {year} {2016})}\BibitemShut {NoStop}%
\bibitem [{\citenamefont {Kadi}\ \emph {et~al.}(2015)\citenamefont {Kadi},
  \citenamefont {Winzer}, \citenamefont {Knorr},\ and\ \citenamefont
  {Malic}}]{kadi2015impact}%
  \BibitemOpen
  \bibfield  {author} {\bibinfo {author} {\bibfnamefont {F.}~\bibnamefont
  {Kadi}}, \bibinfo {author} {\bibfnamefont {T.}~\bibnamefont {Winzer}},
  \bibinfo {author} {\bibfnamefont {A.}~\bibnamefont {Knorr}},\ and\ \bibinfo
  {author} {\bibfnamefont {E.}~\bibnamefont {Malic}},\ }\bibfield  {title}
  {\bibinfo {title} {Impact of doping on the carrier dynamics in graphene},\
  }\href@noop {} {\bibfield  {journal} {\bibinfo  {journal} {Scientific
  reports}\ }\textbf {\bibinfo {volume} {5}},\ \bibinfo {pages} {16841}
  (\bibinfo {year} {2015})}\BibitemShut {NoStop}%
\bibitem [{\citenamefont {Bhimanapati}\ \emph {et~al.}(2015)\citenamefont
  {Bhimanapati}, \citenamefont {Lin}, \citenamefont {Meunier}, \citenamefont
  {Jung}, \citenamefont {Cha}, \citenamefont {Das}, \citenamefont {Xiao},
  \citenamefont {Son}, \citenamefont {Strano}, \citenamefont {Cooper} \emph
  {et~al.}}]{bhimanapati2015recent}%
  \BibitemOpen
  \bibfield  {author} {\bibinfo {author} {\bibfnamefont {G.~R.}\ \bibnamefont
  {Bhimanapati}}, \bibinfo {author} {\bibfnamefont {Z.}~\bibnamefont {Lin}},
  \bibinfo {author} {\bibfnamefont {V.}~\bibnamefont {Meunier}}, \bibinfo
  {author} {\bibfnamefont {Y.}~\bibnamefont {Jung}}, \bibinfo {author}
  {\bibfnamefont {J.}~\bibnamefont {Cha}}, \bibinfo {author} {\bibfnamefont
  {S.}~\bibnamefont {Das}}, \bibinfo {author} {\bibfnamefont {D.}~\bibnamefont
  {Xiao}}, \bibinfo {author} {\bibfnamefont {Y.}~\bibnamefont {Son}}, \bibinfo
  {author} {\bibfnamefont {M.~S.}\ \bibnamefont {Strano}}, \bibinfo {author}
  {\bibfnamefont {V.~R.}\ \bibnamefont {Cooper}}, \emph {et~al.},\ }\bibfield
  {title} {\bibinfo {title} {Recent advances in two-dimensional materials
  beyond graphene},\ }\href@noop {} {\bibfield  {journal} {\bibinfo  {journal}
  {ACS nano}\ }\textbf {\bibinfo {volume} {9}},\ \bibinfo {pages} {11509}
  (\bibinfo {year} {2015})}\BibitemShut {NoStop}%
\bibitem [{\citenamefont {Xia}\ \emph {et~al.}(2014)\citenamefont {Xia},
  \citenamefont {Wang}, \citenamefont {Xiao}, \citenamefont {Dubey},\ and\
  \citenamefont {Ramasubramaniam}}]{xia2014two}%
  \BibitemOpen
  \bibfield  {author} {\bibinfo {author} {\bibfnamefont {F.}~\bibnamefont
  {Xia}}, \bibinfo {author} {\bibfnamefont {H.}~\bibnamefont {Wang}}, \bibinfo
  {author} {\bibfnamefont {D.}~\bibnamefont {Xiao}}, \bibinfo {author}
  {\bibfnamefont {M.}~\bibnamefont {Dubey}},\ and\ \bibinfo {author}
  {\bibfnamefont {A.}~\bibnamefont {Ramasubramaniam}},\ }\bibfield  {title}
  {\bibinfo {title} {Two-dimensional material nanophotonics},\ }\href@noop {}
  {\bibfield  {journal} {\bibinfo  {journal} {Nature Photonics}\ }\textbf
  {\bibinfo {volume} {8}},\ \bibinfo {pages} {899} (\bibinfo {year}
  {2014})}\BibitemShut {NoStop}%
\bibitem [{\citenamefont {Chen}\ \emph
  {et~al.}(2016{\natexlab{b}})\citenamefont {Chen}, \citenamefont {Wen},
  \citenamefont {Zhang}, \citenamefont {Wu}, \citenamefont {Gong},
  \citenamefont {Zhang}, \citenamefont {Yuan}, \citenamefont {Yi},
  \citenamefont {Lou}, \citenamefont {Ajayan} \emph
  {et~al.}}]{chen2016ultrafast}%
  \BibitemOpen
  \bibfield  {author} {\bibinfo {author} {\bibfnamefont {H.}~\bibnamefont
  {Chen}}, \bibinfo {author} {\bibfnamefont {X.}~\bibnamefont {Wen}}, \bibinfo
  {author} {\bibfnamefont {J.}~\bibnamefont {Zhang}}, \bibinfo {author}
  {\bibfnamefont {T.}~\bibnamefont {Wu}}, \bibinfo {author} {\bibfnamefont
  {Y.}~\bibnamefont {Gong}}, \bibinfo {author} {\bibfnamefont {X.}~\bibnamefont
  {Zhang}}, \bibinfo {author} {\bibfnamefont {J.}~\bibnamefont {Yuan}},
  \bibinfo {author} {\bibfnamefont {C.}~\bibnamefont {Yi}}, \bibinfo {author}
  {\bibfnamefont {J.}~\bibnamefont {Lou}}, \bibinfo {author} {\bibfnamefont
  {P.~M.}\ \bibnamefont {Ajayan}}, \emph {et~al.},\ }\bibfield  {title}
  {\bibinfo {title} {Ultrafast formation of interlayer hot excitons in
  atomically thin {MoS$_{2}$}/{WS$_{2}$} heterostructures},\ }\href@noop {}
  {\bibfield  {journal} {\bibinfo  {journal} {Nat. Commun.}\ }\textbf {\bibinfo
  {volume} {7}},\ \bibinfo {pages} {12512} (\bibinfo {year}
  {2016}{\natexlab{b}})}\BibitemShut {NoStop}%
\bibitem [{\citenamefont {Steinhoff}\ \emph {et~al.}(2017)\citenamefont
  {Steinhoff}, \citenamefont {Florian}, \citenamefont {R{\"o}sner},
  \citenamefont {Sch{\"o}nhoff}, \citenamefont {Wehling},\ and\ \citenamefont
  {Jahnke}}]{steinhoff2017excitons}%
  \BibitemOpen
  \bibfield  {author} {\bibinfo {author} {\bibfnamefont {A.}~\bibnamefont
  {Steinhoff}}, \bibinfo {author} {\bibfnamefont {M.}~\bibnamefont {Florian}},
  \bibinfo {author} {\bibfnamefont {M.}~\bibnamefont {R{\"o}sner}}, \bibinfo
  {author} {\bibfnamefont {G.}~\bibnamefont {Sch{\"o}nhoff}}, \bibinfo {author}
  {\bibfnamefont {T.~O.}\ \bibnamefont {Wehling}},\ and\ \bibinfo {author}
  {\bibfnamefont {F.}~\bibnamefont {Jahnke}},\ }\bibfield  {title} {\bibinfo
  {title} {Exciton fission in monolayer transition metal dichalcogenide
  semiconductors},\ }\href@noop {} {\bibfield  {journal} {\bibinfo  {journal}
  {Nature communications}\ }\textbf {\bibinfo {volume} {8}},\ \bibinfo {pages}
  {1166} (\bibinfo {year} {2017})}\BibitemShut {NoStop}%
\bibitem [{\citenamefont {Wall}\ \emph {et~al.}(2012)\citenamefont {Wall},
  \citenamefont {Wegkamp}, \citenamefont {Foglia}, \citenamefont {Appavoo},
  \citenamefont {Nag}, \citenamefont {Haglund}, \citenamefont {St{\"a}hler},\
  and\ \citenamefont {Wolf}}]{wall2012ultrafast}%
  \BibitemOpen
  \bibfield  {author} {\bibinfo {author} {\bibfnamefont {S.}~\bibnamefont
  {Wall}}, \bibinfo {author} {\bibfnamefont {D.}~\bibnamefont {Wegkamp}},
  \bibinfo {author} {\bibfnamefont {L.}~\bibnamefont {Foglia}}, \bibinfo
  {author} {\bibfnamefont {K.}~\bibnamefont {Appavoo}}, \bibinfo {author}
  {\bibfnamefont {J.}~\bibnamefont {Nag}}, \bibinfo {author} {\bibfnamefont
  {R.}~\bibnamefont {Haglund}}, \bibinfo {author} {\bibfnamefont
  {J.}~\bibnamefont {St{\"a}hler}},\ and\ \bibinfo {author} {\bibfnamefont
  {M.}~\bibnamefont {Wolf}},\ }\bibfield  {title} {\bibinfo {title} {Ultrafast
  changes in lattice symmetry probed by coherent phonons},\ }\href@noop {}
  {\bibfield  {journal} {\bibinfo  {journal} {Nature communications}\ }\textbf
  {\bibinfo {volume} {3}},\ \bibinfo {pages} {721} (\bibinfo {year}
  {2012})}\BibitemShut {NoStop}%
\bibitem [{\citenamefont {Mizokawa}\ \emph {et~al.}(2009)\citenamefont
  {Mizokawa}, \citenamefont {Takubo}, \citenamefont {Sudayama}, \citenamefont
  {Wakisaka}, \citenamefont {Takubo}, \citenamefont {Miyano}, \citenamefont
  {Matsumoto}, \citenamefont {Nagata}, \citenamefont {Katayama}, \citenamefont
  {Nohara} \emph {et~al.}}]{mizokawa2009local}%
  \BibitemOpen
  \bibfield  {author} {\bibinfo {author} {\bibfnamefont {T.}~\bibnamefont
  {Mizokawa}}, \bibinfo {author} {\bibfnamefont {K.}~\bibnamefont {Takubo}},
  \bibinfo {author} {\bibfnamefont {T.}~\bibnamefont {Sudayama}}, \bibinfo
  {author} {\bibfnamefont {Y.}~\bibnamefont {Wakisaka}}, \bibinfo {author}
  {\bibfnamefont {N.}~\bibnamefont {Takubo}}, \bibinfo {author} {\bibfnamefont
  {K.}~\bibnamefont {Miyano}}, \bibinfo {author} {\bibfnamefont
  {N.}~\bibnamefont {Matsumoto}}, \bibinfo {author} {\bibfnamefont
  {S.}~\bibnamefont {Nagata}}, \bibinfo {author} {\bibfnamefont
  {T.}~\bibnamefont {Katayama}}, \bibinfo {author} {\bibfnamefont
  {M.}~\bibnamefont {Nohara}}, \emph {et~al.},\ }\bibfield  {title} {\bibinfo
  {title} {{Local Lattice Distortion and Photo-Induced Phase Transition in
  Transition-Metal Compounds with Orbital Degeneracy}},\ }\href@noop {}
  {\bibfield  {journal} {\bibinfo  {journal} {Journal of Superconductivity and
  Novel Magnetism}\ }\textbf {\bibinfo {volume} {22}},\ \bibinfo {pages} {67}
  (\bibinfo {year} {2009})}\BibitemShut {NoStop}%
\bibitem [{\citenamefont {Sykora}\ \emph {et~al.}(2009)\citenamefont {Sykora},
  \citenamefont {Huebsch},\ and\ \citenamefont
  {Becker}}]{sykora2009coexistence}%
  \BibitemOpen
  \bibfield  {author} {\bibinfo {author} {\bibfnamefont {S.}~\bibnamefont
  {Sykora}}, \bibinfo {author} {\bibfnamefont {A.}~\bibnamefont {Huebsch}},\
  and\ \bibinfo {author} {\bibfnamefont {K.~W.}\ \bibnamefont {Becker}},\
  }\bibfield  {title} {\bibinfo {title} {Coexistence of superconductivity and
  charge-density waves in a two-dimensional {Holstein} model at half-filling},\
  }\href@noop {} {\bibfield  {journal} {\bibinfo  {journal} {EPL (Europhysics
  Letters)}\ }\textbf {\bibinfo {volume} {85}},\ \bibinfo {pages} {57003}
  (\bibinfo {year} {2009})}\BibitemShut {NoStop}%
\bibitem [{\citenamefont {Kim}\ \emph {et~al.}(2015)\citenamefont {Kim},
  \citenamefont {Kim},\ and\ \citenamefont {Min}}]{kim2015mechanism}%
  \BibitemOpen
  \bibfield  {author} {\bibinfo {author} {\bibfnamefont {S.}~\bibnamefont
  {Kim}}, \bibinfo {author} {\bibfnamefont {K.}~\bibnamefont {Kim}},\ and\
  \bibinfo {author} {\bibfnamefont {B.}~\bibnamefont {Min}},\ }\bibfield
  {title} {\bibinfo {title} {The mechanism of charge density wave in {Pt-based}
  layered superconductors: {SrPt$_{2}$As$_{2}$} and {LaPt$_{2}$Si$_{2}$}},\
  }\href@noop {} {\bibfield  {journal} {\bibinfo  {journal} {Scientific
  reports}\ }\textbf {\bibinfo {volume} {5}},\ \bibinfo {pages} {15052}
  (\bibinfo {year} {2015})}\BibitemShut {NoStop}%
\bibitem [{\citenamefont {Ritschel}\ \emph {et~al.}(2015)\citenamefont
  {Ritschel}, \citenamefont {Trinckauf}, \citenamefont {Koepernik},
  \citenamefont {B{\"u}chner}, \citenamefont {Zimmermann}, \citenamefont
  {Berger}, \citenamefont {Joe}, \citenamefont {Abbamonte},\ and\ \citenamefont
  {Geck}}]{ritschel2015orbital}%
  \BibitemOpen
  \bibfield  {author} {\bibinfo {author} {\bibfnamefont {T.}~\bibnamefont
  {Ritschel}}, \bibinfo {author} {\bibfnamefont {J.}~\bibnamefont {Trinckauf}},
  \bibinfo {author} {\bibfnamefont {K.}~\bibnamefont {Koepernik}}, \bibinfo
  {author} {\bibfnamefont {B.}~\bibnamefont {B{\"u}chner}}, \bibinfo {author}
  {\bibfnamefont {M.~v.}\ \bibnamefont {Zimmermann}}, \bibinfo {author}
  {\bibfnamefont {H.}~\bibnamefont {Berger}}, \bibinfo {author} {\bibfnamefont
  {Y.}~\bibnamefont {Joe}}, \bibinfo {author} {\bibfnamefont {P.}~\bibnamefont
  {Abbamonte}},\ and\ \bibinfo {author} {\bibfnamefont {J.}~\bibnamefont
  {Geck}},\ }\bibfield  {title} {\bibinfo {title} {Orbital textures and charge
  density waves in transition metal dichalcogenides},\ }\href@noop {}
  {\bibfield  {journal} {\bibinfo  {journal} {Nature physics}\ }\textbf
  {\bibinfo {volume} {11}},\ \bibinfo {pages} {328} (\bibinfo {year}
  {2015})}\BibitemShut {NoStop}%
\bibitem [{\citenamefont {Gu}\ and\ \citenamefont
  {Rondinelli}(2016)}]{gu2016ultrafast}%
  \BibitemOpen
  \bibfield  {author} {\bibinfo {author} {\bibfnamefont {M.}~\bibnamefont
  {Gu}}\ and\ \bibinfo {author} {\bibfnamefont {J.~M.}\ \bibnamefont
  {Rondinelli}},\ }\bibfield  {title} {\bibinfo {title} {Ultrafast band
  engineering and transient spin currents in antiferromagnetic oxides},\
  }\href@noop {} {\bibfield  {journal} {\bibinfo  {journal} {Scientific
  reports}\ }\textbf {\bibinfo {volume} {6}},\ \bibinfo {pages} {25121}
  (\bibinfo {year} {2016})}\BibitemShut {NoStop}%
\bibitem [{\citenamefont {Mathias}\ \emph {et~al.}(2016)\citenamefont
  {Mathias}, \citenamefont {Eich}, \citenamefont {Urbancic}, \citenamefont
  {Michael}, \citenamefont {Carr}, \citenamefont {Emmerich}, \citenamefont
  {Stange}, \citenamefont {Popmintchev}, \citenamefont {Rohwer}, \citenamefont
  {Wiesenmayer} \emph {et~al.}}]{mathias2016self}%
  \BibitemOpen
  \bibfield  {author} {\bibinfo {author} {\bibfnamefont {S.}~\bibnamefont
  {Mathias}}, \bibinfo {author} {\bibfnamefont {S.}~\bibnamefont {Eich}},
  \bibinfo {author} {\bibfnamefont {J.}~\bibnamefont {Urbancic}}, \bibinfo
  {author} {\bibfnamefont {S.}~\bibnamefont {Michael}}, \bibinfo {author}
  {\bibfnamefont {A.}~\bibnamefont {Carr}}, \bibinfo {author} {\bibfnamefont
  {S.}~\bibnamefont {Emmerich}}, \bibinfo {author} {\bibfnamefont
  {A.}~\bibnamefont {Stange}}, \bibinfo {author} {\bibfnamefont
  {T.}~\bibnamefont {Popmintchev}}, \bibinfo {author} {\bibfnamefont
  {T.}~\bibnamefont {Rohwer}}, \bibinfo {author} {\bibfnamefont
  {M.}~\bibnamefont {Wiesenmayer}}, \emph {et~al.},\ }\bibfield  {title}
  {\bibinfo {title} {Self-amplified photo-induced gap quenching in a correlated
  electron material},\ }\href@noop {} {\bibfield  {journal} {\bibinfo
  {journal} {Nature communications}\ }\textbf {\bibinfo {volume} {7}},\
  \bibinfo {pages} {12902} (\bibinfo {year} {2016})}\BibitemShut {NoStop}%
\bibitem [{\citenamefont {Yang}\ \emph {et~al.}(2020)\citenamefont {Yang},
  \citenamefont {Rohde}, \citenamefont {Hanff}, \citenamefont {Stange},
  \citenamefont {Xiong}, \citenamefont {Shi}, \citenamefont {Bauer},\ and\
  \citenamefont {Rossnagel}}]{yang2020bypassing}%
  \BibitemOpen
  \bibfield  {author} {\bibinfo {author} {\bibfnamefont {L.~X.}\ \bibnamefont
  {Yang}}, \bibinfo {author} {\bibfnamefont {G.}~\bibnamefont {Rohde}},
  \bibinfo {author} {\bibfnamefont {K.}~\bibnamefont {Hanff}}, \bibinfo
  {author} {\bibfnamefont {A.}~\bibnamefont {Stange}}, \bibinfo {author}
  {\bibfnamefont {R.}~\bibnamefont {Xiong}}, \bibinfo {author} {\bibfnamefont
  {J.}~\bibnamefont {Shi}}, \bibinfo {author} {\bibfnamefont {M.}~\bibnamefont
  {Bauer}},\ and\ \bibinfo {author} {\bibfnamefont {K.}~\bibnamefont
  {Rossnagel}},\ }\bibfield  {title} {\bibinfo {title} {{Bypassing the
  Structural Bottleneck in the Ultrafast Melting of Electronic Order}},\ }\href
  {https://doi.org/10.1103/PhysRevLett.125.266402} {\bibfield  {journal}
  {\bibinfo  {journal} {Phys. Rev. Lett.}\ }\textbf {\bibinfo {volume} {125}},\
  \bibinfo {pages} {266402} (\bibinfo {year} {2020})}\BibitemShut {NoStop}%
\bibitem [{\citenamefont {Tanabe}\ \emph {et~al.}(2018)\citenamefont {Tanabe},
  \citenamefont {Sugimoto},\ and\ \citenamefont
  {Ohta}}]{tanabe2018nonequilibrium}%
  \BibitemOpen
  \bibfield  {author} {\bibinfo {author} {\bibfnamefont {T.}~\bibnamefont
  {Tanabe}}, \bibinfo {author} {\bibfnamefont {K.}~\bibnamefont {Sugimoto}},\
  and\ \bibinfo {author} {\bibfnamefont {Y.}~\bibnamefont {Ohta}},\ }\bibfield
  {title} {\bibinfo {title} {Nonequilibrium dynamics in the pump-probe
  spectroscopy of excitonic insulators},\ }\href
  {https://doi.org/10.1103/PhysRevB.98.235127} {\bibfield  {journal} {\bibinfo
  {journal} {Phys. Rev. B}\ }\textbf {\bibinfo {volume} {98}},\ \bibinfo
  {pages} {235127} (\bibinfo {year} {2018})}\BibitemShut {NoStop}%
\bibitem [{\citenamefont {Burian}\ \emph {et~al.}(2021)\citenamefont {Burian},
  \citenamefont {Porer}, \citenamefont {Mardegan}, \citenamefont {Esposito},
  \citenamefont {Parchenko}, \citenamefont {Burganov}, \citenamefont {Gurung},
  \citenamefont {Ramakrishnan}, \citenamefont {Scagnoli}, \citenamefont {Ueda},
  \citenamefont {Francoual}, \citenamefont {Fabrizi}, \citenamefont {Tanaka},
  \citenamefont {Togashi}, \citenamefont {Kubota}, \citenamefont {Yabashi},
  \citenamefont {Rossnagel}, \citenamefont {Johnson},\ and\ \citenamefont
  {Staub}}]{burian2020structurally}%
  \BibitemOpen
  \bibfield  {author} {\bibinfo {author} {\bibfnamefont {M.}~\bibnamefont
  {Burian}}, \bibinfo {author} {\bibfnamefont {M.}~\bibnamefont {Porer}},
  \bibinfo {author} {\bibfnamefont {J.~R.~L.}\ \bibnamefont {Mardegan}},
  \bibinfo {author} {\bibfnamefont {V.}~\bibnamefont {Esposito}}, \bibinfo
  {author} {\bibfnamefont {S.}~\bibnamefont {Parchenko}}, \bibinfo {author}
  {\bibfnamefont {B.}~\bibnamefont {Burganov}}, \bibinfo {author}
  {\bibfnamefont {N.}~\bibnamefont {Gurung}}, \bibinfo {author} {\bibfnamefont
  {M.}~\bibnamefont {Ramakrishnan}}, \bibinfo {author} {\bibfnamefont
  {V.}~\bibnamefont {Scagnoli}}, \bibinfo {author} {\bibfnamefont
  {H.}~\bibnamefont {Ueda}}, \bibinfo {author} {\bibfnamefont {S.}~\bibnamefont
  {Francoual}}, \bibinfo {author} {\bibfnamefont {F.}~\bibnamefont {Fabrizi}},
  \bibinfo {author} {\bibfnamefont {Y.}~\bibnamefont {Tanaka}}, \bibinfo
  {author} {\bibfnamefont {T.}~\bibnamefont {Togashi}}, \bibinfo {author}
  {\bibfnamefont {Y.}~\bibnamefont {Kubota}}, \bibinfo {author} {\bibfnamefont
  {M.}~\bibnamefont {Yabashi}}, \bibinfo {author} {\bibfnamefont
  {K.}~\bibnamefont {Rossnagel}}, \bibinfo {author} {\bibfnamefont {S.~L.}\
  \bibnamefont {Johnson}},\ and\ \bibinfo {author} {\bibfnamefont
  {U.}~\bibnamefont {Staub}},\ }\bibfield  {title} {\bibinfo {title}
  {Structural involvement in the melting of the charge density wave in
  {1T}$-${TiSe$_{2}$}},\ }\href
  {https://doi.org/10.1103/PhysRevResearch.3.013128} {\bibfield  {journal}
  {\bibinfo  {journal} {Phys. Rev. Research}\ }\textbf {\bibinfo {volume}
  {3}},\ \bibinfo {pages} {013128} (\bibinfo {year} {2021})}\BibitemShut
  {NoStop}%
\bibitem [{\citenamefont {Ch\'avez-Cervantes}\ \emph
  {et~al.}(2019)\citenamefont {Ch\'avez-Cervantes}, \citenamefont {Topp},
  \citenamefont {Aeschlimann}, \citenamefont {Krause}, \citenamefont {Sato},
  \citenamefont {Sentef},\ and\ \citenamefont {Gierz}}]{chavez2019charge}%
  \BibitemOpen
  \bibfield  {author} {\bibinfo {author} {\bibfnamefont {M.}~\bibnamefont
  {Ch\'avez-Cervantes}}, \bibinfo {author} {\bibfnamefont {G.~E.}\ \bibnamefont
  {Topp}}, \bibinfo {author} {\bibfnamefont {S.}~\bibnamefont {Aeschlimann}},
  \bibinfo {author} {\bibfnamefont {R.}~\bibnamefont {Krause}}, \bibinfo
  {author} {\bibfnamefont {S.~A.}\ \bibnamefont {Sato}}, \bibinfo {author}
  {\bibfnamefont {M.~A.}\ \bibnamefont {Sentef}},\ and\ \bibinfo {author}
  {\bibfnamefont {I.}~\bibnamefont {Gierz}},\ }\bibfield  {title} {\bibinfo
  {title} {{Charge Density Wave Melting in One-Dimensional Wires with
  Femtosecond Subgap Excitation}},\ }\href
  {https://doi.org/10.1103/PhysRevLett.123.036405} {\bibfield  {journal}
  {\bibinfo  {journal} {Phys. Rev. Lett.}\ }\textbf {\bibinfo {volume} {123}},\
  \bibinfo {pages} {036405} (\bibinfo {year} {2019})}\BibitemShut {NoStop}%
\bibitem [{\citenamefont {Werdehausen}\ \emph {et~al.}(2018)\citenamefont
  {Werdehausen}, \citenamefont {Takayama}, \citenamefont {Albrecht},
  \citenamefont {Lu}, \citenamefont {Takagi},\ and\ \citenamefont
  {Kaiser}}]{werdehausen2018photo}%
  \BibitemOpen
  \bibfield  {author} {\bibinfo {author} {\bibfnamefont {D.}~\bibnamefont
  {Werdehausen}}, \bibinfo {author} {\bibfnamefont {T.}~\bibnamefont
  {Takayama}}, \bibinfo {author} {\bibfnamefont {G.}~\bibnamefont {Albrecht}},
  \bibinfo {author} {\bibfnamefont {Y.}~\bibnamefont {Lu}}, \bibinfo {author}
  {\bibfnamefont {H.}~\bibnamefont {Takagi}},\ and\ \bibinfo {author}
  {\bibfnamefont {S.}~\bibnamefont {Kaiser}},\ }\bibfield  {title} {\bibinfo
  {title} {Photo-excited dynamics in the excitonic insulator
  {Ta$_{2}$NiSe$_{5}$}},\ }\href@noop {} {\bibfield  {journal} {\bibinfo
  {journal} {Journal of Physics: Condensed Matter}\ }\textbf {\bibinfo {volume}
  {30}},\ \bibinfo {pages} {305602} (\bibinfo {year} {2018})}\BibitemShut
  {NoStop}%
\bibitem [{\citenamefont {Perfetto}\ \emph {et~al.}(2020)\citenamefont
  {Perfetto}, \citenamefont {Bianchi},\ and\ \citenamefont
  {Stefanucci}}]{perfetto2020time}%
  \BibitemOpen
  \bibfield  {author} {\bibinfo {author} {\bibfnamefont {E.}~\bibnamefont
  {Perfetto}}, \bibinfo {author} {\bibfnamefont {S.}~\bibnamefont {Bianchi}},\
  and\ \bibinfo {author} {\bibfnamefont {G.}~\bibnamefont {Stefanucci}},\
  }\bibfield  {title} {\bibinfo {title} {{Time-resolved ARPES spectra of
  nonequilibrium excitonic insulators: Revealing macroscopic coherence with
  ultrashort pulses}},\ }\href {https://doi.org/10.1103/PhysRevB.101.041201}
  {\bibfield  {journal} {\bibinfo  {journal} {Phys. Rev. B}\ }\textbf {\bibinfo
  {volume} {101}},\ \bibinfo {pages} {041201} (\bibinfo {year}
  {2020})}\BibitemShut {NoStop}%
\bibitem [{\citenamefont {K\"ohler}\ \emph {et~al.}(2020)\citenamefont
  {K\"ohler}, \citenamefont {Paeckel}, \citenamefont {Meyer},\ and\
  \citenamefont {Manmana}}]{kohler2020formation}%
  \BibitemOpen
  \bibfield  {author} {\bibinfo {author} {\bibfnamefont {T.}~\bibnamefont
  {K\"ohler}}, \bibinfo {author} {\bibfnamefont {S.}~\bibnamefont {Paeckel}},
  \bibinfo {author} {\bibfnamefont {C.}~\bibnamefont {Meyer}},\ and\ \bibinfo
  {author} {\bibfnamefont {S.~R.}\ \bibnamefont {Manmana}},\ }\bibfield
  {title} {\bibinfo {title} {Formation of spatial patterns by spin-selective
  excitations of interacting fermions},\ }\href
  {https://doi.org/10.1103/PhysRevB.102.235166} {\bibfield  {journal} {\bibinfo
   {journal} {Phys. Rev. B}\ }\textbf {\bibinfo {volume} {102}},\ \bibinfo
  {pages} {235166} (\bibinfo {year} {2020})}\BibitemShut {NoStop}%
\bibitem [{\citenamefont {Cohen-Stead}\ \emph {et~al.}(2019)\citenamefont
  {Cohen-Stead}, \citenamefont {Costa}, \citenamefont {Khatami},\ and\
  \citenamefont {Scalettar}}]{cohen2019effect}%
  \BibitemOpen
  \bibfield  {author} {\bibinfo {author} {\bibfnamefont {B.}~\bibnamefont
  {Cohen-Stead}}, \bibinfo {author} {\bibfnamefont {N.~C.}\ \bibnamefont
  {Costa}}, \bibinfo {author} {\bibfnamefont {E.}~\bibnamefont {Khatami}},\
  and\ \bibinfo {author} {\bibfnamefont {R.~T.}\ \bibnamefont {Scalettar}},\
  }\bibfield  {title} {\bibinfo {title} {Effect of strain on charge density
  wave order in the {Holstein} model},\ }\href
  {https://doi.org/10.1103/PhysRevB.100.045125} {\bibfield  {journal} {\bibinfo
   {journal} {Phys. Rev. B}\ }\textbf {\bibinfo {volume} {100}},\ \bibinfo
  {pages} {045125} (\bibinfo {year} {2019})}\BibitemShut {NoStop}%
\bibitem [{\citenamefont {Wei}\ \emph {et~al.}(2019)\citenamefont {Wei},
  \citenamefont {Tian}, \citenamefont {Yang}, \citenamefont {Yang},
  \citenamefont {Ma}, \citenamefont {Guo},\ and\ \citenamefont
  {Qiu}}]{wei2019ultrafast}%
  \BibitemOpen
  \bibfield  {author} {\bibinfo {author} {\bibfnamefont {R.}~\bibnamefont
  {Wei}}, \bibinfo {author} {\bibfnamefont {X.}~\bibnamefont {Tian}}, \bibinfo
  {author} {\bibfnamefont {L.}~\bibnamefont {Yang}}, \bibinfo {author}
  {\bibfnamefont {D.}~\bibnamefont {Yang}}, \bibinfo {author} {\bibfnamefont
  {Z.}~\bibnamefont {Ma}}, \bibinfo {author} {\bibfnamefont {H.}~\bibnamefont
  {Guo}},\ and\ \bibinfo {author} {\bibfnamefont {J.}~\bibnamefont {Qiu}},\
  }\bibfield  {title} {\bibinfo {title} {Ultrafast and large optical
  nonlinearity of a {TiSe$_{2}$} saturable absorber in the 2 $\mu$m wavelength
  region},\ }\href@noop {} {\bibfield  {journal} {\bibinfo  {journal}
  {Nanoscale}\ }\textbf {\bibinfo {volume} {11}},\ \bibinfo {pages} {22277}
  (\bibinfo {year} {2019})}\BibitemShut {NoStop}%
\bibitem [{\citenamefont {Karlsson}\ \emph {et~al.}(2018)\citenamefont
  {Karlsson}, \citenamefont {van Leeuwen}, \citenamefont {Perfetto},\ and\
  \citenamefont {Stefanucci}}]{karlsson2018generalized}%
  \BibitemOpen
  \bibfield  {author} {\bibinfo {author} {\bibfnamefont {D.}~\bibnamefont
  {Karlsson}}, \bibinfo {author} {\bibfnamefont {R.}~\bibnamefont {van
  Leeuwen}}, \bibinfo {author} {\bibfnamefont {E.}~\bibnamefont {Perfetto}},\
  and\ \bibinfo {author} {\bibfnamefont {G.}~\bibnamefont {Stefanucci}},\
  }\bibfield  {title} {\bibinfo {title} {The generalized {Kadanoff-Baym} ansatz
  with initial correlations},\ }\href
  {https://doi.org/10.1103/PhysRevB.98.115148} {\bibfield  {journal} {\bibinfo
  {journal} {Phys. Rev. B}\ }\textbf {\bibinfo {volume} {98}},\ \bibinfo
  {pages} {115148} (\bibinfo {year} {2018})}\BibitemShut {NoStop}%
\bibitem [{\citenamefont {Tuovinen}\ \emph {et~al.}(2019)\citenamefont
  {Tuovinen}, \citenamefont {Gole{\v{z}}}, \citenamefont {Sch{\"u}ler},
  \citenamefont {Werner}, \citenamefont {Eckstein},\ and\ \citenamefont
  {Sentef}}]{tuovinen2019adiabatic}%
  \BibitemOpen
  \bibfield  {author} {\bibinfo {author} {\bibfnamefont {R.}~\bibnamefont
  {Tuovinen}}, \bibinfo {author} {\bibfnamefont {D.}~\bibnamefont
  {Gole{\v{z}}}}, \bibinfo {author} {\bibfnamefont {M.}~\bibnamefont
  {Sch{\"u}ler}}, \bibinfo {author} {\bibfnamefont {P.}~\bibnamefont {Werner}},
  \bibinfo {author} {\bibfnamefont {M.}~\bibnamefont {Eckstein}},\ and\
  \bibinfo {author} {\bibfnamefont {M.~A.}\ \bibnamefont {Sentef}},\ }\bibfield
   {title} {\bibinfo {title} {{Adiabatic Preparation of a Correlated
  Symmetry-Broken Initial State with the Generalized {Kadanoff--Baym}
  Ansatz}},\ }\href@noop {} {\bibfield  {journal} {\bibinfo  {journal} {physica
  status solidi (b)}\ }\textbf {\bibinfo {volume} {256}},\ \bibinfo {pages}
  {1800469} (\bibinfo {year} {2019})}\BibitemShut {NoStop}%
\bibitem [{\citenamefont {Freericks}\ \emph {et~al.}(2017)\citenamefont
  {Freericks}, \citenamefont {Matveev}, \citenamefont {Shen}, \citenamefont
  {Shvaika},\ and\ \citenamefont {Devereaux}}]{freericks2017theoretical}%
  \BibitemOpen
  \bibfield  {author} {\bibinfo {author} {\bibfnamefont {J.}~\bibnamefont
  {Freericks}}, \bibinfo {author} {\bibfnamefont {O.}~\bibnamefont {Matveev}},
  \bibinfo {author} {\bibfnamefont {W.}~\bibnamefont {Shen}}, \bibinfo {author}
  {\bibfnamefont {A.}~\bibnamefont {Shvaika}},\ and\ \bibinfo {author}
  {\bibfnamefont {T.}~\bibnamefont {Devereaux}},\ }\bibfield  {title} {\bibinfo
  {title} {Theoretical description of pump/probe experiments in
  electron-mediated charge-density-wave insulators},\ }\href@noop {} {\bibfield
   {journal} {\bibinfo  {journal} {Physica Scripta}\ }\textbf {\bibinfo
  {volume} {92}},\ \bibinfo {pages} {034007} (\bibinfo {year}
  {2017})}\BibitemShut {NoStop}%
\bibitem [{\citenamefont {Lenk}\ and\ \citenamefont
  {Eckstein}(2020)}]{lenk2020collective}%
  \BibitemOpen
  \bibfield  {author} {\bibinfo {author} {\bibfnamefont {K.}~\bibnamefont
  {Lenk}}\ and\ \bibinfo {author} {\bibfnamefont {M.}~\bibnamefont
  {Eckstein}},\ }\bibfield  {title} {\bibinfo {title} {Collective excitations
  of the {$U$(1)}-symmetric exciton insulator in a cavity},\ }\href
  {https://doi.org/10.1103/PhysRevB.102.205129} {\bibfield  {journal} {\bibinfo
   {journal} {Phys. Rev. B}\ }\textbf {\bibinfo {volume} {102}},\ \bibinfo
  {pages} {205129} (\bibinfo {year} {2020})}\BibitemShut {NoStop}%
\bibitem [{\citenamefont {Zhao}\ \emph {et~al.}(2020)\citenamefont {Zhao},
  \citenamefont {Shang}, \citenamefont {Li}, \citenamefont {Liang},
  \citenamefont {Li},\ and\ \citenamefont {Zhang}}]{zhao2020strong}%
  \BibitemOpen
  \bibfield  {author} {\bibinfo {author} {\bibfnamefont {L.}~\bibnamefont
  {Zhao}}, \bibinfo {author} {\bibfnamefont {Q.}~\bibnamefont {Shang}},
  \bibinfo {author} {\bibfnamefont {M.}~\bibnamefont {Li}}, \bibinfo {author}
  {\bibfnamefont {Y.}~\bibnamefont {Liang}}, \bibinfo {author} {\bibfnamefont
  {C.}~\bibnamefont {Li}},\ and\ \bibinfo {author} {\bibfnamefont
  {Q.}~\bibnamefont {Zhang}},\ }\bibfield  {title} {\bibinfo {title} {Strong
  exciton-photon interaction and lasing of two-dimensional transition metal
  dichalcogenide semiconductors},\ }\href@noop {} {\bibfield  {journal}
  {\bibinfo  {journal} {Nano Research}\ }\textbf {\bibinfo {volume} {14(6)}},\
  \bibinfo {pages} {1937} (\bibinfo {year} {2020})}\BibitemShut {NoStop}%
\bibitem [{\citenamefont {Wang}\ \emph {et~al.}(2020)\citenamefont {Wang},
  \citenamefont {Liu}, \citenamefont {Pang}, \citenamefont {Song},
  \citenamefont {Tang}, \citenamefont {Ren},\ and\ \citenamefont
  {Xia}}]{wang2020threshold}%
  \BibitemOpen
  \bibfield  {author} {\bibinfo {author} {\bibfnamefont {J.}~\bibnamefont
  {Wang}}, \bibinfo {author} {\bibfnamefont {S.}~\bibnamefont {Liu}}, \bibinfo
  {author} {\bibfnamefont {J.}~\bibnamefont {Pang}}, \bibinfo {author}
  {\bibfnamefont {P.}~\bibnamefont {Song}}, \bibinfo {author} {\bibfnamefont
  {W.}~\bibnamefont {Tang}}, \bibinfo {author} {\bibfnamefont {Y.}~\bibnamefont
  {Ren}},\ and\ \bibinfo {author} {\bibfnamefont {W.}~\bibnamefont {Xia}},\
  }\bibfield  {title} {\bibinfo {title} {Threshold decrease and output-power
  improvement in dual-loss {Q-switched} laser based on a few-layer {WTe$_{2}$}
  saturable absorber},\ }\href@noop {} {\bibfield  {journal} {\bibinfo
  {journal} {Applied Physics Express}\ }\textbf {\bibinfo {volume} {13}},\
  \bibinfo {pages} {052004} (\bibinfo {year} {2020})}\BibitemShut {NoStop}%
\bibitem [{\citenamefont {Wu}\ \emph {et~al.}(2020)\citenamefont {Wu},
  \citenamefont {Ma}, \citenamefont {Yin}, \citenamefont {Ge}, \citenamefont
  {Zhang}, \citenamefont {Li}, \citenamefont {Zhang},\ and\ \citenamefont
  {Lin}}]{wutwo}%
  \BibitemOpen
  \bibfield  {author} {\bibinfo {author} {\bibfnamefont {J.}~\bibnamefont
  {Wu}}, \bibinfo {author} {\bibfnamefont {H.}~\bibnamefont {Ma}}, \bibinfo
  {author} {\bibfnamefont {P.}~\bibnamefont {Yin}}, \bibinfo {author}
  {\bibfnamefont {Y.}~\bibnamefont {Ge}}, \bibinfo {author} {\bibfnamefont
  {Y.}~\bibnamefont {Zhang}}, \bibinfo {author} {\bibfnamefont
  {L.}~\bibnamefont {Li}}, \bibinfo {author} {\bibfnamefont {H.}~\bibnamefont
  {Zhang}},\ and\ \bibinfo {author} {\bibfnamefont {H.}~\bibnamefont {Lin}},\
  }\bibfield  {title} {\bibinfo {title} {{Two-dimensional Materials for
  Integrated Photonics: Recent Advances and Future Challenges}},\ }\href@noop
  {} {\bibfield  {journal} {\bibinfo  {journal} {Small Science}\ }\textbf
  {\bibinfo {volume} {1}},\ \bibinfo {pages} {2000053} (\bibinfo {year}
  {2020})}\BibitemShut {NoStop}%
\bibitem [{\citenamefont {Perea-Caus{\'\i}n}\ \emph {et~al.}(2020)\citenamefont
  {Perea-Caus{\'\i}n}, \citenamefont {Brem},\ and\ \citenamefont
  {Malic}}]{perea2020microscopic}%
  \BibitemOpen
  \bibfield  {author} {\bibinfo {author} {\bibfnamefont {R.}~\bibnamefont
  {Perea-Caus{\'\i}n}}, \bibinfo {author} {\bibfnamefont {S.}~\bibnamefont
  {Brem}},\ and\ \bibinfo {author} {\bibfnamefont {E.}~\bibnamefont {Malic}},\
  }\bibfield  {title} {\bibinfo {title} {{Microscopic Modeling of Pump--Probe
  Spectroscopy and Population Inversion in Transition Metal Dichalcogenides}},\
  }\href@noop {} {\bibfield  {journal} {\bibinfo  {journal} {physica status
  solidi (b)}\ }\textbf {\bibinfo {volume} {257}},\ \bibinfo {pages} {2000223}
  (\bibinfo {year} {2020})}\BibitemShut {NoStop}%
\bibitem [{\citenamefont {Meckbach}\ \emph {et~al.}(2020)\citenamefont
  {Meckbach}, \citenamefont {Hader}, \citenamefont {Huttner}, \citenamefont
  {Neuhaus}, \citenamefont {Steiner}, \citenamefont {Stroucken}, \citenamefont
  {Moloney},\ and\ \citenamefont {Koch}}]{meckbach2020ultrafast}%
  \BibitemOpen
  \bibfield  {author} {\bibinfo {author} {\bibfnamefont {L.}~\bibnamefont
  {Meckbach}}, \bibinfo {author} {\bibfnamefont {J.}~\bibnamefont {Hader}},
  \bibinfo {author} {\bibfnamefont {U.}~\bibnamefont {Huttner}}, \bibinfo
  {author} {\bibfnamefont {J.}~\bibnamefont {Neuhaus}}, \bibinfo {author}
  {\bibfnamefont {J.~T.}\ \bibnamefont {Steiner}}, \bibinfo {author}
  {\bibfnamefont {T.}~\bibnamefont {Stroucken}}, \bibinfo {author}
  {\bibfnamefont {J.~V.}\ \bibnamefont {Moloney}},\ and\ \bibinfo {author}
  {\bibfnamefont {S.~W.}\ \bibnamefont {Koch}},\ }\bibfield  {title} {\bibinfo
  {title} {Ultrafast band-gap renormalization and build-up of optical gain in
  monolayer {MoTe$_{2}$}},\ }\href
  {https://doi.org/10.1103/PhysRevB.101.075401} {\bibfield  {journal} {\bibinfo
   {journal} {Phys. Rev. B}\ }\textbf {\bibinfo {volume} {101}},\ \bibinfo
  {pages} {075401} (\bibinfo {year} {2020})}\BibitemShut {NoStop}%
\bibitem [{\citenamefont {Chen}\ \emph
  {et~al.}(2015{\natexlab{b}})\citenamefont {Chen}, \citenamefont {Zhang},
  \citenamefont {Wu}, \citenamefont {Wang}, \citenamefont {Wang},\ and\
  \citenamefont {Chen}}]{chen2015q}%
  \BibitemOpen
  \bibfield  {author} {\bibinfo {author} {\bibfnamefont {B.}~\bibnamefont
  {Chen}}, \bibinfo {author} {\bibfnamefont {X.}~\bibnamefont {Zhang}},
  \bibinfo {author} {\bibfnamefont {K.}~\bibnamefont {Wu}}, \bibinfo {author}
  {\bibfnamefont {H.}~\bibnamefont {Wang}}, \bibinfo {author} {\bibfnamefont
  {J.}~\bibnamefont {Wang}},\ and\ \bibinfo {author} {\bibfnamefont
  {J.}~\bibnamefont {Chen}},\ }\bibfield  {title} {\bibinfo {title} {Q-switched
  fiber laser based on transition metal dichalcogenides {MoS$_{2}$},
  {MoSe$_{2}$}, {WS$_{2}$}, and {WSe$_{2}$}},\ }\href@noop {} {\bibfield
  {journal} {\bibinfo  {journal} {Optics express}\ }\textbf {\bibinfo {volume}
  {23}},\ \bibinfo {pages} {26723} (\bibinfo {year}
  {2015}{\natexlab{b}})}\BibitemShut {NoStop}%
\bibitem [{\citenamefont {Gies}\ and\ \citenamefont
  {Steinhoff}(2021)}]{gies2021atomically}%
  \BibitemOpen
  \bibfield  {author} {\bibinfo {author} {\bibfnamefont {C.}~\bibnamefont
  {Gies}}\ and\ \bibinfo {author} {\bibfnamefont {A.}~\bibnamefont
  {Steinhoff}},\ }\bibfield  {title} {\bibinfo {title} {{Atomically Thin van
  der Waals Semiconductors - A Theoretical Perspective}},\ }\href@noop {}
  {\bibfield  {journal} {\bibinfo  {journal} {Laser \& Photonics Reviews}\
  }\textbf {\bibinfo {volume} {15}},\ \bibinfo {pages} {2000482} (\bibinfo
  {year} {2021})}\BibitemShut {NoStop}%
\bibitem [{\citenamefont {Hahn}\ \emph {et~al.}(2021)\citenamefont {Hahn},
  \citenamefont {Kasprzak}, \citenamefont {Machnikowski}, \citenamefont
  {Kuhn},\ and\ \citenamefont {Wigger}}]{hahn2021influence}%
  \BibitemOpen
  \bibfield  {author} {\bibinfo {author} {\bibfnamefont {T.}~\bibnamefont
  {Hahn}}, \bibinfo {author} {\bibfnamefont {J.}~\bibnamefont {Kasprzak}},
  \bibinfo {author} {\bibfnamefont {P.}~\bibnamefont {Machnikowski}}, \bibinfo
  {author} {\bibfnamefont {T.}~\bibnamefont {Kuhn}},\ and\ \bibinfo {author}
  {\bibfnamefont {D.}~\bibnamefont {Wigger}},\ }\bibfield  {title} {\bibinfo
  {title} {{Influence of local fields on the dynamics of four-wave mixing
  signals from 2D semiconductor systems}},\ }\href@noop {} {\bibfield
  {journal} {\bibinfo  {journal} {New Journal of Physics}\ }\textbf {\bibinfo
  {volume} {23}},\ \bibinfo {pages} {023036} (\bibinfo {year}
  {2021})}\BibitemShut {NoStop}%
\bibitem [{\citenamefont {Lohof}\ \emph {et~al.}(2018)\citenamefont {Lohof},
  \citenamefont {Steinhoff}, \citenamefont {Florian}, \citenamefont {Lorke},
  \citenamefont {Erben}, \citenamefont {Jahnke},\ and\ \citenamefont
  {Gies}}]{lohof2018prospects}%
  \BibitemOpen
  \bibfield  {author} {\bibinfo {author} {\bibfnamefont {F.}~\bibnamefont
  {Lohof}}, \bibinfo {author} {\bibfnamefont {A.}~\bibnamefont {Steinhoff}},
  \bibinfo {author} {\bibfnamefont {M.}~\bibnamefont {Florian}}, \bibinfo
  {author} {\bibfnamefont {M.}~\bibnamefont {Lorke}}, \bibinfo {author}
  {\bibfnamefont {D.}~\bibnamefont {Erben}}, \bibinfo {author} {\bibfnamefont
  {F.}~\bibnamefont {Jahnke}},\ and\ \bibinfo {author} {\bibfnamefont
  {C.}~\bibnamefont {Gies}},\ }\bibfield  {title} {\bibinfo {title} {Prospects
  and limitations of transition metal dichalcogenide laser gain materials},\
  }\href@noop {} {\bibfield  {journal} {\bibinfo  {journal} {Nano letters}\
  }\textbf {\bibinfo {volume} {19}},\ \bibinfo {pages} {210} (\bibinfo {year}
  {2018})}\BibitemShut {NoStop}%
\bibitem [{\citenamefont {Li}\ \emph {et~al.}(2019)\citenamefont {Li},
  \citenamefont {Sun}, \citenamefont {Gan}, \citenamefont {Zhang},
  \citenamefont {Feng}, \citenamefont {Zhang},\ and\ \citenamefont
  {Ning}}]{li2019optical}%
  \BibitemOpen
  \bibfield  {author} {\bibinfo {author} {\bibfnamefont {Y.}~\bibnamefont
  {Li}}, \bibinfo {author} {\bibfnamefont {H.}~\bibnamefont {Sun}}, \bibinfo
  {author} {\bibfnamefont {L.}~\bibnamefont {Gan}}, \bibinfo {author}
  {\bibfnamefont {J.}~\bibnamefont {Zhang}}, \bibinfo {author} {\bibfnamefont
  {J.}~\bibnamefont {Feng}}, \bibinfo {author} {\bibfnamefont {D.}~\bibnamefont
  {Zhang}},\ and\ \bibinfo {author} {\bibfnamefont {C.-Z.}\ \bibnamefont
  {Ning}},\ }\bibfield  {title} {\bibinfo {title} {Optical properties and
  light-emission device applications of 2--d layered semiconductors},\
  }\href@noop {} {\bibfield  {journal} {\bibinfo  {journal} {Proceedings of the
  IEEE}\ }\textbf {\bibinfo {volume} {108}},\ \bibinfo {pages} {676} (\bibinfo
  {year} {2019})}\BibitemShut {NoStop}%
\bibitem [{\citenamefont {Coleman}\ \emph {et~al.}(1988)\citenamefont
  {Coleman}, \citenamefont {Giambattista}, \citenamefont {Hansma},
  \citenamefont {Johnson}, \citenamefont {McNairy},\ and\ \citenamefont
  {Slough}}]{coleman1988scanning}%
  \BibitemOpen
  \bibfield  {author} {\bibinfo {author} {\bibfnamefont {R.}~\bibnamefont
  {Coleman}}, \bibinfo {author} {\bibfnamefont {B.}~\bibnamefont
  {Giambattista}}, \bibinfo {author} {\bibfnamefont {P.}~\bibnamefont
  {Hansma}}, \bibinfo {author} {\bibfnamefont {A.}~\bibnamefont {Johnson}},
  \bibinfo {author} {\bibfnamefont {W.}~\bibnamefont {McNairy}},\ and\ \bibinfo
  {author} {\bibfnamefont {C.}~\bibnamefont {Slough}},\ }\bibfield  {title}
  {\bibinfo {title} {Scanning tunnelling microscopy of charge-density waves in
  transition metal chalcogenides},\ }\href@noop {} {\bibfield  {journal}
  {\bibinfo  {journal} {Advances in Physics}\ }\textbf {\bibinfo {volume}
  {37}},\ \bibinfo {pages} {559} (\bibinfo {year} {1988})}\BibitemShut
  {NoStop}%
\bibitem [{\citenamefont {Yoshida}\ and\ \citenamefont
  {Motizuki}(1980)}]{yoshida1980electron}%
  \BibitemOpen
  \bibfield  {author} {\bibinfo {author} {\bibfnamefont {Y.}~\bibnamefont
  {Yoshida}}\ and\ \bibinfo {author} {\bibfnamefont {K.}~\bibnamefont
  {Motizuki}},\ }\bibfield  {title} {\bibinfo {title} {Electron lattice
  interaction and lattice instability of {1T}$-${TiSe$_{2}$}},\ }\href@noop {}
  {\bibfield  {journal} {\bibinfo  {journal} {Journal of the Physical Society
  of Japan}\ }\textbf {\bibinfo {volume} {49}},\ \bibinfo {pages} {898}
  (\bibinfo {year} {1980})}\BibitemShut {NoStop}%
\bibitem [{\citenamefont {Mattuck}\ and\ \citenamefont
  {Johansson}(1968)}]{mattuck1968quantum}%
  \BibitemOpen
  \bibfield  {author} {\bibinfo {author} {\bibfnamefont {R.~D.}\ \bibnamefont
  {Mattuck}}\ and\ \bibinfo {author} {\bibfnamefont {B.}~\bibnamefont
  {Johansson}},\ }\bibfield  {title} {\bibinfo {title} {{Quantum field theory
  of phase transitions in Fermi systems}},\ }\href@noop {} {\bibfield
  {journal} {\bibinfo  {journal} {Advances in Physics}\ }\textbf {\bibinfo
  {volume} {17}},\ \bibinfo {pages} {509} (\bibinfo {year} {1968})}\BibitemShut
  {NoStop}%
\bibitem [{\citenamefont {Kohn}\ and\ \citenamefont
  {Sherrington}(1970)}]{kohn1970two}%
  \BibitemOpen
  \bibfield  {author} {\bibinfo {author} {\bibfnamefont {W.}~\bibnamefont
  {Kohn}}\ and\ \bibinfo {author} {\bibfnamefont {D.}~\bibnamefont
  {Sherrington}},\ }\bibfield  {title} {\bibinfo {title} {{Two Kinds of Bosons
  and Bose Condensates}},\ }\href {https://doi.org/10.1103/RevModPhys.42.1}
  {\bibfield  {journal} {\bibinfo  {journal} {Rev. Mod. Phys.}\ }\textbf
  {\bibinfo {volume} {42}},\ \bibinfo {pages} {1} (\bibinfo {year}
  {1970})}\BibitemShut {NoStop}%
\bibitem [{\citenamefont {Binder}\ and\ \citenamefont
  {Koch}(1995)}]{Binder-Koch}%
  \BibitemOpen
  \bibfield  {author} {\bibinfo {author} {\bibfnamefont {R.}~\bibnamefont
  {Binder}}\ and\ \bibinfo {author} {\bibfnamefont {S.}~\bibnamefont {Koch}},\
  }\bibfield  {title} {\bibinfo {title} {Nonequilibrium semiconductor
  dynamics},\ }\href {https://doi.org/10.1016/0079-6727(95)00001-S} {\bibfield
  {journal} {\bibinfo  {journal} {Progress in Quantum Electronics}\ }\textbf
  {\bibinfo {volume} {19}},\ \bibinfo {pages} {307} (\bibinfo {year}
  {1995})}\BibitemShut {NoStop}%
\bibitem [{\citenamefont {Michael}\ and\ \citenamefont
  {Schneider}(2019)}]{PhysRevB.100.035431}%
  \BibitemOpen
  \bibfield  {author} {\bibinfo {author} {\bibfnamefont {S.}~\bibnamefont
  {Michael}}\ and\ \bibinfo {author} {\bibfnamefont {H.~C.}\ \bibnamefont
  {Schneider}},\ }\bibfield  {title} {\bibinfo {title} {Impact ionization
  dynamics in small band-gap two-dimensional materials from a coherent phonon
  mechanism},\ }\href {https://doi.org/10.1103/PhysRevB.100.035431} {\bibfield
  {journal} {\bibinfo  {journal} {Phys. Rev. B}\ }\textbf {\bibinfo {volume}
  {100}},\ \bibinfo {pages} {035431} (\bibinfo {year} {2019})}\BibitemShut
  {NoStop}%
\bibitem [{\citenamefont {Rossi}\ and\ \citenamefont
  {Kuhn}(2002)}]{rossi2002theory}%
  \BibitemOpen
  \bibfield  {author} {\bibinfo {author} {\bibfnamefont {F.}~\bibnamefont
  {Rossi}}\ and\ \bibinfo {author} {\bibfnamefont {T.}~\bibnamefont {Kuhn}},\
  }\bibfield  {title} {\bibinfo {title} {Theory of ultrafast phenomena in
  photoexcited semiconductors},\ }\href
  {https://doi.org/10.1103/RevModPhys.74.895} {\bibfield  {journal} {\bibinfo
  {journal} {Rev. Mod. Phys.}\ }\textbf {\bibinfo {volume} {74}},\ \bibinfo
  {pages} {895} (\bibinfo {year} {2002})}\BibitemShut {NoStop}%
\bibitem [{\citenamefont {Snow}\ \emph {et~al.}(2003)\citenamefont {Snow},
  \citenamefont {Karpus}, \citenamefont {Cooper}, \citenamefont {Kidd},\ and\
  \citenamefont {Chiang}}]{PhysRevLett.91.136402}%
  \BibitemOpen
  \bibfield  {author} {\bibinfo {author} {\bibfnamefont {C.~S.}\ \bibnamefont
  {Snow}}, \bibinfo {author} {\bibfnamefont {J.~F.}\ \bibnamefont {Karpus}},
  \bibinfo {author} {\bibfnamefont {S.~L.}\ \bibnamefont {Cooper}}, \bibinfo
  {author} {\bibfnamefont {T.~E.}\ \bibnamefont {Kidd}},\ and\ \bibinfo
  {author} {\bibfnamefont {T.-C.}\ \bibnamefont {Chiang}},\ }\bibfield  {title}
  {\bibinfo {title} {{Quantum Melting of the Charge-Density-Wave State in
  {1T}$-${TiSe$_{2}$}}},\ }\href
  {https://doi.org/10.1103/PhysRevLett.91.136402} {\bibfield  {journal}
  {\bibinfo  {journal} {Phys. Rev. Lett.}\ }\textbf {\bibinfo {volume} {91}},\
  \bibinfo {pages} {136402} (\bibinfo {year} {2003})}\BibitemShut {NoStop}%
\bibitem [{\citenamefont {Boykin}\ and\ \citenamefont
  {Klimeck}(2005)}]{PhysRevB.71.115215}%
  \BibitemOpen
  \bibfield  {author} {\bibinfo {author} {\bibfnamefont {T.~B.}\ \bibnamefont
  {Boykin}}\ and\ \bibinfo {author} {\bibfnamefont {G.}~\bibnamefont
  {Klimeck}},\ }\bibfield  {title} {\bibinfo {title} {Practical application of
  zone-folding concepts in tight-binding calculations},\ }\href
  {https://doi.org/10.1103/PhysRevB.71.115215} {\bibfield  {journal} {\bibinfo
  {journal} {Phys. Rev. B}\ }\textbf {\bibinfo {volume} {71}},\ \bibinfo
  {pages} {115215} (\bibinfo {year} {2005})}\BibitemShut {NoStop}%
\bibitem [{\citenamefont {Boykin}\ \emph {et~al.}(2007)\citenamefont {Boykin},
  \citenamefont {Kharche}, \citenamefont {Klimeck},\ and\ \citenamefont
  {Korkusinski}}]{boykin2007approximate}%
  \BibitemOpen
  \bibfield  {author} {\bibinfo {author} {\bibfnamefont {T.~B.}\ \bibnamefont
  {Boykin}}, \bibinfo {author} {\bibfnamefont {N.}~\bibnamefont {Kharche}},
  \bibinfo {author} {\bibfnamefont {G.}~\bibnamefont {Klimeck}},\ and\ \bibinfo
  {author} {\bibfnamefont {M.}~\bibnamefont {Korkusinski}},\ }\bibfield
  {title} {\bibinfo {title} {Approximate bandstructures of semiconductor alloys
  from tight-binding supercell calculations},\ }\href@noop {} {\bibfield
  {journal} {\bibinfo  {journal} {Journal of Physics: Condensed Matter}\
  }\textbf {\bibinfo {volume} {19}},\ \bibinfo {pages} {036203} (\bibinfo
  {year} {2007})}\BibitemShut {NoStop}%
\bibitem [{\citenamefont {Medeiros}\ \emph {et~al.}(2015)\citenamefont
  {Medeiros}, \citenamefont {Tsirkin}, \citenamefont {Stafstr\"om},\ and\
  \citenamefont {Bj\"ork}}]{medeiros2015unfolding}%
  \BibitemOpen
  \bibfield  {author} {\bibinfo {author} {\bibfnamefont {P.~V.~C.}\
  \bibnamefont {Medeiros}}, \bibinfo {author} {\bibfnamefont {S.~S.}\
  \bibnamefont {Tsirkin}}, \bibinfo {author} {\bibfnamefont {S.}~\bibnamefont
  {Stafstr\"om}},\ and\ \bibinfo {author} {\bibfnamefont {J.}~\bibnamefont
  {Bj\"ork}},\ }\bibfield  {title} {\bibinfo {title} {Unfolding spinor wave
  functions and expectation values of general operators: Introducing the
  unfolding-density operator},\ }\href
  {https://doi.org/10.1103/PhysRevB.91.041116} {\bibfield  {journal} {\bibinfo
  {journal} {Phys. Rev. B}\ }\textbf {\bibinfo {volume} {91}},\ \bibinfo
  {pages} {041116} (\bibinfo {year} {2015})}\BibitemShut {NoStop}%
\bibitem [{\citenamefont {Medeiros}\ \emph {et~al.}(2014)\citenamefont
  {Medeiros}, \citenamefont {Stafstr\"om},\ and\ \citenamefont
  {Bj\"ork}}]{medeiros2014effects}%
  \BibitemOpen
  \bibfield  {author} {\bibinfo {author} {\bibfnamefont {P.~V.~C.}\
  \bibnamefont {Medeiros}}, \bibinfo {author} {\bibfnamefont {S.}~\bibnamefont
  {Stafstr\"om}},\ and\ \bibinfo {author} {\bibfnamefont {J.}~\bibnamefont
  {Bj\"ork}},\ }\bibfield  {title} {\bibinfo {title} {Effects of extrinsic and
  intrinsic perturbations on the electronic structure of graphene: {Retaining}
  an effective primitive cell band structure by band unfolding},\ }\href
  {https://doi.org/10.1103/PhysRevB.89.041407} {\bibfield  {journal} {\bibinfo
  {journal} {Phys. Rev. B}\ }\textbf {\bibinfo {volume} {89}},\ \bibinfo
  {pages} {041407} (\bibinfo {year} {2014})}\BibitemShut {NoStop}%
\bibitem [{\citenamefont {Popescu}\ and\ \citenamefont
  {Zunger}(2012)}]{popescu2012extracting}%
  \BibitemOpen
  \bibfield  {author} {\bibinfo {author} {\bibfnamefont {V.}~\bibnamefont
  {Popescu}}\ and\ \bibinfo {author} {\bibfnamefont {A.}~\bibnamefont
  {Zunger}},\ }\bibfield  {title} {\bibinfo {title} {{Extracting $E$ versus
  $\vec{k}$ effective band structure from supercell calculations on alloys and
  impurities}},\ }\href {https://doi.org/10.1103/PhysRevB.85.085201} {\bibfield
   {journal} {\bibinfo  {journal} {Phys. Rev. B}\ }\textbf {\bibinfo {volume}
  {85}},\ \bibinfo {pages} {085201} (\bibinfo {year} {2012})}\BibitemShut
  {NoStop}%
\bibitem [{\citenamefont {Lucovsky}\ \emph {et~al.}(1976)\citenamefont
  {Lucovsky}, \citenamefont {Liang}, \citenamefont {White},\ and\ \citenamefont
  {Pisharody}}]{lucovsky1976reflectivity}%
  \BibitemOpen
  \bibfield  {author} {\bibinfo {author} {\bibfnamefont {G.}~\bibnamefont
  {Lucovsky}}, \bibinfo {author} {\bibfnamefont {W.}~\bibnamefont {Liang}},
  \bibinfo {author} {\bibfnamefont {R.}~\bibnamefont {White}},\ and\ \bibinfo
  {author} {\bibfnamefont {K.}~\bibnamefont {Pisharody}},\ }\bibfield  {title}
  {\bibinfo {title} {Reflectivity studies of {Ti-} and {Ta-}dichalcogenides:
  {Phonons}},\ }\href@noop {} {\bibfield  {journal} {\bibinfo  {journal} {Solid
  State Communications}\ }\textbf {\bibinfo {volume} {19}},\ \bibinfo {pages}
  {303} (\bibinfo {year} {1976})}\BibitemShut {NoStop}%
\bibitem [{\citenamefont {Mahan}(1990)}]{mahan1990many}%
  \BibitemOpen
  \bibfield  {author} {\bibinfo {author} {\bibfnamefont {G.~D.}\ \bibnamefont
  {Mahan}},\ }\href@noop {} {\emph {\bibinfo {title} {Many-particle physics}}}\
  (\bibinfo  {publisher} {Springer Science \& Business Media},\ \bibinfo {year}
  {1990})\BibitemShut {NoStop}%
\bibitem [{\citenamefont {Chow}\ and\ \citenamefont
  {Koch}(1999)}]{chow1999semiconductor}%
  \BibitemOpen
  \bibfield  {author} {\bibinfo {author} {\bibfnamefont {W.~W.}\ \bibnamefont
  {Chow}}\ and\ \bibinfo {author} {\bibfnamefont {S.~W.}\ \bibnamefont
  {Koch}},\ }\href@noop {} {\emph {\bibinfo {title} {{Semiconductor-Laser
  Fundamentals: Physics of the Gain Materials}}}}\ (\bibinfo  {publisher}
  {Springer Science \& Business Media},\ \bibinfo {year} {1999})\BibitemShut
  {NoStop}%
\bibitem [{\citenamefont {Slater}\ and\ \citenamefont
  {Koster}(1954)}]{slater1954simplified}%
  \BibitemOpen
  \bibfield  {author} {\bibinfo {author} {\bibfnamefont {J.~C.}\ \bibnamefont
  {Slater}}\ and\ \bibinfo {author} {\bibfnamefont {G.~F.}\ \bibnamefont
  {Koster}},\ }\bibfield  {title} {\bibinfo {title} {{Simplified {LCAO} Method
  for the Periodic Potential Problem}},\ }\href
  {https://doi.org/10.1103/PhysRev.94.1498} {\bibfield  {journal} {\bibinfo
  {journal} {Phys. Rev.}\ }\textbf {\bibinfo {volume} {94}},\ \bibinfo {pages}
  {1498} (\bibinfo {year} {1954})}\BibitemShut {NoStop}%
\end{thebibliography}%

\end{document}